\documentclass[]{aastex631}
\turnoffeditone

\usepackage{comment}
\usepackage{chngcntr}
\usepackage{multirow}
\received{September 25, 2024}
\revised{March 6, 2025}
\accepted{\today}

\begin{document}

\title{Investigating the emission mechanism in the spatially-resolved jet of two $z \approx 3$ radio-loud quasars}

\author[0000-0002-4423-4584]{Jaya Maithil}
\affiliation{Center for Astrophysics $\vert$ Harvard \& Smithsonian, 60 Garden St, Cambridge, MA 02138, USA}

\author[0000-0001-8252-4753]{Daniel A. Schwartz}
\affiliation{Center for Astrophysics $\vert$ Harvard \& Smithsonian, 60 Garden St, Cambridge, MA 02138, USA}

\author[0000-0002-0905-7375]{Aneta Siemiginowska} 
\affiliation{Center for Astrophysics $\vert$ Harvard \& Smithsonian, 60 Garden St, Cambridge, MA 02138, USA} 

\author[0000-0003-3203-1613]{Preeti Kharb}
\affiliation{National Centre for Radio Astrophysics, Pune, Maharashtra 411007, India}

\author[0000-0002-1516-0336]{Diana M. Worrall} 
\affiliation{H.H. Wills Physics Laboratory, University of Bristol, Tyndall Ave., Bristol BS8 1TL, UK} 

\author[0000-0002-8960-2942]{John F. C. Wardle} 
\affiliation{Physics Department, Brandeis University, Waltham, MA 02454, USA} 

\author[0000-0003-0216-8053]{Giulia Migliori} 
\affiliation{INAF-Institute of Radio Astronomy, Bologna, Via Gobetti 101, I-40129 Bologna, Italy} 

\author[0000-0002-4377-0174]{Chi C. Cheung} 
\affiliation{Space Science Division, Naval Research Laboratory, Washington, DC 20375, USA} 

\author[0000-0002-4900-928X]{Bradford Snios} 
\affiliation{Center for Astrophysics $\vert$ Harvard \& Smithsonian, 60 Garden St, Cambridge, MA 02138, USA} 

\author{Doug B. Gobeille} 
\affiliation{Physics Department, University of Rhode Island, Kingston, RI 02881, USA}

\author[0000-0002-6492-1293]{Herman L. Marshall} 
\affiliation{Kavli Institute for Astrophysics and Space Research, Massachusetts Institute of Technology, Cambridge, MA 02139, USA}

\author[0000-0002-1858-277X]{Mark Birkinshaw}
\altaffiliation{Deceased}
\affiliation{H.H. Wills Physics Laboratory, University of Bristol, Tyndall Ave., Bristol BS8 1TL, UK}



\begin{abstract}

This study focuses on high-redshift, $z >3$, quasars where resolved X-ray jets remain underexplored in comparison to nearby sources. Building upon previous work, we identify and confirm extended kpc-scale jets emission in two quasars  (J1405+0415, $z =$ 3.215 ; J1610+1811, $z =$ 3.122) through meticulous analysis of Chandra X-ray data. To deepen our understanding, high-resolution radio follow-up observations were conducted to constrain relativistic parameters, providing valuable insights into the enthalpy flux of these high-redshift AGN jets. The investigation specifically aims to test the X-ray emission mechanism in these quasars by exploring the inverse Compton scattering of cosmic microwave background (IC/CMB) photons by synchrotron-emitting electrons. Our novel method uses a prior to make a Bayesian estimate of the unknown angle of the jet to our line of sight, thus breaking the usual degeneracy of the bulk Lorentz factor and the Doppler beaming factor. 

\end{abstract}



\section{Introduction} \label{sec:intro}
Supermassive black-hole (SMBH) masses are correlated with the bulge masses of their host galaxies \citep{Kormendy1995}. This implies that black-hole formation and evolution are connected to how galaxies are formed and evolve. This connection is likely due to the feedback mediated by the gas that is accreted onto the SMBH and the energy that is expelled (as shown in \citealt{Fabian2012} and \citealt{Heckman+2014}). Therefore, the study of the power of these jetted outflows at various redshifts is crucial in understanding the formation and growth of galaxies and galaxy clusters. In powerful active galactic nuclei (AGN) at $z \leq 2$, mechanical energy outflow through winds or collimated jets can be as strong as, or even stronger than, the Eddington luminosity \citep[][]{Gofford+2015,Schwartz+2006}. These outflows can also exceed the total radiation output of the SMBH \citep[e.g.,][]{Ghisellini+2014}. Models of accretion flows need to be able to incorporate those powers. To address feedback and evolution, numerous observational and theoretical studies have been carried out \citep{Fabian2012, Hardcastle+Croston2020}. 
If the X-ray jets are directly detected and assumed to be produced by inverse Compton (IC) scattering of the cosmic microwave background (CMB) radiation, as proposed by \cite{Tavecchio2000} and \cite{Celotti2001}, this enables estimates of the enthalpy flux carried by the relativistic jet, in contrast to the usual deduction from age estimates and energy content of lobes \citep{Odea2009}, or proxies using the total radio luminosity \citep{Daly2012}.

The dominant mechanism for producing high-energy emission at X-ray energies ($\sim ~ 1$ keV) depends on the energy density of the target field. For jets with a bulk Lorentz factor, $\Gamma \sim 4$ at $z > 3$, the CMB energy density dominates the synchrotron energy density when the magnetic field strength is below several hundred $\mu G$ \citep[][]{Schwartz+2019}. Hence the dominant mechanism is inverse Compton scattering of the CMB photons by the relativistic electrons. This is due to the enhancement of the CMB energy density by $(1+z)^4$. If the magnetic field is higher, synchrotron emission becomes the dominant mechanism.
When bulk relativistic motion is present, estimating the minimum-energy magnetic field strength, $B$, and the relativistic electron density using radio data alone becomes challenging due to uncertain beaming parameters. However, X-ray observations can help constrain these relativistic parameters by using the measured X-ray surface brightness and the known energy density of the CMB. Under common assumptions, this approach allows for estimates of $B$ and the Doppler factor, $\delta = 1/(\Gamma(1 - \beta \cos \theta))$, which can then be used in the formalism of \cite{Bicknell1994} to calculate the enthalpy flux (see \citealt{Schwartz+2006} for examples).

While previous Chandra observations have successfully resolved X-ray jets up to $z=6.1$ \citep{Ighina2022,Migliori2023}, X-ray jets at $z>3$ remain vastly undersampled compared to nearby sources \citep{Worrall+2020}. This paucity poses a challenge in determining population trends of jets over a broad redshift range \citep[e.g.][]{McKeough2016}, making each detection of a high-redshift jet a significant discovery. \cite{Simionescu+2016} serendipitously discovered the first dramatic example of an X-ray jet at $z=2.5$ resolved on arcsecond scale without a corresponding radio jet detection, which, as we will show in this paper, is expected from a scenario of dominant IC/CMB emission. This remarkable finding prompted efforts to identify more high-redshift jet candidates. 

\cite{Schwartz+2019, Schwartz2020} conducted an exploratory Chandra survey for X-ray jets associated with high-redshift radio-loud quasars. Their survey targeted 16 quasars with redshifts z $>$ 3, selected from the complete survey of \citet{Gobeille2011,  Gobeille2014}. The parent sample of quasars was chosen based on their total flux density ($>$70 mJy at either 1.4 or 5 GHz) and the presence of spectroscopically measured redshifts. \cite{Schwartz+2019, Schwartz2020} identified two candidates: quasars J1405+0415 (B1402+044, $z=3.215$) and J1610+1811 (B1607+183, $z=3.122$) that showed X-ray jets with an absence of radio jets. 
\citet{Snios2021} presented the deeper Chandra X-ray observations of the two targets and confirmed the presence of an extended jet with high significance.

Our work delves into high-resolution radio follow-up observations of these two quasars, aiming to test the mechanism of X-ray emission as inverse Compton scattering of CMB photons by synchrotron emitting electrons. In Section 2, we present the analysis of new 6 GHz VLA radio data, along with archival Chandra X-ray data. Section 3 details the flux densities and spectral indices of various components from the new 6 GHz images. It includes a spectroscopic analysis of the X-ray core, examines the morphology of the extended X-ray emission, and provides spatial extent and flux densities of the jet in both radio and X-rays. Although we used the same \textit{Chandra} observations as \cite{Snios+2022}, we reanalyzed the quasar core X-ray emission for two primary reasons. First, we employed more recent calibration files to incorporate the latest instrument corrections. Second, our investigation of the extended X-ray emission required simulating the quasar core’s X-ray emission, which necessitated an accurately determined core spectrum as input. Consequently, we performed an independent spectral analysis of the core X-ray emission to ensure accurate determination of the extended components. Section 4 outlines the assumptions and details of the IC/CMB model, presents constraints on relativistic and jet parameters, and introduces a novel method for estimating the jet’s angle to our line of sight. It also compares these results with other models and techniques. Section 5 summarizes and concludes the findings. Appendices A and B provide details on selecting blur parameters for simulating X-ray cores and comparing simulated PSFs with observed images. Appendix \ref{Appendix_Historic} presents the detailed analysis and results of historical VLA radio continuum and polarization observations at 1.5 GHz and 4.9 GHz, including flux measurements, spectral indices, and polarization properties.

Throughout this paper, the spectral index $\alpha$ is defined such that flux density at a frequency $\nu$, $S_\nu\propto\nu^{-\alpha}$. The X-ray photon index $\Gamma$ is defined by the relationship between the photon number density ($N(E)$) and the photon energy ($E$), such that $N(E) \propto E^{-\Gamma}$.  Photon index relates to spectral index by the relation $\Gamma = \alpha + 1$. We have adopted a flat cosmology with H$_0=67.8$~km~s$^{-1}$~Mpc$^{-1}$, $\Omega_\mathrm{mat}=0.308$, $\Omega_\mathrm{vac} = 0.692$ \citep{Planck2016}. All uncertainties in the measured quantities are reported as 1$\sigma$ unless otherwise specified.

\section{Observations \& Data Analysis}
\subsection{New Radio Observations}
Our radio observations were conducted using the Karl G. Jansky Very Large Array (VLA) at the C band (4-8 GHz) with the A configuration to achieve a resolution of 0.3$\arcsec$. The VLA project ID for these observations is SC220006. We utilized 3-bit digital samplers that are appropriate for C-band wideband observations with basebands of 2 × 2 GHz and a total bandwidth of 4 GHz. To ensure adequate coverage of the UV plane, we allocated 3 hours for J1405+0415 and 2 hours for J1610+1811. 
Our dynamical observations were conducted in two scheduling blocks (SBs) on July 11, 2023, SBID 43929403 for J1405+0415 and SBID 44169909 for J1610+1811. 

\begin{deluxetable}{crccccc}[h!]
\tablecaption{ Observation log for the radio data. \label{tab:table1}}
\tablehead{
\colhead{Object name} & \colhead{Telescope} & \colhead{Date} & \colhead{Project Id} & \colhead{Frequency} & \colhead{Time}  \\\colhead{} & \colhead{} & \colhead{} & \colhead{} & \colhead{(GHz)} & \colhead{(min)} \\
\colhead{(1)} & \colhead{(2)} & \colhead{(3)} & \colhead{(4)} & \colhead{(5)}  & \colhead{(6)}
}
\startdata
J1405+0415	& {VLA-A}  & {2023 Jul 11}   & {SC220006}    & {6} & {134}    \\
{ }	        & {VLA-A}  & {1987 Aug 16}   & {AB449}       & {1.49} & {15} \\
{ }	        & {VLA-A}  & {1984 Dec 10}   & {AW122}       & {4.86} & {12.5} \\
\hline
J1610+1811	& {VLA-A} & {2023 Jul 11}   & {SC220006}  & {6}  & {85}   \\
{ }	        & {VLA-A} & {1987 Aug 16}   & {AB449}     & {1.49} & {8.8} \\
{ }	        & {VLA-A} & {1995 Sep 03}   & {AB750}     & {4.86}  & {18.2}\\
\enddata
\tablecomments{(1) Source name. (2) Name of the telescope and array configuration for VLA data. (3) Date of observation. (4) Project Id. (5)  Frequency of observation. (6) On-Source Time}
\end{deluxetable}

To mitigate the effects of radio frequency interference (RFI) for near equatorial source J1405+0415, we created a new instrument setup (NRAO-C32f2rfi) by centering the 2 GHz-wide basebands from the 3-bit sampler at 5.25 GHz and 7.2 GHz. Following the guideline for observing in the presence of strong RFI, our first block started with a 4.5~min dummy scan using J1415+1320 (outside the geosynchronous satellite zone, also the phase calibrator) as a source with hardware setup X16f2A (X-band, 8-bit). We then followed it up with a 1m attenuator setup scan and a 10s requantization setup scan on the same source but with the new hardware setup NRAO-C32f2rfi.
The hardware setup was fixed to the NRAO-C32f2rfi for the rest of the scans. For the second block for J1610+1811, we used X16f2A (X-band 8-bit) for a dummy scan of 3~min on J1608+1029 (which is also the phase calibrator), followed by a 1~min Attenuator setup scan and a 10s Requantization Setup scan on the same source but with the C32f2A hardware setup (2$\times$2 GHz basebands centered at 5 and 7 GHz). For both blocks, we used 3C 286 as our flux calibrator, and each block included a loop of a phase calibrator and science target, spending 1.5-2 min on the phase calibrator each time. The total on-source time amounts to 2 hours and 14 minutes for J1405+0415 and 1 hour and 25 minutes for J1610+1811. 

The dataset underwent automatic preprocessing through the VLA Common Astronomy Software Applications \citep[CASA][]{CASA} calibration pipeline using CASA version 6.4.1. The pipeline is designed for Stokes I continuum data and carries out standard flagging and calibration procedures. For further data reduction and image synthesis after splitting we used CASA  version 6.5.2.26. We utilized the SPLIT task to separate the corrected data from the main data set. The imaging process involved using tclean in the mt-mfs (multi-scale) multi-frequency synthesis deconvolution algorithm (Rau \& Cornwell 2011). The mt-mfs algorithm utilizes a Taylor series expansion to model the spectrum of each flux component (pixel) about the reference frequency. For imaging J1610+1811, we use mt-mfs with 2 Taylor coefficients, applied the ``Briggs" weighting scheme with $\rm robust=0.1$, and an additional tapering of visibilities \citep{Briggs1995, Briggs+1999}. The final outputs of the cleaning process included a TT0 map that highlights specific intensities at 6 GHz, a corresponding TT1 and $\alpha$ map, as well as an $\Delta \alpha$ map that describes the empirical error estimate based on the errors of the TT0 and TT1 residual images. The uncertainties in our spectral index measurements are derived directly from the $\Delta \alpha$ maps. The resulting images were improved through multiple rounds of phase and amplitude self-calibration. Using mt-mfs with 4 Taylor coefficients better models the spectrum of J1405+0415 and helps reduce the artifacts in the final intensity image. The final maps presented in this paper are at a central frequency of 6 GHz and covers 4.0-6.0 GHz and 6.0-8.0 GHz bands.

\subsection{Chandra X-ray Observations and Data Reduction}
J1405+0415 and J1610+1811 were observed with the Chandra Advanced CCD Imaging Spectrometer (ACIS) S-3 during Cycle 22 in timed exposure operation mode with faint telemetry. The observations aimed to reduce pile-up issues arising from the AGN core by utilizing the 1/4 subarray type and preventing overlap between the extended radio feature and the ACIS readout streak by introducing roll direction constraints. Table~\ref{tab:table2}  presents the observation log for these targets. We analyzed the X-ray data using the Chandra Interactive Analysis of Observations software (CIAO) version 4.15. For each observation, we reprocessed the data using the \texttt{chandra\_repro} script with CALDB version 4.10.2, thereby generating level 2 event files. We determined the centroid position of the quasar core using \texttt{dmstat} with a circular region of radius $1\farcs 2$ the center of the X-ray source and aligned it with the radio position through the \texttt{wcs\_update} routine. This gives us an image alignment at subpixel accuracy. So, the X-ray cores are centered at 14:05:01.10, +04:15:35.80, and 16:10:05.30, +18:11:43.50 for J1405+0415 and J1610+1811, respectively.

\section{Results}
\subsection{New VLA 6 GHz images}
\begin{figure}[h!]
\includegraphics[width=0.5\columnwidth,trim=0cm 0cm 0.cm 0.cm,clip]{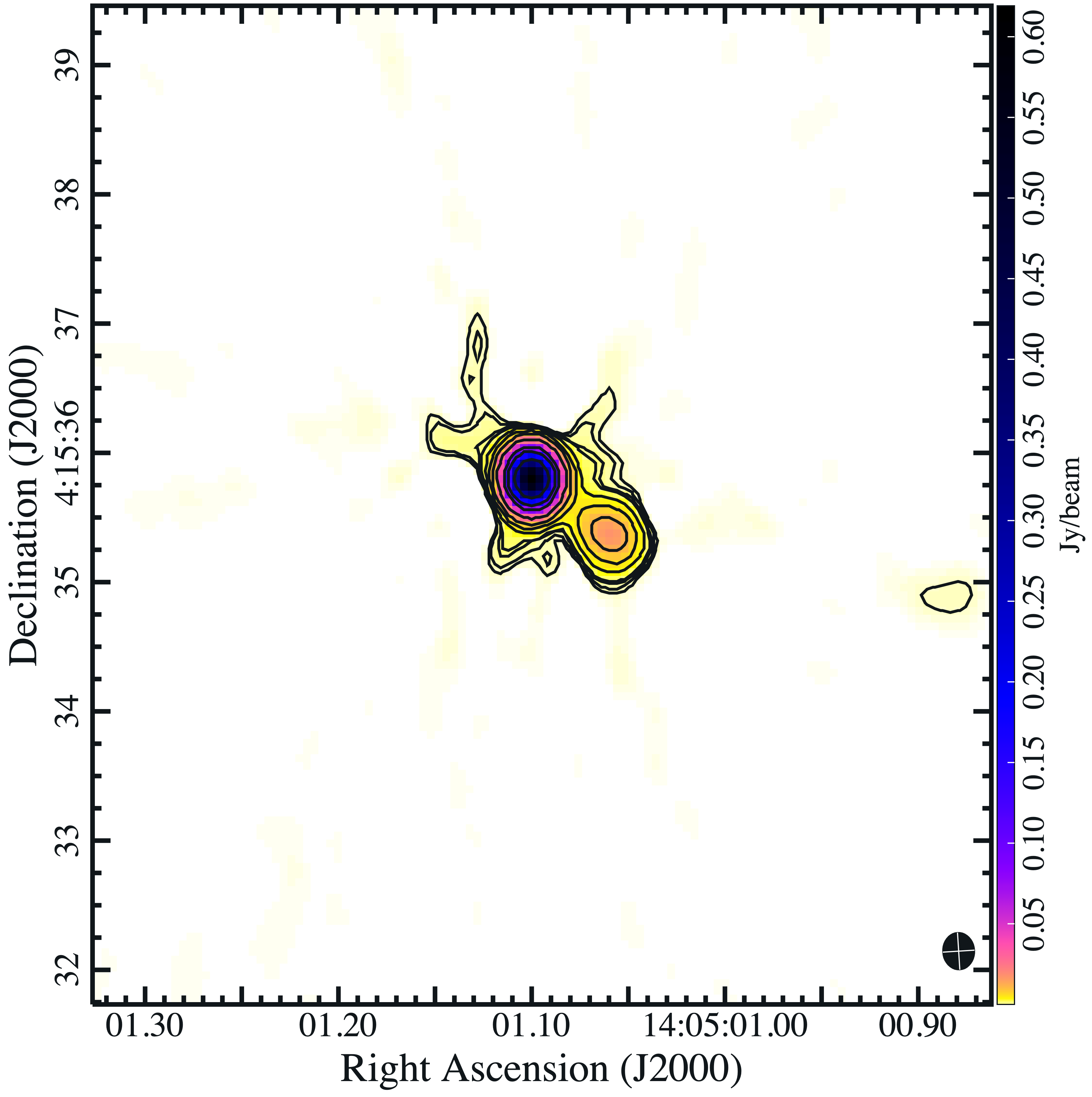}
\includegraphics[width=0.5\columnwidth,trim=0.cm 0.cm 0.cm 0.cm,clip]{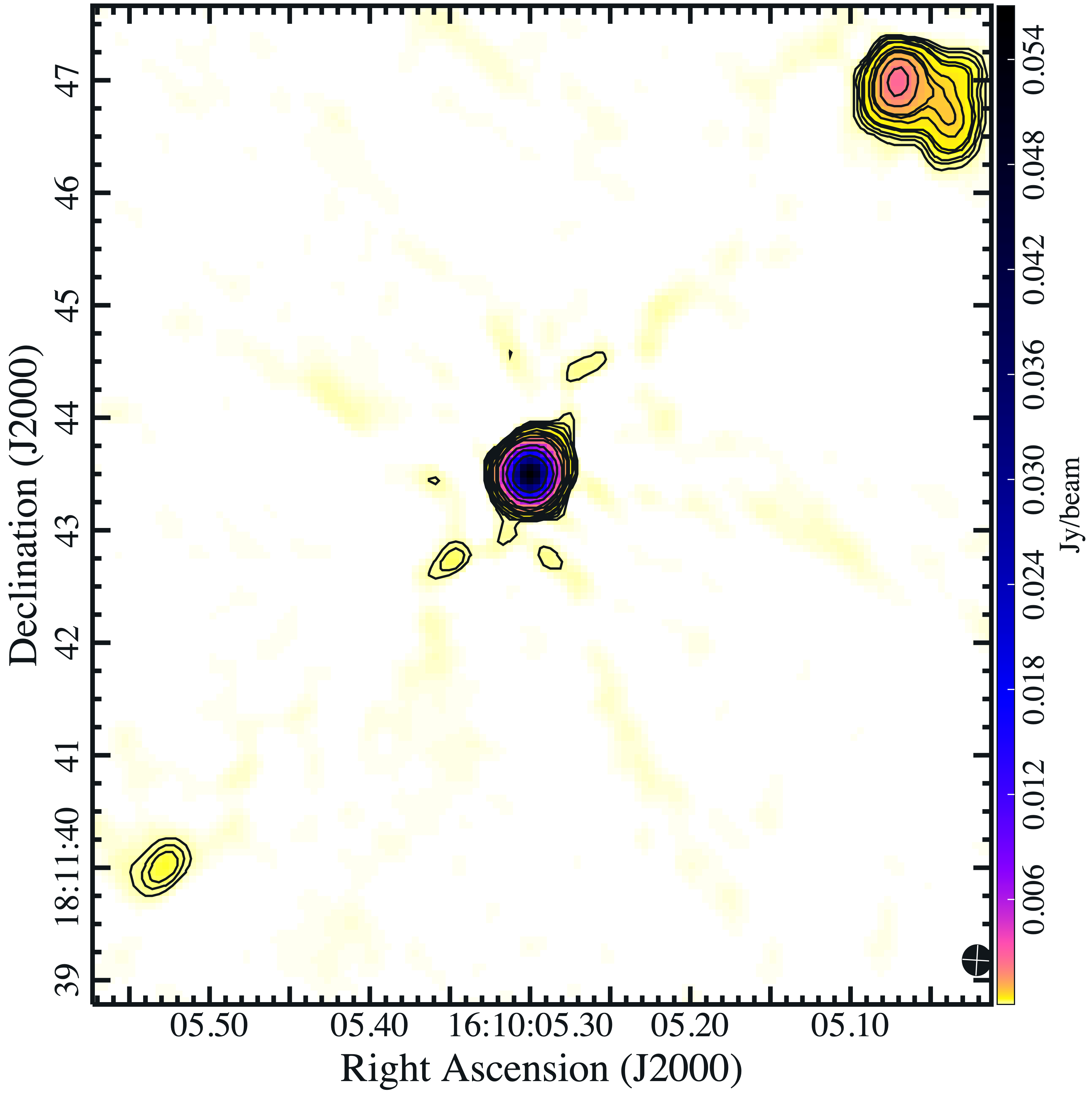}
\caption{\small Left: 6 GHz VLA image of J1405+0415. The synthesized beam shown in the bottom right corner is of size $0 \farcs 29 \times 0 \farcs 25$ at a PA of $3.59\degr$. The contour levels are [$3\sqrt{2}, 4\sqrt{2}, 6\sqrt{2}, 10, 20, 40, 80, 200, 600, 1000, 2000] \times ~ 141~ \rm \mu Jy~beam^{-1}$.} Right: VLA image of J1610+1811 at 6 GHz. The synthesized beam shown in the bottom right corner is of size $0 \farcs 27 \times 0 \farcs 26$ at a PA of $-3.53\degr$. The contour levels are [$2\sqrt{2}, 3\sqrt{2}, 4\sqrt{2}, 6\sqrt{2}, 10, 18, 26, 30, 50, 80, 200, 400, 800] \times ~ 27.5~ \rm \mu Jy~beam^{-1}$.
\label{VLA6Ghz}
\end{figure}

The newly synthesized VLA images are shown in Figure \ref{VLA6Ghz}. We used the CASA task \texttt{imfit} to measure the 6 GHz core peak flux densities as the cores remain unresolved. The flux density uncertainties were calculated by \texttt{imfit} following the methodology established by \citep{Condon1997}. The root mean square (rms) noises were derived from the standard deviations in the residual images produced by the \texttt{CLEAN} task. The rms noise levels are 23.5 $\rm \mu Jy$ and 5.5 $\rm \mu Jy$ for J1405+0415 and J1610+1811, respectively. The core flux density for J1405+0415 is $617.4\pm2.6$ mJy beam $^{-1}$, while for J1610+1811, it is an order of magnitude lower at  $57.7\pm0.1$ mJy beam $^{-1}$. 
Both quasars have a flatter core spectrum with mean value of spectral index in a $0 \farcs 475$ circular core region being $0.160\pm0.001$ and $0.42$ for J1405+0415 and J1610+1811, respectively. The spatial resolution of $0 \farcs 3$ allowed for the separation of the knot from the core at the position angle of 234$\degr$ in J1405+0415 (Figure \ref{VLA6Ghz}, left). The flux density of the knot using $0 \farcs 3$ circular region is $29.6\pm2.2$ mJy and its spectral index is $0.72\pm0.02$. We also discern a $1.6\pm0.4$ mJy hotspot $3 \farcs 2$ away at a position angle of 254$\degr$. 

The known hotspot $4 \farcs 76$ from the core in the north-west direction (PA $=$ 317$\degr$) of J1610+1811 also shows extended lobe-like structure in our 6 GHz image. It has a flux density of $7.3\pm0.4$ mJy and $\alpha$ of $1.09\pm0.06$. We also detect a $0.56\pm0.07$ mJy feature in the south-east direction at a similar distance from the core as the counter hotspot/lobe. 
Even at $\mu$Jy sensitivity, we did not detect any continuous kpc-scale radio jet connecting the core and hotspot/lobe structure in either quasar.

\subsection{X-ray Core Emission}
\begin{figure*}[h!]
\includegraphics[width=0.5\columnwidth,trim=0.0cm 0cm 0.0cm 0.cm,clip]{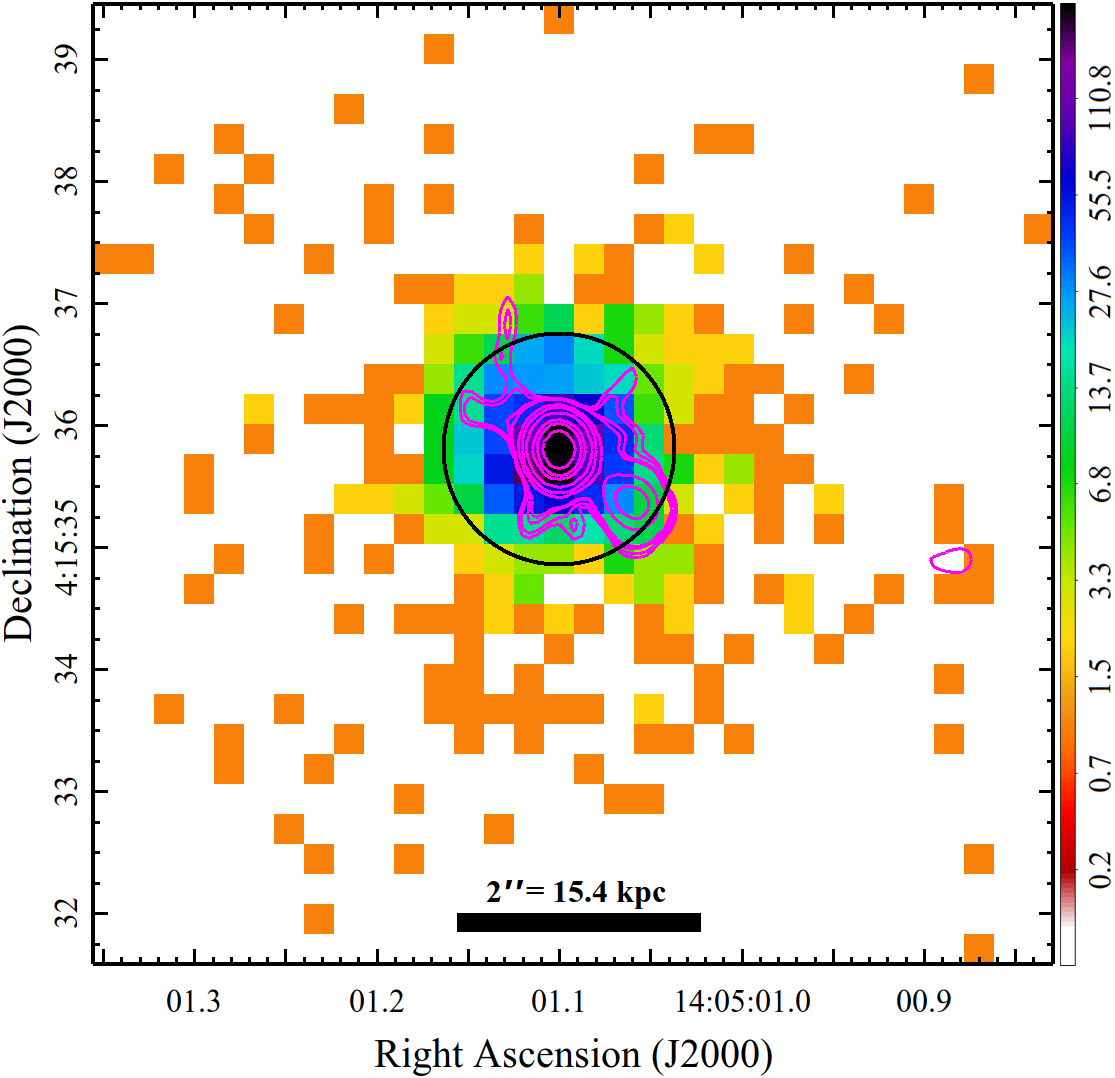}
\includegraphics[width=0.51\columnwidth,trim=0.0cm 0cm 0.0cm 0.cm,clip]{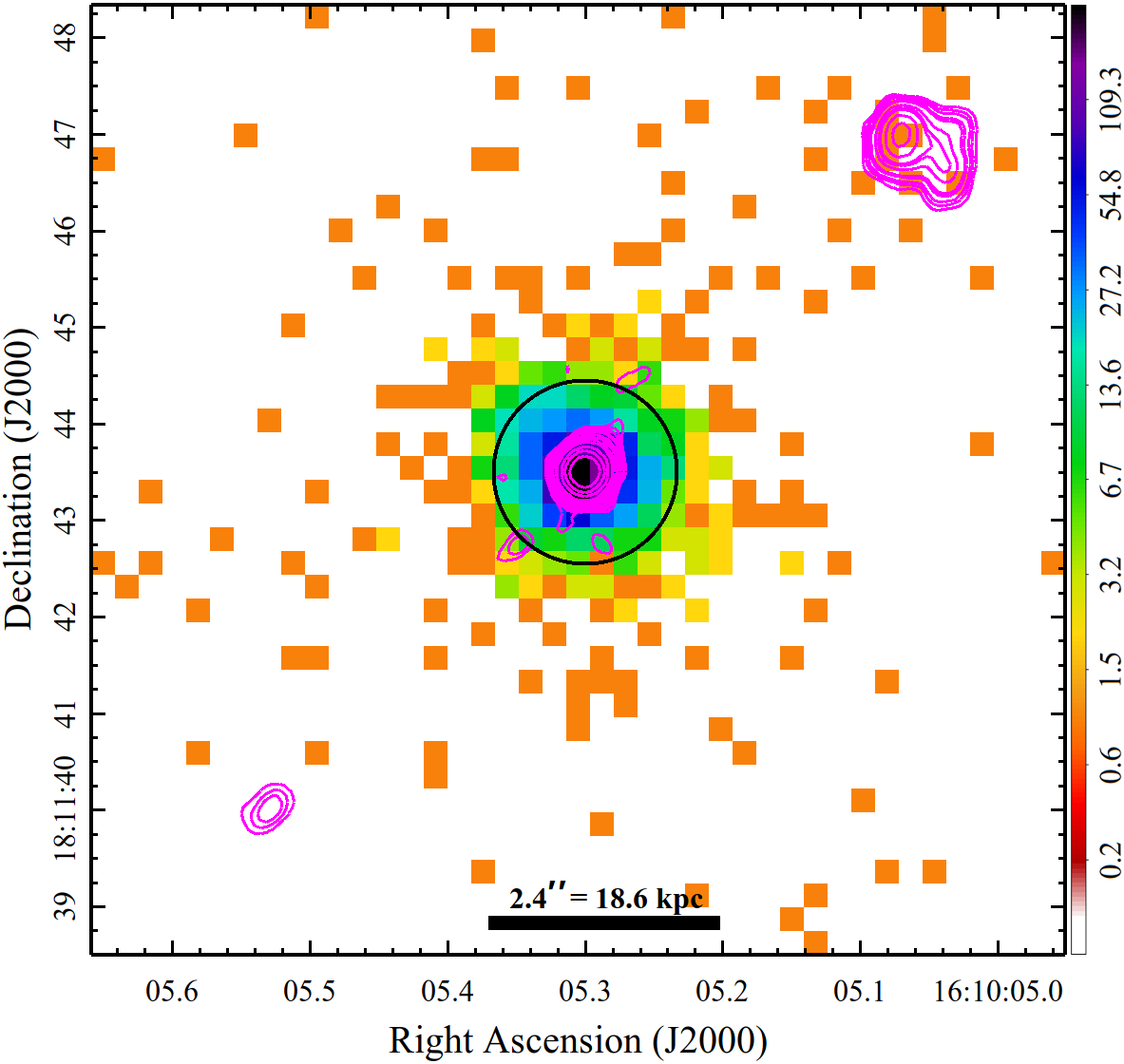}
\caption{\small Chandra ACIS-S images of J1405+0415 (left) and J1610+1811 (right) at 0.5-7 keV range are shown with $0 \farcs 246$ pixels to match the spatial resolution of the radio data. Both images are displayed in a logarithmic scale, and the color bar indicates counts per pixel. The black circle represents the $0 \farcs 95$ radius circular region used to extract core spectrum. The radio contours from the new 6 GHz VLA observations are overlaid on the X-ray images. }
\label{ChandraImage}
\end{figure*}
\begin{figure*}[h!]
\includegraphics[width=0.5\columnwidth,trim=0.cm 0cm 1cm 0cm,clip]{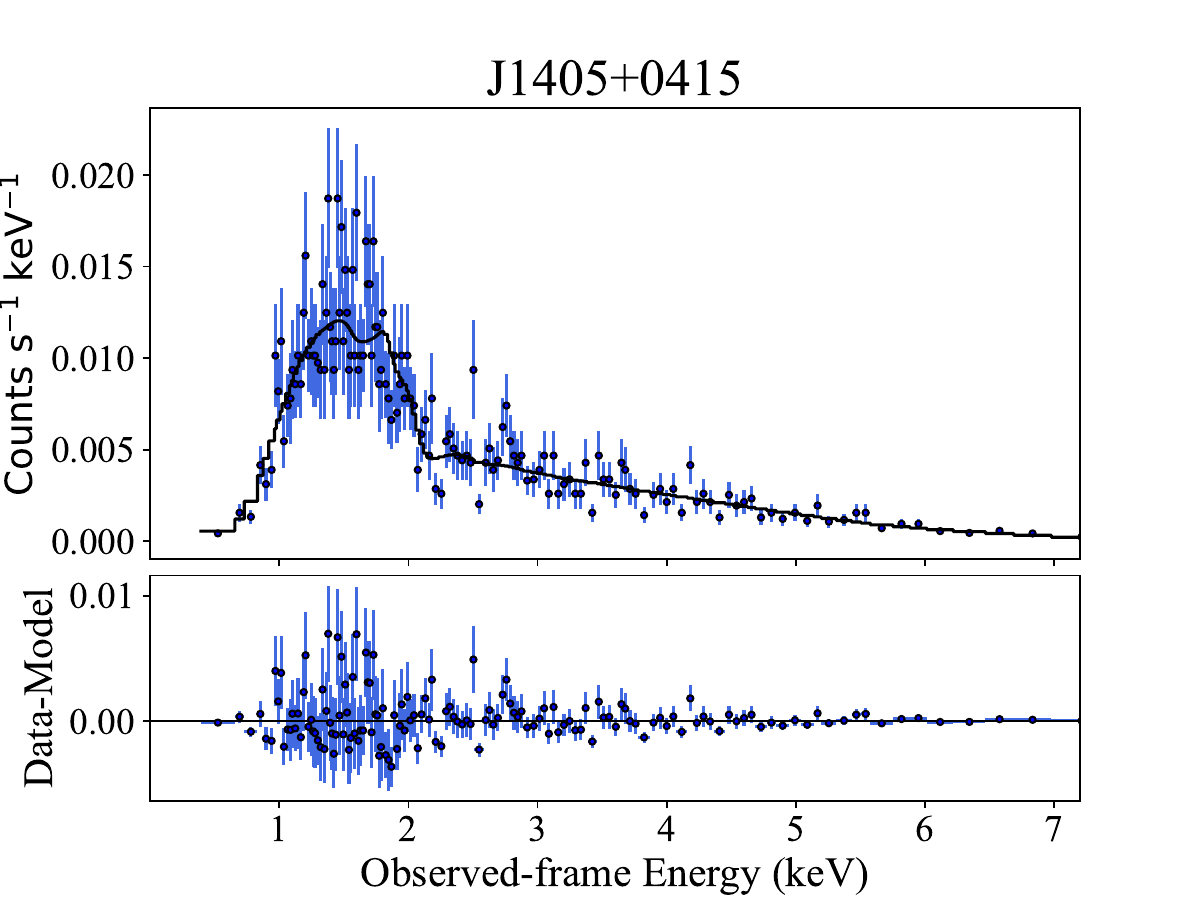}
\includegraphics[width=0.5\columnwidth,trim=0.cm 0cm 1cm 0cm,clip]{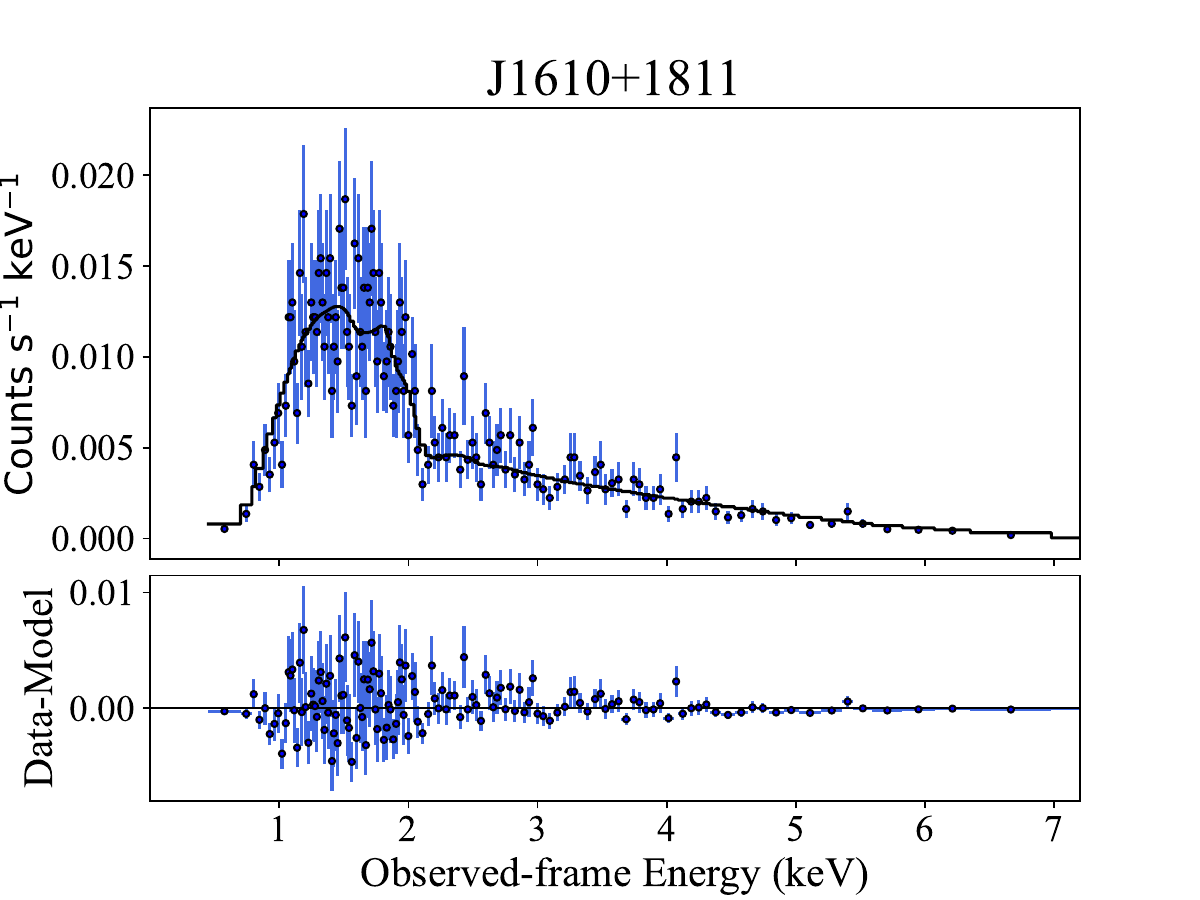}
\caption{\small Chandra X-ray spectrum of the combined observations. For better illustration, we have grouped the minimum counts per bin to be 10. For J1405 the best-fit is given by a power-law model with Galactic absorption. In case of J1610, best-fit includes an additional intrinsic absorption component.}
\label{Xraycorespectrum}
\end{figure*}

For each observation, we utilized the \texttt{specextract} script in CIAO to extract source and background spectra, using a $0 \farcs 95$ circular aperture for the source and an $8\arcsec \times 5\arcsec$ rectangular region away from the source as the background. A $0 \farcs 95$ radius corresponds to the 90\% enclosed count fraction i.e., 90\% of the X-ray counts fall within this radius. Figure \ref{ChandraImage} shows the Chandra images with overlayed radio contours and core region marked by the black circle. Table~\ref{tab:table3} presents the measured  background-subtracted core counts in broad (0.5-7 keV), soft (0.5-2 keV) and hard (2-7 keV) energy bands as well as the hardness ratio (HR). 

We used Sherpa \citep{Freeman2001} to simultaneously fit the core and background spectrum over the 0.5-7 keV energy band using a power-law with Galactic absorption shown in Figure \ref{Xraycorespectrum}. The best-fit value for the photon index, measured flux and measured 1 keV flux density are also presented in Table~\ref{tab:table3}. We then tested the addition of an intrinsic absorption component to this baseline model.

To evaluate the statistical significance of adding intrinsic absorption, we employed the Maximum Likelihood Ratio (MLR) test using Sherpa's \texttt{calc\_mlr} function. For J1405+0415, the MLR test indicates that including an intrinsic absorber does not significantly improve the fit. However, for merged spectrum of J1610+1811, a model incorporating both Galactic and intrinsic absorption provides a statistically superior fit (1 extra degrees of freedom and a statistic value that is larger by 8.01), yielding best-fit values of $\Gamma_{0.5-7} = 1.96\pm0.08$ and intrinsic absorption column density $N_{\rm H} = 4.93^{+2.33}_{-2.02} \times 10^{22} \rm cm^{-2}$. The MLR test yields a p-value of 0.005, strongly supporting the inclusion of the intrinsic absorption component. Notably, this derived intrinsic absorption of $\sim5\times 10^{22}~ \mathrm{cm}^{-2}$ falls within the typical range observed in Type 1 quasars, considering the measurement uncertainties.  By contrast, Compton-thick AGNs, which are viewed at larger inclination angles where the torus provides significant obscuration, typically exhibit much higher intrinsic absorption ($> 1.5 \times 10^{24}~ \mathrm{cm}^{-2}$; \citealt{Comastri2004}) due to obscuration by the torus. For consistency in comparison between sources, Table~\ref{tab:table3} presents the best-fit values from the simpler model without intrinsic absorption for both objects.


\begin{deluxetable*}{crrrcccccc}[h!]
\tablecaption{ X-ray properties of the core. \label{tab:table3}}
\tablehead{
\colhead{Object} & \colhead{ObsID} & \multicolumn{3}{c}{Net Count} & \colhead{HR} & \colhead{$\Gamma_{\rm 0.5-7~  keV}$} & \colhead{$f_{\rm 0.5-7~  keV}$} & \colhead{$f_{\rm 1~  keV}$} & \colhead{C-stat}\\
\cline{3-5}
\colhead{} & \colhead{} & \colhead{{\footnotesize($\rm 0.5-7~  keV$)}} & \colhead{{\footnotesize($\rm 0.5-2~  keV$)}} & \colhead{{\footnotesize($\rm 2-7~  keV$)}} & \colhead{} & \colhead{} & \colhead{}\\
\colhead{(1)} & \colhead{(2)} & \colhead{(3)} & \colhead{(4)} & \colhead{(5)} & \colhead{(6)} & \colhead{(7)} & \colhead{(8)} & \colhead{(9)} & \colhead{(10)}
}
\startdata
J1405+0415  & 23649    & $357.7\pm 18.9$  & $190.8\pm 13.8$ & $169.9\pm13.0$ &  $-0.06\pm0.05$ &  1.55$\pm$0.10 & $33.46^{+1.68}_{-1.76}$ & $36.74^{+3.15}_{-3.39}$  &  606.0 \\
{ }	        & 24316    & $326.8\pm 18.0$  & $169.9\pm 13.0$ & $159.9\pm 12.6$ &  $-0.03^{+0.06}_{-0.05}$ &  1.46$\pm$0.11 & $32.71^{+1.89}_{-1.75}$ & $32.97^{+3.27}_{-3.25}$  &  681.9 \\
{ }	        & 24317    & $567.7\pm 23.8$  & $287.9\pm 16.0$ & $283.8\pm 16.8$ &  $-0.01\pm0.04$ &  1.54$\pm$0.08 & 32.63$\pm$1.35& $35.30^{+2.65}_{-2.77}$  &  442.0 \\
{ }	        & 25011    & $437.8\pm20.9 $  & $226.9\pm 15.1$ & $213.9\pm 14.6$ &  $-0.03^{+0.05}_{-0.04}$ &  1.70$\pm$0.09 & $30.50^{+1.54}_{-1.35}$ & $37.87^{+3.13}_{-2.94}$  &  561.7 \\
{ }	        & 25014    & $324.7\pm 18.0$  & $160.0\pm 12.6$ & $164.7\pm 12.8$  &  $0.01^{+0.06}_{-0.05}$ &  $1.49^{+0.07}_{-0.16}$ & $34.68^{+5.79}_{-4.72}$ & $35.37^{+3.70}_{-3.24}$  &  628.7 \\
{ }	        & merged &$2014.8\pm 44.9$  &$1035.6\pm32.2 $ & $992.2\pm 31.5$   & $-0.02\pm0.02$  & 1.56$\pm$0.04  & $32.71^{+0.07}_{-0.08}$ & $35.67^{+1.49}_{-1.46}$ & -3514.6 \\
\hline
J1610+1811	&  23648    & 575.6$\pm$24.0  & $310.9\pm 17.6$  & $268.7\pm16.4$  &  $-0.07\pm0.04$ & 1.74$\pm$0.08  & $30.07^{+1.22}_{-1.34}$  & $37.49^{+2.84}_{-2.51}$   & 353.0  \\
{ }	        &  24486    & $294.9\pm17.2$  & $146.9\pm 12.1$  & $147.9\pm 12.2$  & $0.004\pm0.06$  & 1.62$\pm$0.11  & $28.96^{+4.31}_{-3.49}$  & $32.94^{+3.14}_{-3.42}$  & 599.2  \\
{ }	        &  24487    & $717.9\pm26.8$  & $399.9\pm 20.0$  & $321.0\pm 17.9$  & $-0.11^{+0.03}_{-0.04}$  & 1.83$\pm$0.07  & $34.93^{+1.19}_{-1.35}$  & $46.92^{+3.06}_{-3.09}$  & 32.5  \\
{ }	        &  25035    & $322.8\pm18.0$  & $183.9\pm 13.6$  & $140.9\pm  11.9$  & $-0.13\pm0.05$ & 1.80$\pm$0.11  & $30.40^{+1.74}_{-1.72}$  & $40.17^{+3.83}_{-3.92}$  &  579.1 \\
{ }	        & merged & $1911.1\pm43.7$ & $1041.6\pm 32.3$  & $878.6\pm 29.6$  & $-0.08\pm0.02$  & 1.77$\pm$0.04  & $31.59^{+0.69}_{-0.67}$ & $40.22^{+1.71}_{-1.58}$ & -3498 \\
\enddata
\tablecomments{(1) Source name. (2) Observation Id. (3) Background-subtracted counts in the core region over the 0.5-7 keV energy band. (4) Background-subtracted counts in the core region over the 0.5-2 keV energy band. (5) Background-subtracted counts in the core region over the 2-7 keV energy band. (6) Hardness ratio, HR = (Hard-Soft)/(Hard+Soft). (7) Power-law photon index 0.5-7 keV energy band. (8) Flux in the 0.5-7 keV energy band in units of $\rm 10^{-14}~ erg~ s^{-1}~ cm^{-2}$. (9) Flux density at 1 keV in the units of nJy which is equal to $\rm 10^{-32}~ erg~ s^{-1}~ cm^{-2}~ Hz^{-1}$. (10) Best-fit Cash-statistics. The Galactic neutral hydrogen column density was fixed to $2.2 \times 10^{20} \rm cm^{-2}$ and $3.2 \times 10^{20} \rm cm^{-2}$ for J1405+0415 and J1610+1811, respectively, adopted from \cite{DickeyLockman1990}}.
\end{deluxetable*}

\subsection{Extended X-ray Emission}
Figure \ref{ChandraImage} presents the X-ray images of J1405+0415 and J1610+1811 with overlaid radio contours highlighting the direction of extended jet emission. To access the morphology of extended X-ray jet structure, we simulated images of the quasar cores using SAOTrace and MARX and compared them with the observed images. For each Chandra observation, we simulated 1000 ray-tracing files using SAOTrace given the spectrum of the core. These files were then projected on to the detector plane using the ACIS Energy-Dependent Subpixel Event Repositioning algorithm in MARX. We tested different AspectBlur parameters due to uncertainties in determining the aspect solution. We found a best fit the data using an AspectBlur of $0 \farcs 20$ up to a radius of 1.5 pixels (i.e., 0 \farcs 74) and $0 \farcs 288$ beyond (see Appendix \ref{Appendix1} for more detals). We suggest that the change is due to the High Resolution Mirror Assembly (HRMA) artifact\footnote{\href{https://cxc.harvard.edu/ciao/caveats/psf_artifact.html}{https://cxc.harvard.edu/ciao/caveats/psf\_artifact.html}}, which is not modelled in SAOTrace.  

The Chandra HRMA can exhibit a stable “hook-like” artifact in its point-spread function (PSF). Typically seen around $0\farcs 6$ to $0 \farcs 8$ off center, this feature appears in both HRC and ACIS data and accounts for around 5\% of the total detected brightness. It remains fixed in spacecraft coordinates and was first noted in AR Lac observations taken sometime between late 1999 and December 2000. Subsequent data confirm that this asymmetry has persisted without significant change. Various analysis techniques, such as PSF fitting and deconvolution, have consistently revealed this artifact in multiple observations.

We analyzed X-ray counts in various sectors around the quasars to identify regions of significant excess emission. Assuming the count distributions per sector follow Poisson statistics, we generated a predicted probability distribution for each annular sector. To examine extended X-ray emission, we compared observed and simulated counts, normalizing the simulations to the observed core counts, and identified significant emission with a probability threshold of $p<0.001$. 
The details are presented in Appendix \ref{Appendix2}. For J1405+0415 (Figure \ref{J1405PA}), we found significant extended emission in one sector at position angle (PA) of 225$\degr$, with 36 observed counts versus 16 predicted, yielding a significance level of $p<5\times 10^{-6}$. Other sectors showed no significant excess. For quasar J1610+1811 (Figure \ref{J1610PA}), we identified significant excess emission in two sectors, although only one associated with the radio structure. It coincides with the radio hotspot at PA=315$\degr$, with 14 observed counts versus 2 predicted ($p< 4\times 10^{-9}$). The other one lies close to the core at PA= 45$\degr$ with 39 observed counts versus 21 predicted ($p=0.00014$) and is associated with the HRMA artifact \citep{Kashyap2010}. Our detection of extended X-ray emission is consistent with that determined by \cite{Snios+2022}, \cite{Snios2021} \& \cite{Schwartz2020}. Moreover, the radial surface brightness profiles of these quasars indicate excess X-ray emission in the soft energy band, confirming the presence of extended X-ray emission (see Figure \ref{J1405radialprofile}, \ref{J610radialprofile}, \ref{J1405jetprofile} and \ref{J610jetprofile}).

\subsection{Estimates on Extended X-ray and Radio Flux densities}
\begin{figure*}[h!]
\includegraphics[width=0.5\columnwidth,trim=0.0cm 0cm 0.0cm 0.cm,clip]{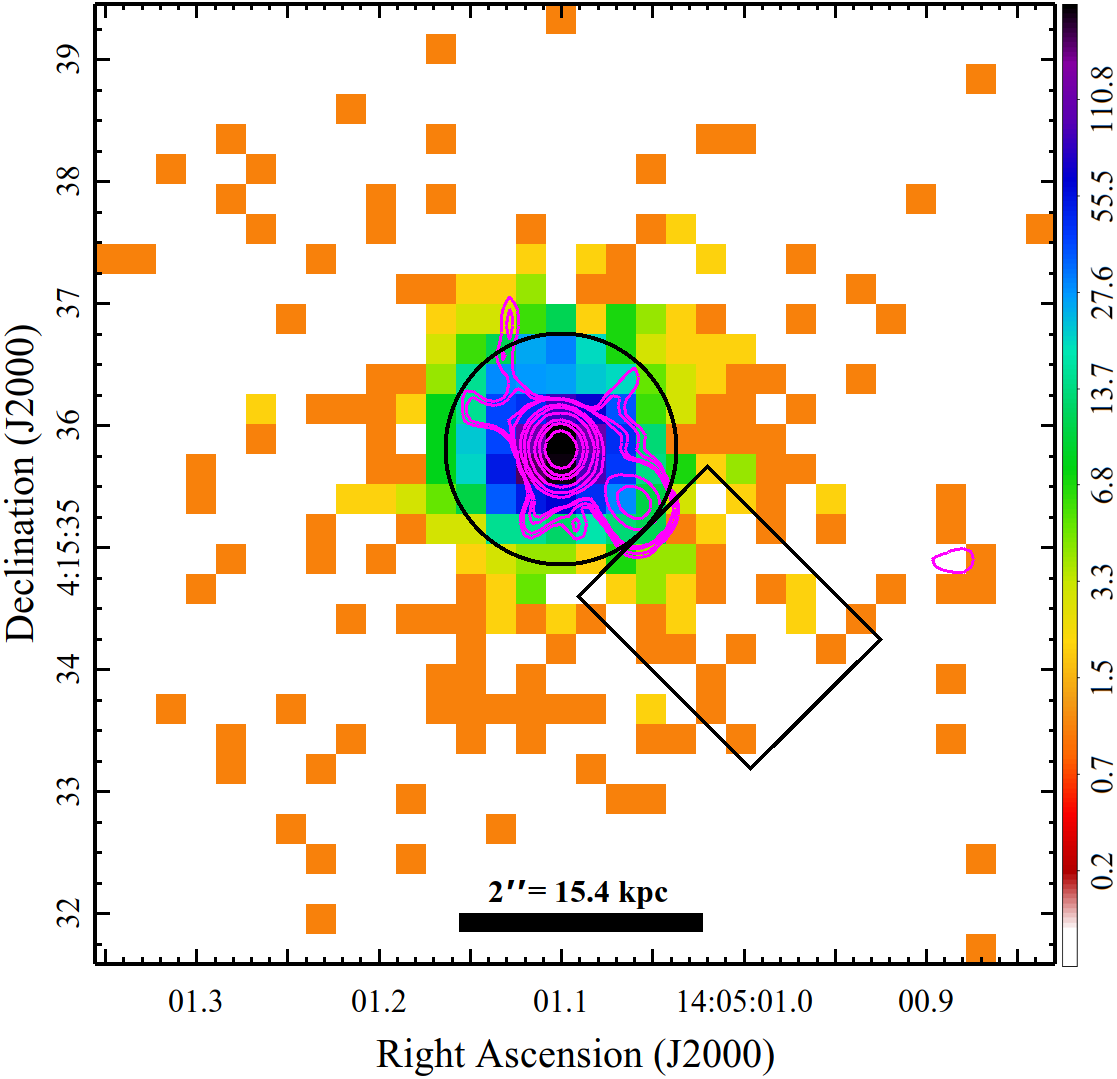}
\includegraphics[width=0.5\columnwidth,trim=0.0cm 0cm 0.0cm 0.cm,clip]{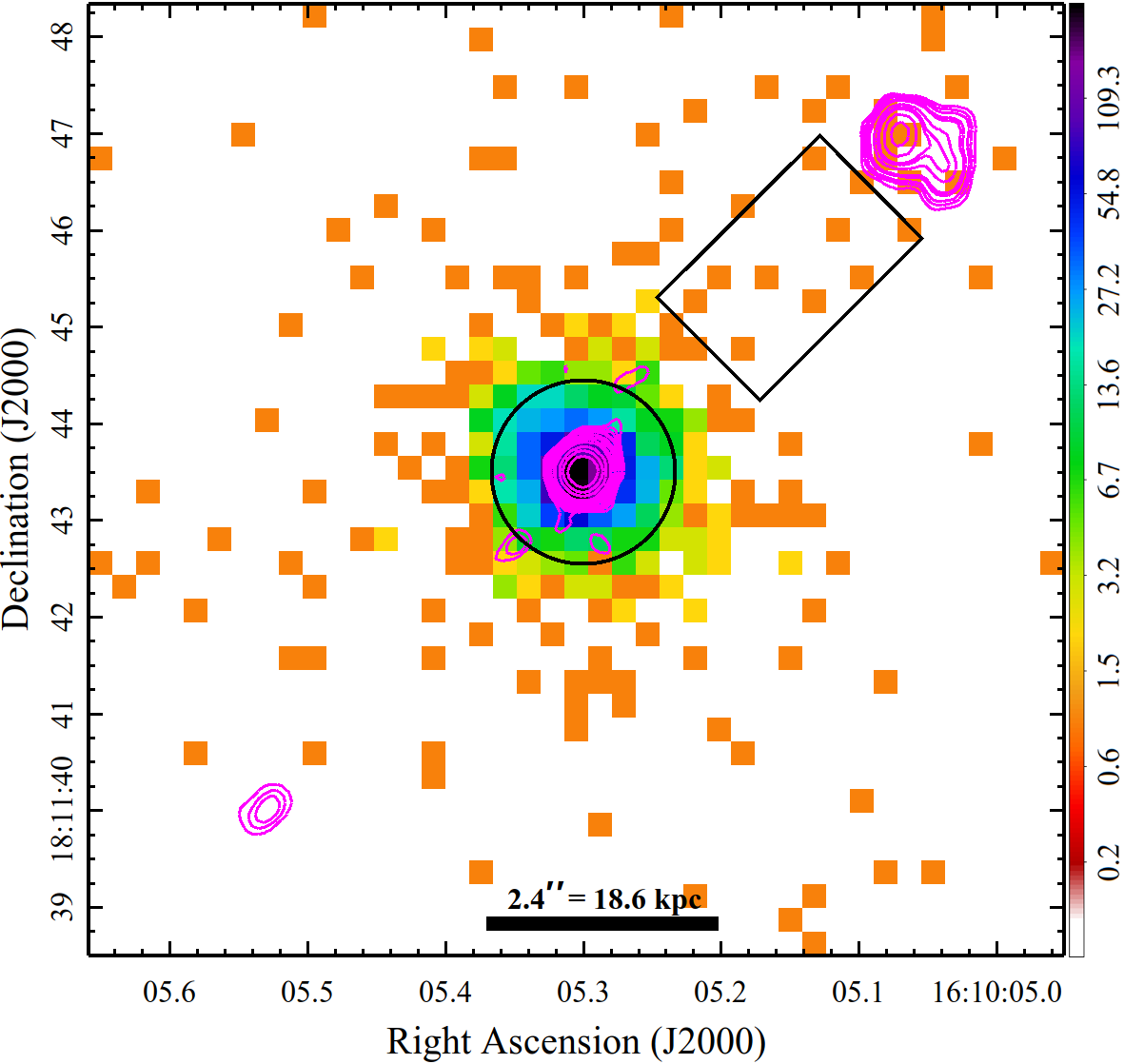}
\caption{\small Chandra ACIS-S images of J1405+0415 (left) and J1610+1811 (right) at 0.5-7 keV range are shown with $0 \farcs 246$ pixels. Both images are displayed in a logarithmic scale. The black box represents the $2\arcsec \times 1 \farcs 5$ ($2\farcs 4 \times 1 \farcs 5$) extended region used to extract jet spectrum of J1405+0415 (J1610+1811). The radio contours from the new 6 GHz VLA observations are overlaid on the X-ray images.} 
\label{ChandraImageJet}
\end{figure*}

We extracted the jet spectra of J1405+0415 using a $2\arcsec \times 1 \farcs 5$ box (indicated by the black box in Figure \ref{ChandraImageJet}), starting beyond the radio knot and extending $2\arcsec$ to where significant excess X-ray emission is observed. For J1610+1811, we used a $2.4\arcsec \times 1 \farcs 5$ box (also shown by the black box in Figure \ref{ChandraImageJet}), covering the region beyond the X-ray core's 90\% enclosed count fraction radius (i.e., the radius enclosing 90\% of the X-ray core counts) to just before the radio hotspot/lobe structure in the northwest direction, to minimize contamination from the non-jet components. The width of both jet boxes in X-ray analysis was set to $1 \farcs 5$ accounting for the poorer X-ray resolution (approximately twice that of radio). Table~\ref{tab:Jetprops} gives the net counts in the jet regions in the 0.5-7 keV energy band and the hardness ratio. We modeled both the jet and background spectra simultaneously using a power-law with Galactic absorption, employing an $8\arcsec \times 5\arcsec$ background region away from the jet. The table also lists the best-fit values for the 0.5-7 keV photon index, flux in the 0.5-7 keV energy band, and the flux density at 1 keV. The photon indices are $\Gamma_{\rm 0.5-7 keV} = 2.37_{-0.32}^{+0.34}$ for J1405$+$0415 and $\Gamma_{\rm 0.5-7 keV} = 1.47_{-0.73}^{+0.75}$ for J1610$+$1811, which overlap within the errors. Note that the sector coinciding with the jet box region in J1610+1811 does not show a significant excess in X-ray counts, according to our azimuthal distribution analysis (see Figure \ref{J1405PA} \& \ref{J1610PA}). Therefore, the jet X-ray flux density in J1610+1811 is considered an upper limit. 

In the absence of a continuous radio jet, we estimate an upper limit on the radio jet flux density using the image rms of 23.5 $\mu$Jy $\rm beam^{-1}$ and 5.5 $\mu$Jy $\rm beam^{-1}$ for J1405+0415 and J1610+1811, respectively. We determine the number of radio beams in the jet region, measuring of $2\arcsec \times 0 \farcs 75$ for J1405+0415 and $2.4\arcsec \times 0 \farcs 75$ for J1610+1811, to match the length of the X-ray extraction regions. The width of $0 \farcs 75$ in radio analysis corresponds to an assumed jet radius of 3 kpc (yielding a diameter of approximately $0 \farcs 75$). The radio flux limit is then calculated as three times the rms multiplied by the number of beams. Results are given in Table~\ref{tab:Jetprops}.

\begin{table}[h!]
    \centering
\caption{X-ray and radio properties of the jet.}
\label{tab:Jetprops}
    \begin{tabular}{|l|c|c|} \hline 
         Jet Properties &  J1405+0415 & J1610+1811\\ \hline 
         Observed Counts (0.5-7 keV) & 48 & 9\\
         Background Counts (0.5-7 keV) &  &  \\ 
         ~~   Observed X-ray background & 1.3 &  1.0 \\ 
         ~~   Scattered from core & 16 & 6 \\ 
         Net Counts (0.5-7 keV) &  46.7$\pm$6.9 & 7.9$\pm$3.0\\ 
         Hardness Ratio &  $-0.44\pm0.13$ & $-0.13\pm0.33$\\ 
         $\Gamma_{\rm 0.5-7~  keV}$ &  $2.37^{+0.34}_{-0.32}$ & $1.47^{+0.75}_{-0.73}$\\ 
         $f_{\rm 0.5-7~  keV}~ \rm (10^{-14}~ erg~ s^{-1}~ cm^{-2})$ &  $0.76^{+0.12}_{-0.11}$ & $0.13\pm0.05$\\ 
         $f_{\rm 1~  keV}~ (\rm nJy)$ & 	$1.39^{+0.23}_{-0.33}$	& $<0.14\pm 0.08$\\ 
         6 GHz Flux density upper-limit (mJy) & $<$0.39 & $<$0.13\\\hline
    \end{tabular}
\end{table}

For J1610+1811, our main analysis employs a conservative jet region that deliberately excludes the distant lobe/hotspot complex, maintaining focus on the pure jet emission. While we detect X-ray counts above background in this region (though not at statistically significant levels), we still extracted a spectrum to derive meaningful constraints. In Appendix \ref{Appendix3}, we present supplementary results using a $3 \farcs 7$ long jet box that includes the lobe/hotspot region, which shows a more significant X-ray excess (p$<$4e-9) along the jet axis.

\section{Discussion}
The most plausible explanation of X-ray emission from quasar jets at large redshift involves the IC/CMB process, where relativistic electrons moving at bulk relativistic speeds within the jet scatter low-energy CMB photons, boosting them to X-ray energies through the inverse-Compton process \citep{Simionescu+2016, Worrall+2020}. The IC/CMB mechanism is favored in relativistic kpc-scale jets, where the energy density of the CMB becomes comparable to or even exceeds that of the magnetic fields, making inverse-Compton scattering a dominant process. \cite{Tavecchio2000} show that IC/CMB mechanism in kpc-scale jets successfully reproduces X-ray properties of blazars, a subclass of quasars. Blazars are favored because efficient IC/CMB scattering at low redshifts requires aligned jets with highly relativistic motions on kiloparsec scales. Additionally, \cite{Celotti2001} discuss implications for the location of gamma-ray emission in blazars. Several other studies also support the IC/CMB model \citep{Sambruna2002, Sambruna2004, Siemiginowska+2003, Schwartz+2006, Meyer+2019}.

Despite the widespread application  of the IC/CMB model, it fails in some cases. The model predicts significant gamma-ray emission, which exceeds upper limits observed from some X-ray jets \citep{Meyer2015, Meyer2017, Breiding+2017, Meyer2023}. 
Other mechanisms, like synchrotron emission, contribute to X-ray emission in jets, needing detailed multi-wavelength studies for analysis. Nonetheless, the IC/CMB model remains a viable and widely discussed explanation for X-ray emission in high redshift quasar kpc jets, and we focus on it here.

In this section, we describe our IC/CMB model and provide constraints on relativistic and jet parameters. We introduce a method to determine the median and confidence interval on predicted quantities. Our IC/CMB model fits well with observed radio and X-ray data for relativistic jets in quasars, showing $\Gamma \sim 3-4$, magnetic fields of a few tens of $\mu G$, and indicating that the jet of J1405+0415 is oriented closer to our line of sight than that of J1610+1811. Finally, we compare our results with other techniques and find them consistent within the uncertainties.

\subsection{IC/CMB Model}
We adopt a cylindrical jet geometry, assuming a typical jet radii ($a_r$) of 3 kpc \citep[e.g.,][]{Marshall2018} for both J1405+0415 and J1610+1811 due to the absence of direct radius measurements. Jet lengths ($a_l$) are estimated using radio and X-ray images as discussed in Section 3.4. For J1610+1811, where both X-ray and radio emissions are upper limits, we assumed a standard radio spectral index of $\alpha = 0.7$ \citep[e.g.,][]{Readhead+1979, Marscher+1988, Pushkarev+2012} for jet modeling.  In contrast, for J1405+0415, we detected X-ray jet emission with a photon index of $2.37^{+0.34}_{-0.32}$ but obtained only an upper limit on radio emission. We adopted a steeper radio-spectral index of $\alpha=1$ for modeling its jet--a value within the $1\sigma$ error of the X-ray slope. With only 46 counts, we probably don't have a good estimate of the slope error. 

To model the synchrotron spectrum, we assume a power-law distribution of electrons, $N(\gamma) d\gamma = \kappa \gamma^{-m} d\gamma$, with a minimum Lorentz factor $\gamma_{\rm min}=30$ and a maximum Lorentz factor $\gamma_{\rm max} = 10^5$. Here, the slope of the electron energy spectrum, $m=(2\alpha+1) = 2.4$. Assuming a uniform magnetic field, an  isotropic pitch angle distribution, a volume filling factor ($\phi$) of one, and a baryon-to-electron energy density ratio ($k$) of one, we calculate the minimum energy magnetic field ($B_{\rm{min}}$) from the radio luminosity per unit frequency ($L_{\nu}$) following \cite{Worrall+Birkinshaw2006} as
\begin{equation}
    B_{\rm{min}} ({\rm{G}})  = (4 \pi~ 8.1871 \times 10^{-7})^{1/(\alpha+3)} \left[  \frac{(1+k)}{\phi V} C_1 L_{\nu} (\alpha + 1) \nu^{\alpha} \left ( \frac{\gamma_{\rm{max}}^{1-2\alpha}}{1-2\alpha} -  \frac{\gamma_{\rm{min}}^{1-2\alpha}}{1-2\alpha} \right) \right]^{1/(\alpha+3)}.
\end{equation}
Here, $C_1$ includes all the constants like charge of electron ($e$), mass of electron ($m_e$), speed of light ($c$), classical electron radius ($r_e$) and Gamma functions, $\mathsf{\Gamma}(\alpha)$, coming from the synchrotron theory
\begin{equation}
    C_1{^{-1}} = \left( \frac{e}{2 \pi c m_e} \right)^{\alpha+1} \frac{\sqrt{\pi}~ \mathsf{\Gamma} \left((\alpha+3)/2\right)}{2~ \mathsf{\Gamma} \left ( (\alpha+4)/2 \right)} m_e~ c~ r_e ~ \frac{2 \pi 3^{(\alpha+\frac{1}{2})}}{2\alpha+2}~ \mathsf{\Gamma} \left(\frac{2\alpha+1}{4}+\frac{19}{12}\right)~ \mathsf{\Gamma} \left(\frac{2\alpha+1}{4}-\frac{1}{12}\right).
\end{equation}
We transform this magnetic field to the rest-frame of the jet moving with bulk Lorentz factor, $\Gamma$, and bulk jet speed $\beta = \sqrt{1-(1/\Gamma^2)}$, as
\begin{equation}
    B_{\rm{min}}^{\rm source} ({\rm{G}})  = (4 \pi~ 8.1871 \times 10^{-7})^{1/(\alpha+3)} \left[  \frac{(1+k)}{\phi V^{\rm source}} C_1 L_{\nu}^{\rm source} (\alpha + 1) \left (\frac{\nu}{\delta}\right )^{\alpha} \left ( \frac{\gamma_{\rm{max}}^{1-2\alpha}}{1-2\alpha} -  \frac{\gamma_{\rm{min}}^{1-2\alpha}}{1-2\alpha} \right) \right]^{1/(\alpha+3)}.
\end{equation}
Here, $\delta = \left[ \Gamma \left(1-\sqrt{1-\frac{1}{\Gamma^2}} \cos \theta \right)\right]^{-1}$ is the Doppler factor, $V^{\rm source} = \pi a_l a_r^2/\delta~ \sin \theta $ is the volume of the cylindrical jet in source frame, and $L_{\nu}^{\rm source} =  4 \pi D_{\rm{L}}^2 F_{\nu} (1+z)^{\alpha-1}/\delta^{3}$ is the radio luminosity of the jet in source frame. Note that we do not assume $\delta= \Gamma$, therefore, the minimum energy magnetic field depends on both $\Gamma$ and $\theta$.

\begin{figure*}[ht!]
\gridline{
\includegraphics[width=0.33\columnwidth,trim=0cm 0.0cm 0.cm 0.cm,clip]{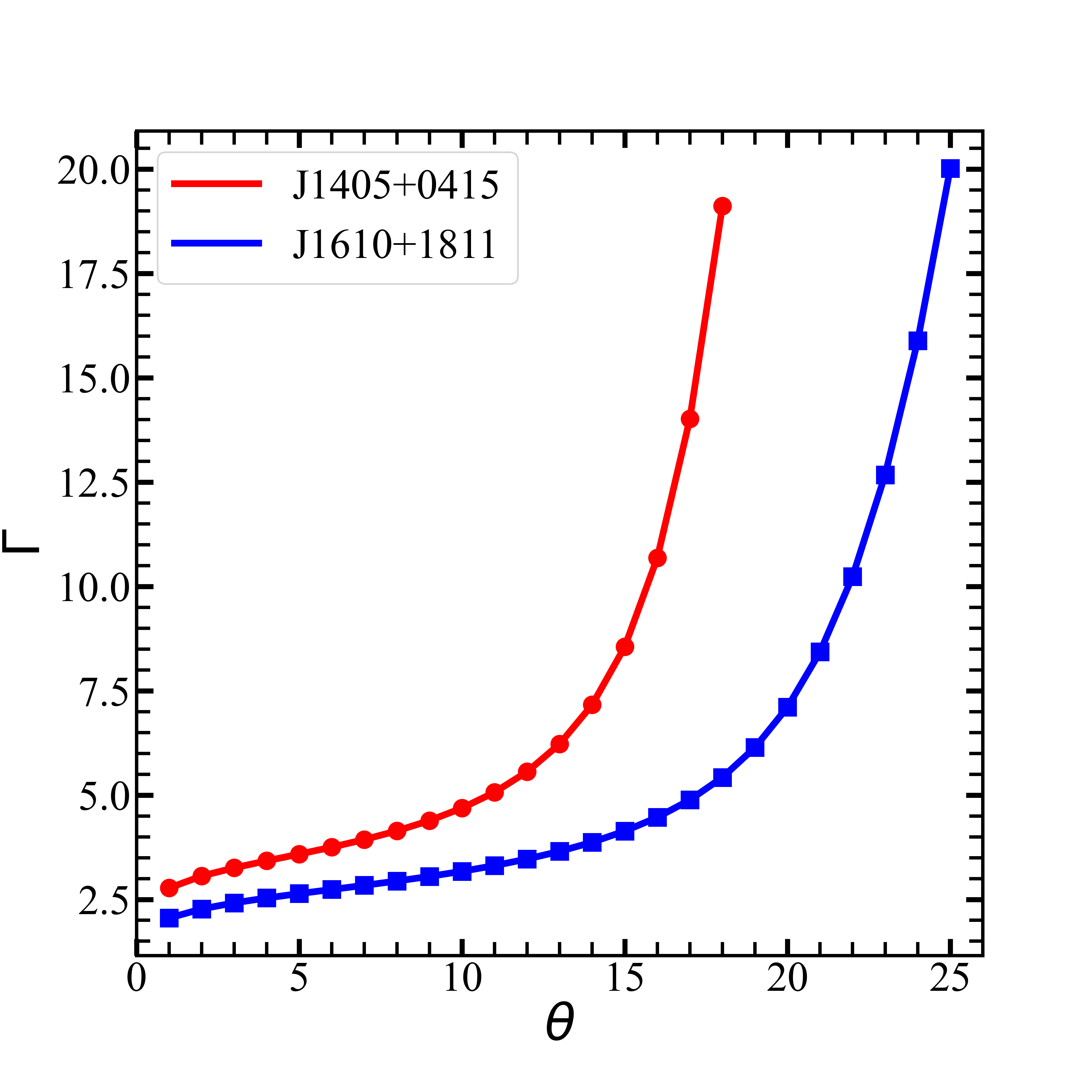}
\includegraphics[width=0.33\columnwidth,trim=0.cm 0cm 0.0cm 0.5cm]{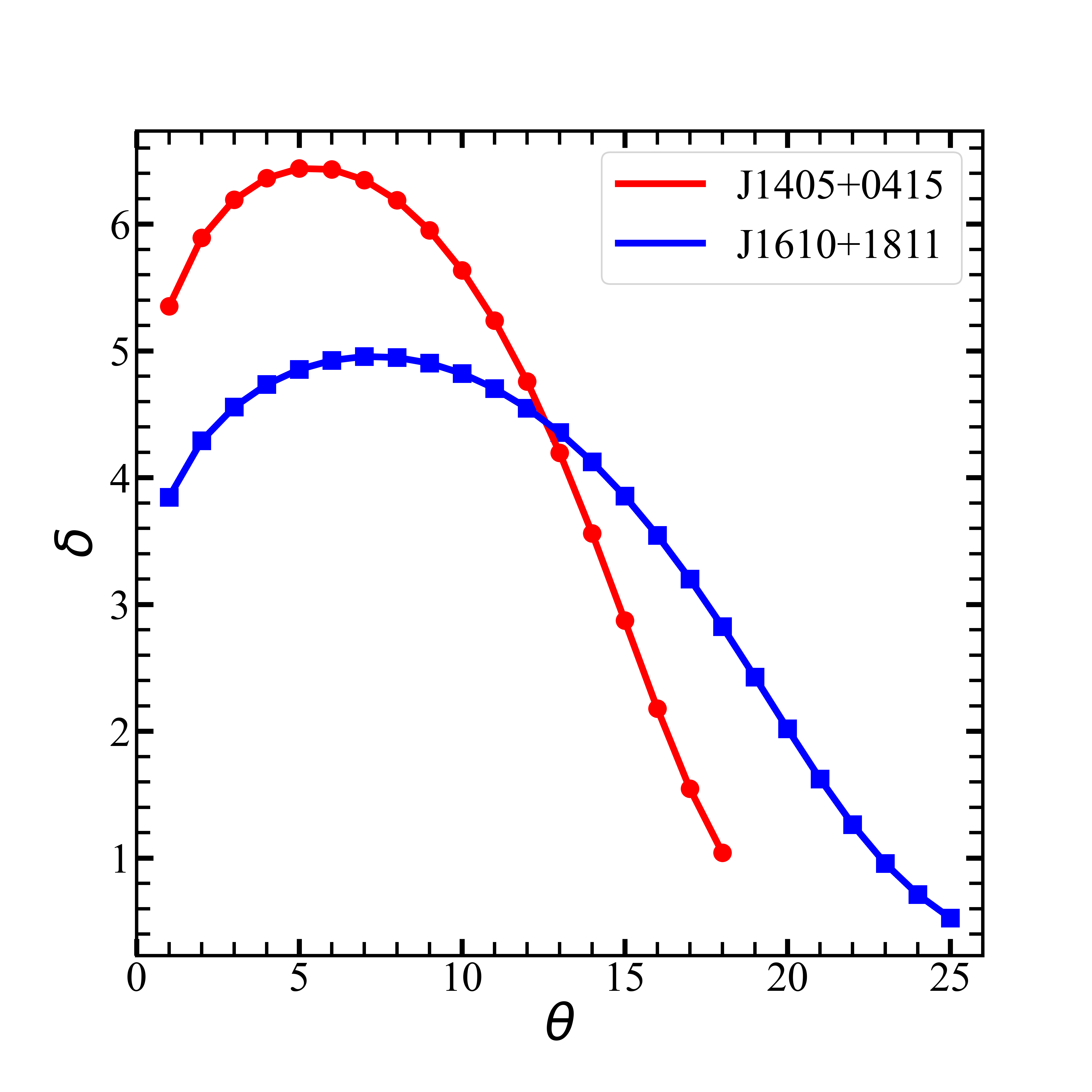}
\includegraphics[width=0.33\columnwidth,trim=0.cm 0.cm 0.0cm 0.cm,clip]{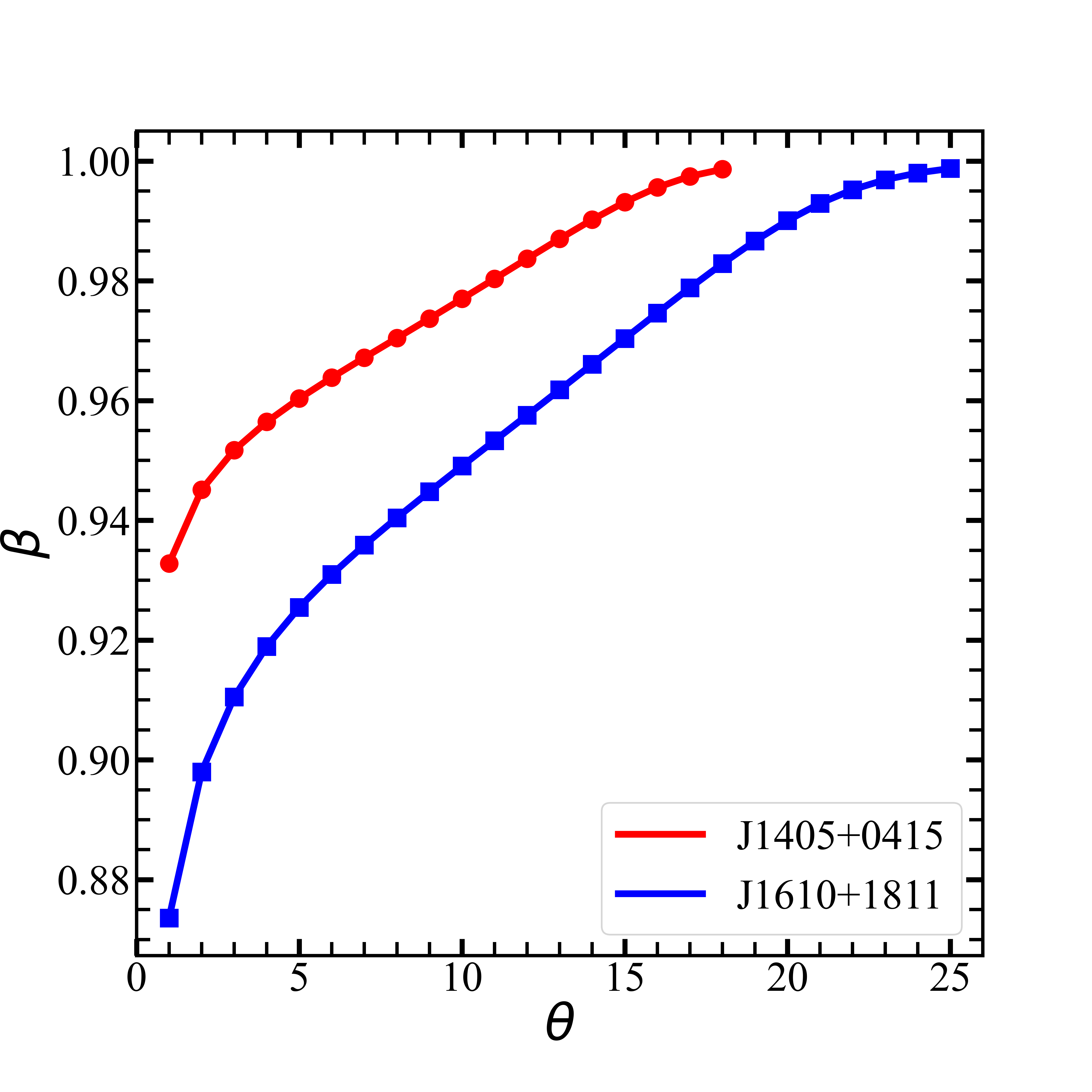}}
\gridline{
\includegraphics[width=0.33\columnwidth,trim=0cm 0.0cm 0cm 0cm,clip]{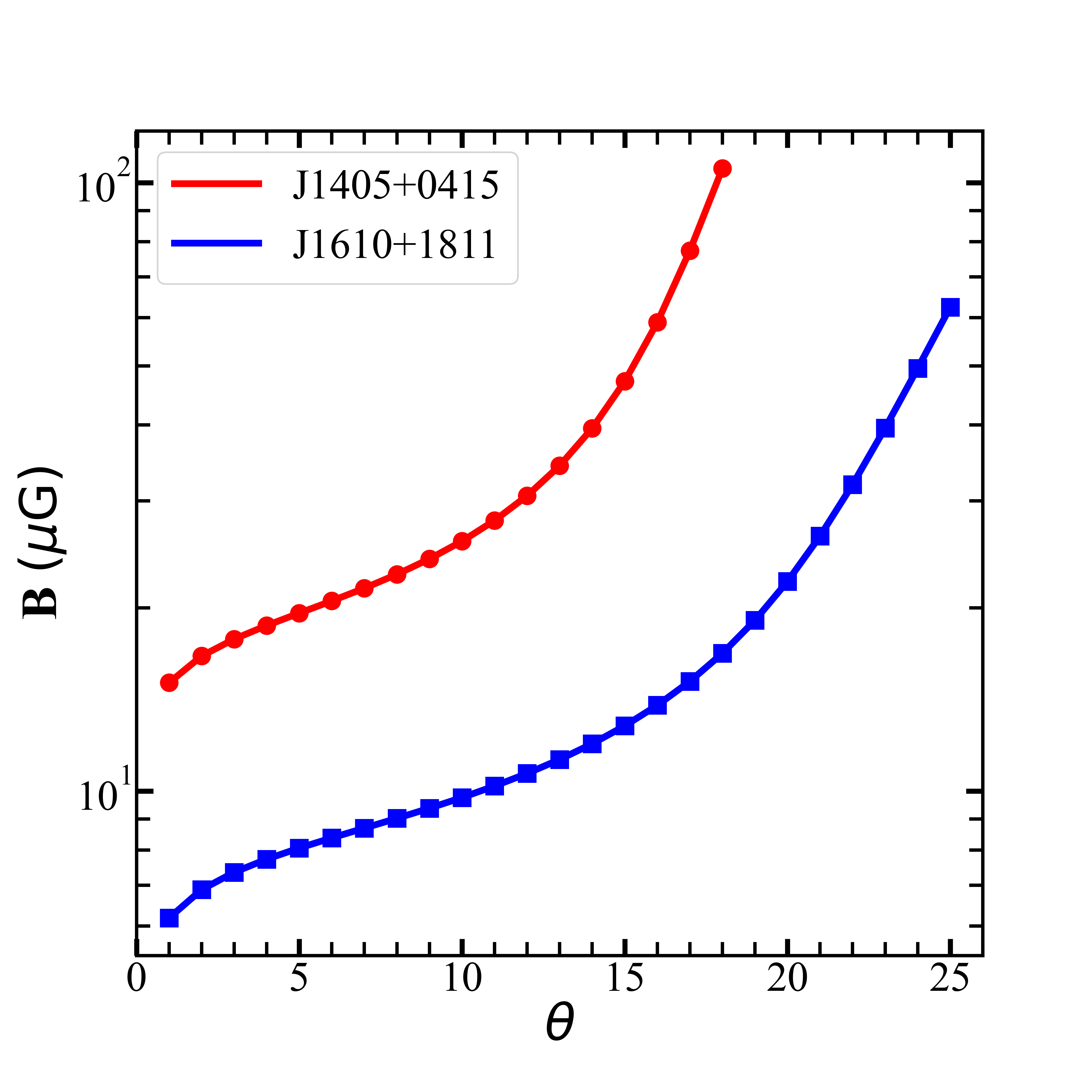}
\includegraphics[width=0.33\columnwidth,trim=0.cm 0.0cm 0cm 0cm,clip]{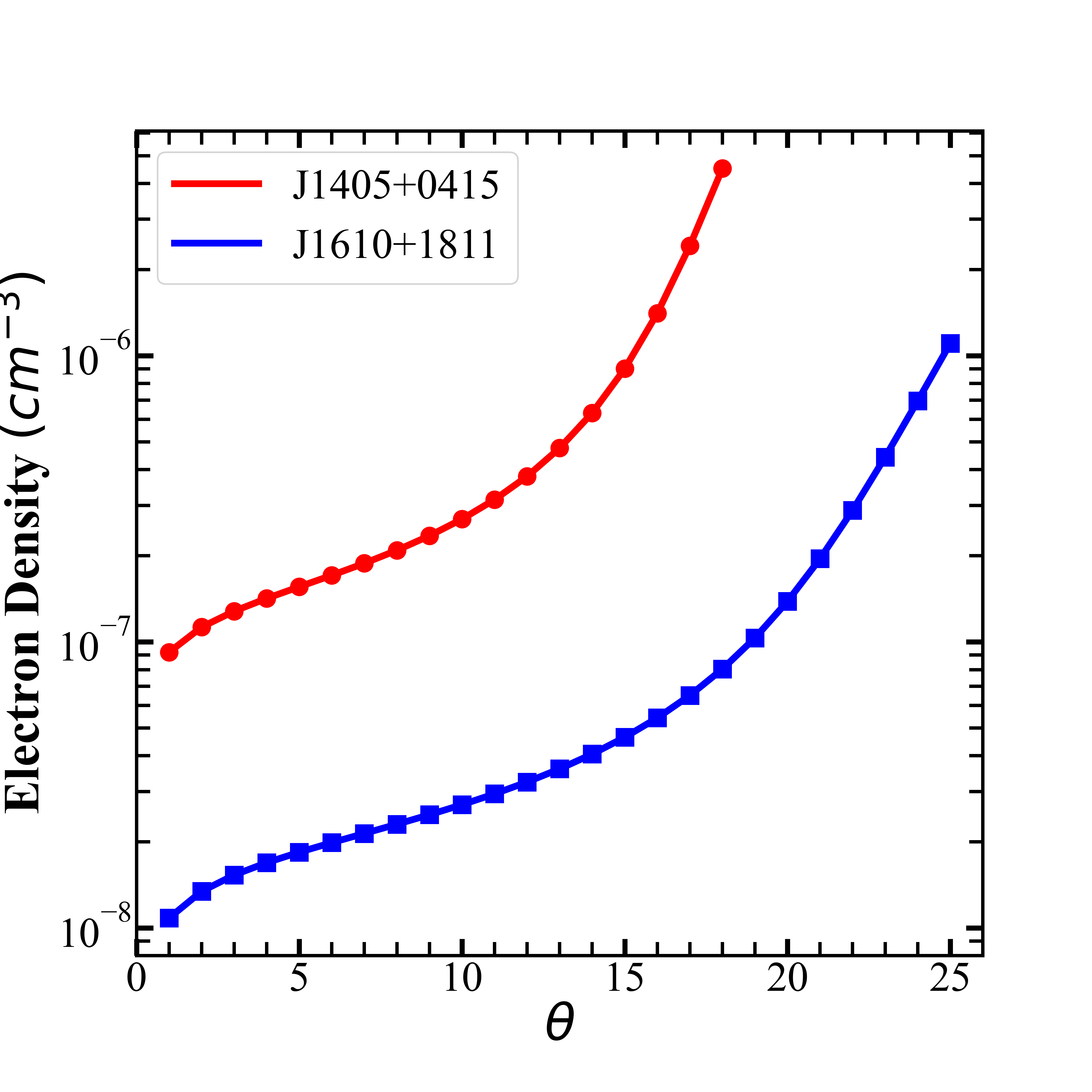}
\includegraphics[width=0.33\columnwidth,trim=0cm 0.0cm 0cm 0cm,clip]{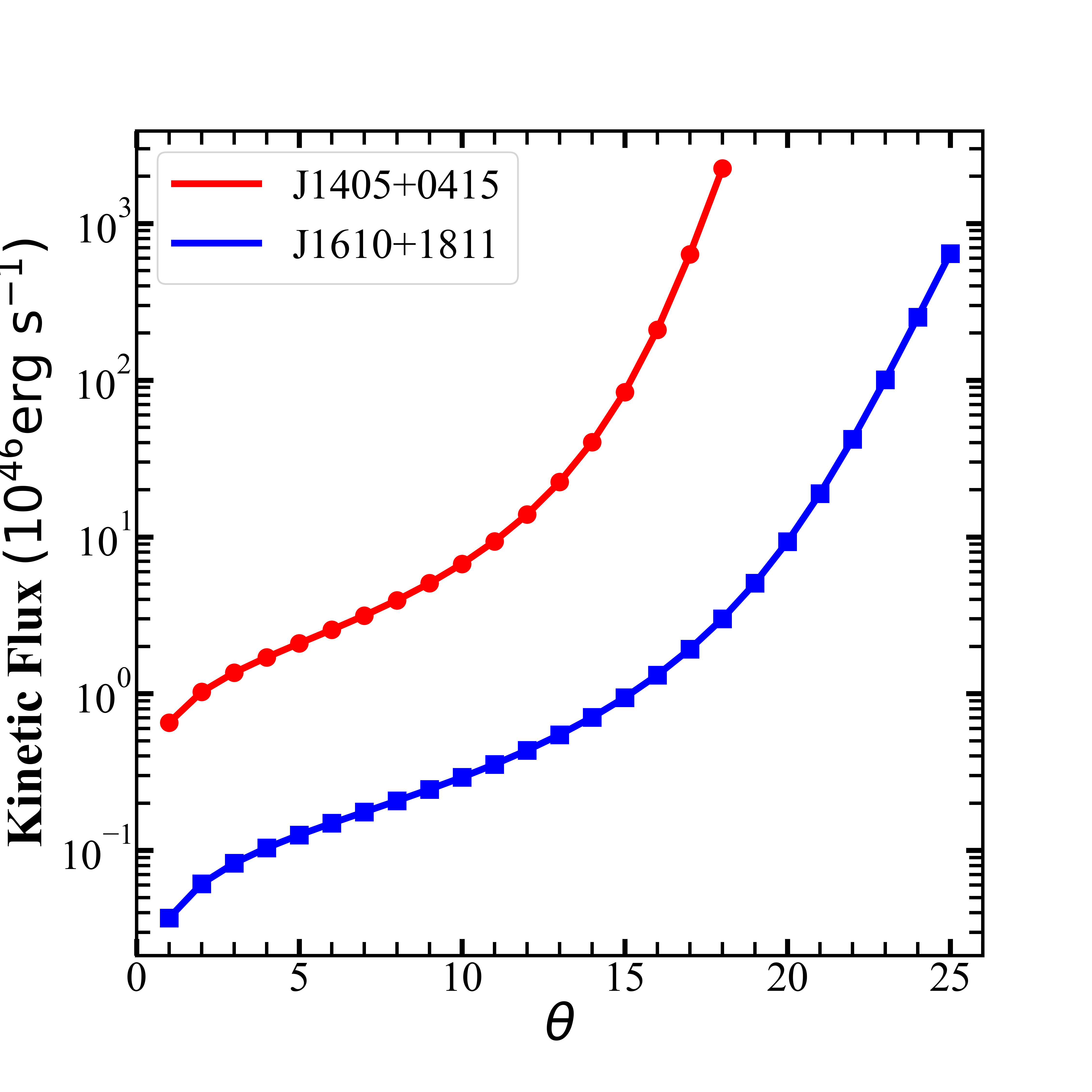}}
\gridline{
\includegraphics[width=0.33\columnwidth,trim=0cm 0.0cm 0cm 0cm,clip]{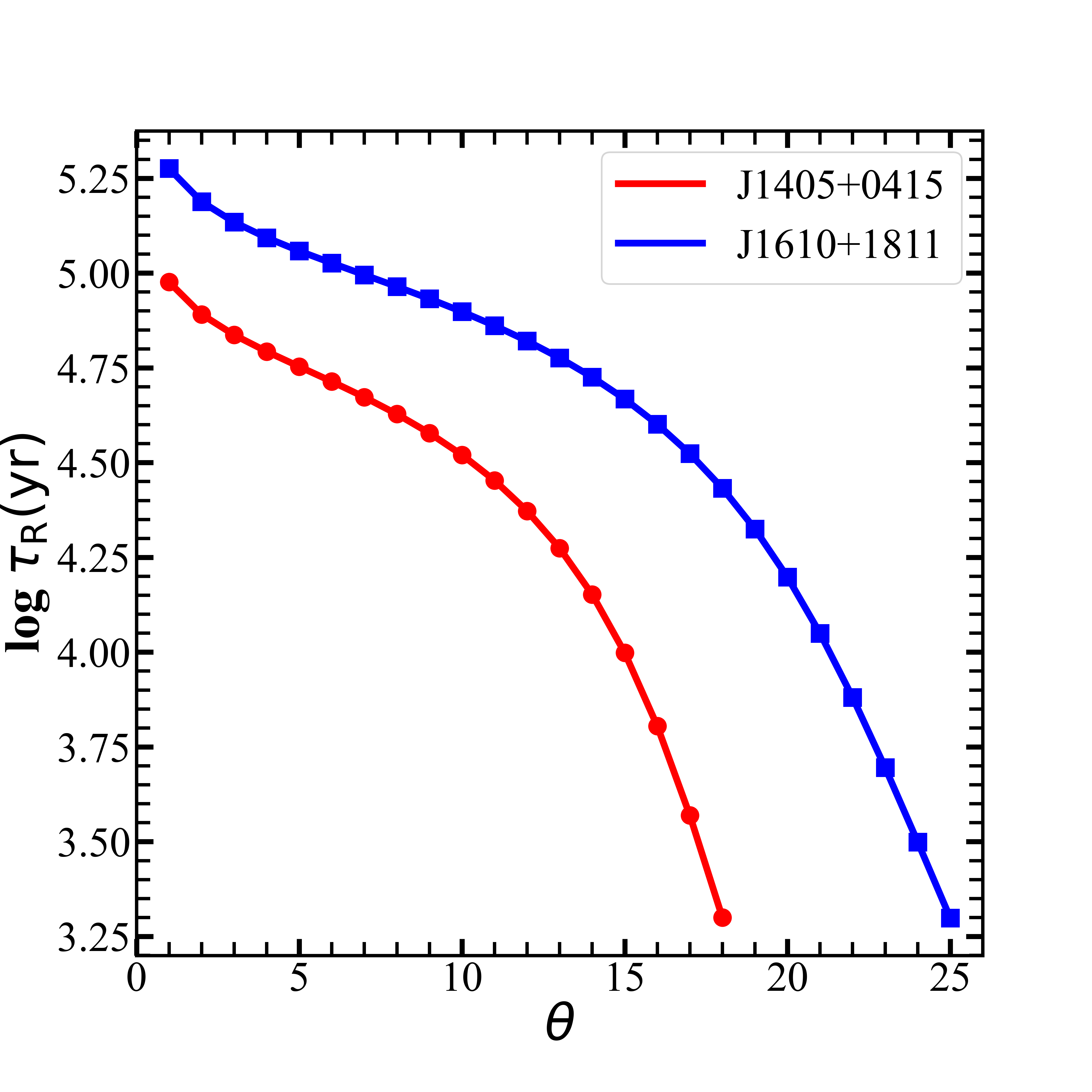}
\includegraphics[width=0.33\columnwidth,trim=0.cm 0.0cm 0cm 0cm,clip]{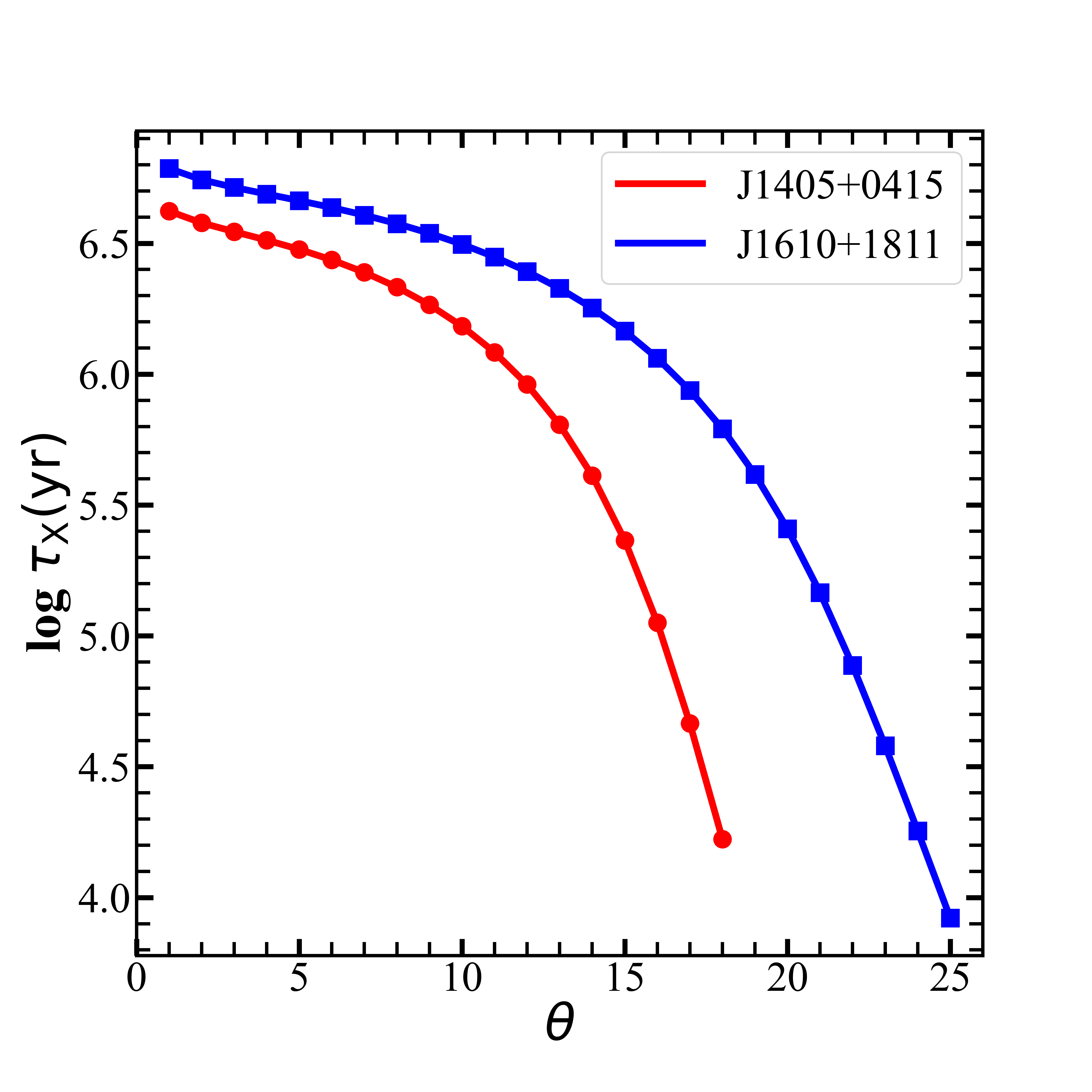}
\includegraphics[width=0.33\columnwidth,trim=0cm 0.0cm 0cm 1cm,clip]{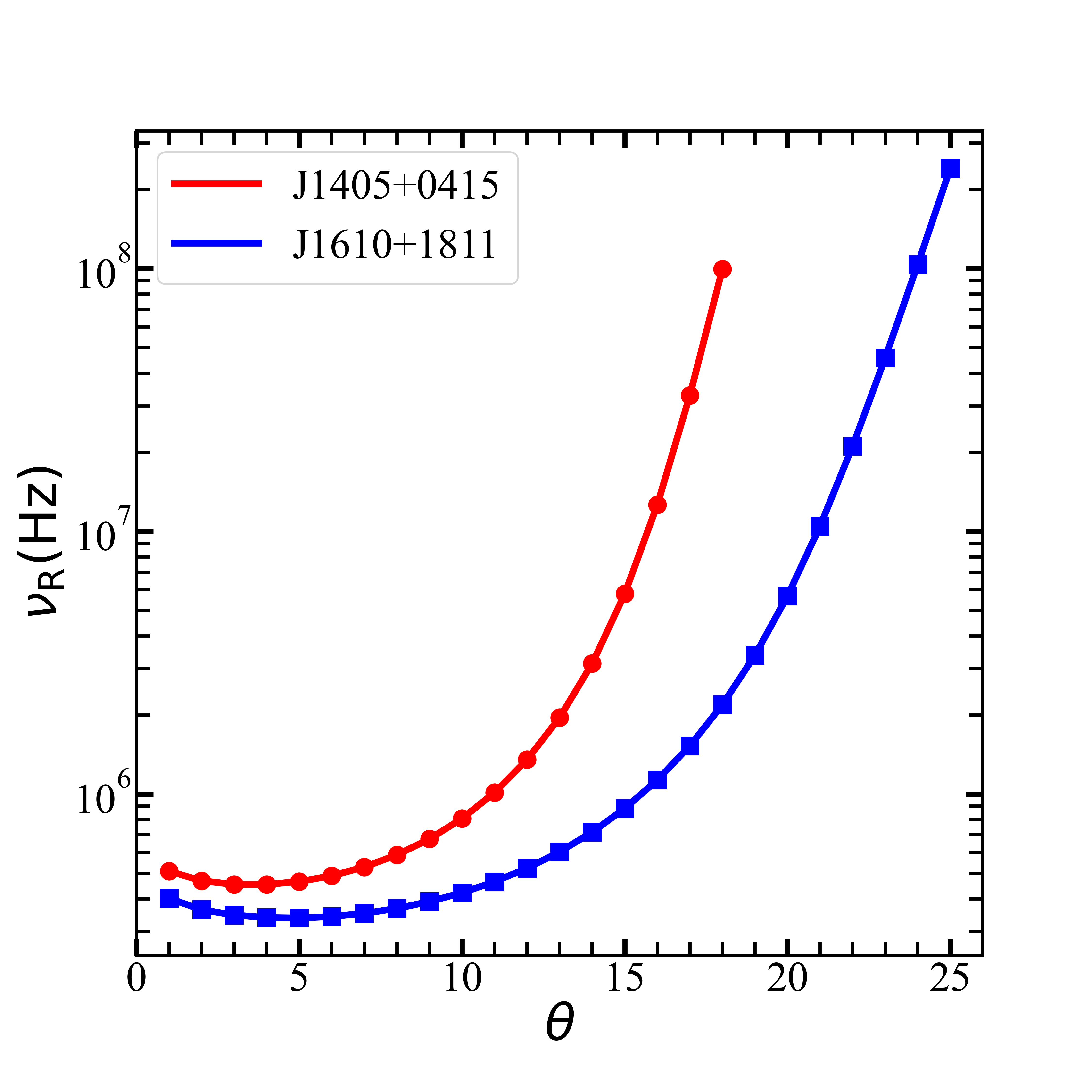}}
\caption{\small IC/CMB model results for J1405+0415 is shown in red, and for J1610+1811 in blue, highlighting the differences between the two sources. \textbf{Top:} The relativistic parameters $\Gamma$ (bulk Lorentz factor), $\delta$ (Doppler factor), and $\beta$ (bulk jet speed in units of the speed of light) as functions of the unknown viewing angle $\theta$. \textbf{Middle:} The jet parameters, including the magnetic field strength (B), the number density of synchrotron-emitting electrons, and the kinetic energy of the jet, plotted as functions of $\theta$. \textbf{Bottom:} $\tau_R$ (lifetime of radio-emitting electrons at $\gamma_{\rm max}=5000$), $\tau_X$ (lifetime of X-ray-emitting electrons via inverse-Compton scattering of CMB at 1 keV), and $\nu_R$ (radio-synchrotron emission frequency by 1 keV emitting electrons) as a function of $\theta$.}
\label{ICCMB}
\end{figure*}

We assume that the dominant mechanism for producing X-rays is the inverse-Compton of Cosmic Microwave Background (IC/CMB) from the same power-law distribution of synchrotron emitting electrons. The observed ratio of radio to X-ray flux density is determined by the ratio of energy density of the magnetic field to the energy density of the CMB photons.  
The calculation of synchrotron and inverse-Compton flux densities is simplified by assuming that each radiation peaks at its own characteristic frequency, as detailed by \cite{FeltenMorrison1966}. By applying this approach, we calculate the magnetic field required to produce the observed radio to X-ray flux densities ratio ($S_{\rm{R}}/S_{\rm{X}}$)
\begin{equation}
    B_{\rm{IC/CMB}} ({\rm{G}}) = \frac{[8\pi \times 4.19 \times 10^{-13}]^{\frac{1}{1+\alpha}}}{[5.45 \times 10^4]^{\frac{1-\alpha}{1+\alpha}}} (1+z)^{\frac{3+\alpha}{1+\alpha}} \left[\frac{S_{\rm{R}}}{S_{\rm{X}}} \left( \frac{\nu_R}{\nu_X} \right)^{\alpha} \right]^{\frac{1}{1+\alpha}}.
\end{equation}
Here, the upper limit from the 6 GHz radio maps (discussed in Section 3.4) is used as a detection to measure the jet's radio flux density. Additionally, the observed jet X-ray flux density in J1610+1811 is considered an upper limit (as discussed in previous section).
By plugging the redshift, measured radio to X-ray flux density ratio and assumed radio spectral index into equation 3, we obtain $B_{\rm{IC/CMB}}$ in the observed frame. Transforming $B_{\rm{IC/CMB}}$ to the jet rest frame,
\begin{equation}
    B_{\rm{IC/CMB}}^{\rm source} ({\rm{G}}) = \Gamma \left[ 1+ \frac{ \left ( \sqrt{1-\frac{1}{\Gamma^2}} \right )^2}{3} \right]^{1/(\alpha+1)} B_{\rm{IC/CMB}} (\rm G),
\end{equation}
making it a function of $\Gamma$. Our model iteratively determines $\Gamma$ for various line-of-sight angles $\theta$, where the minimum energy magnetic field matches the field required to produce the observed flux density ratio. We achive this by finding the root of the following equation
\begin{equation}
    B_{\rm{min}}^{\rm source} (\Gamma, \theta) - B_{\rm{IC/CMB}}^{\rm source} (\Gamma) = 0.
\end{equation}
In Figure \ref{ICCMB}(top panel), we illustrate the relativistic parameters $\Gamma$, $\delta$, and $\beta$ as functions of $\theta$, identifying reasonable ranges for both quasars. For viewing angles of $1\degr \leq \theta \leq 18\degr$, J1405+0415 exhibits  bulk Lorentz factors of $\Gamma=3-19$, while for J1610+1811 shows $\Gamma =2-20$ across a wider viewing angle range of $1\degr \leq \theta \leq 25\degr$. The Doppler factor ($\delta$), a function of the bulk Lorentz factor and the line of sight angle, first increases with $\theta$, peaks, and then decreases. For J1405+0415, $\delta$ increases from 5, peaks to 6 at $\theta \sim 5\degr$, and falls to 1 at $\theta = 18\degr$ due to the large Lorentz factor. For J1610+1811, $\delta$ rises slowly from 4 to 5 at $\theta \sim 8\degr$ and decreases more gradually to 1 at $\theta =\sim  23\degr$. Notably, the condition $\Gamma = \delta$ is met at $\theta =\sim 11\degr$ for J1405+1811, while for J1610+1811, it occurs at $\theta= \sim 15\degr$. Both quasars exhibit relativistic speeds, necessitating higher bulk velocities when viewed at larger angles. The bulk velocity ranges between 0.93c-0.99c for J1405+0415, and 0.87c-0.99c for J1610+1811.

Without further information on the angle to our line of sight, we plot the jet parameters such as magnetic field (B) strength, number density of synchrotron-emitting electrons, and kinetic energy of the jet, as functions of $\theta$ in Figure \ref{ICCMB}(middle panel). If we consider $1\degr \leq \theta \leq 18\degr$ for J1405+0415, our model predicts B values between 15-105 $\mu G$, and for J1610+1811, it predicts much lower B values ranging between 6-17 $\mu G$. The magnetic field strength gradually increase to 62 $\mu G$ for a viewing angle of $25\degr$ for J1610+1811. The electron number density ($n_e = \kappa \times(\gamma_{\rm min}^{-2 \alpha}-\gamma_{\rm max}^{-2*\alpha}) / (2 \alpha)$) ranges from $0.9 \times 10^{-7}$ to $45.1 \times 10^{-7} \mathrm{cm^{-3}}$ for J1405+0415, and from $0.1 \times 10^{-7}$ to $11.0.1\times 10^{-7} \mathrm{cm^{-3}}$ for J1610+1811. Using equation B17 from \cite{Schwartz+2006}, our model finds that J1405+0415 has a higher kinetic flux jet ($0.65 \times 10^{46}$ to $22.4 \times 10^{48} \mathrm{erg~ s^{-1}}$) compared to J1610+1811 ($0.04 \times 10^{46}$ to $6.4\times 10^{48} \mathrm{erg~ s^{-1}}$).

We determine the lifetime of radio-emitting electrons at $\gamma_{\rm max}=5000$ ($\tau_R$), the lifetime of electrons emitting IC/CMB in the X-ray at 1 keV ($\tau_X$), and the frequency at which these electrons produce radio-synchrotron emission ($\nu_R$), as shown in Figure \ref{ICCMB}(bottom panel). Using equation (4) from \cite{Worrall+2020} for cases where inverse Compton losses dominate over synchrotron losses, we find that for both quasars, the lifetime of radio-emitting electrons is shorter than that of X-ray-emitting electrons via IC/CMB. Our model suggests that electrons producing X-ray emission via IC/CMB would emit radio at 0.5-99.5 MHz for J1405+0415 and 0.4-250 MHz for J1610+1811.

\newpage 

\newpage
\subsection{Median value and confidence range of estimated parameters}
 We utilize a bayesian analysis to constrain the angle between jet axis and the line of sight ($\theta$). and We can then solve for jet parameters for an assumed range of $\theta$. As priors we take the distribution of $\theta$ to be proportional to the solid angle for the given jet direction, namely  $\propto \sin(\theta)$. We take the prior for the flux to be distributed proportional to the Doppler factor $\delta^3 \propto (1-\beta \cos \theta)^{-3}$. \cite{Murphy1988} approximates the probability density function of $\theta$ for sources selected based on their core flux density as  
\begin{equation}
    f(\theta) = \frac{2(1-\beta)^2}{(2-\beta)} \sin \theta (1-\beta \cos \theta)^{-3}.
\end{equation}
This approximation assumes that the number of sources as a function of flux follows the relation $N \propto S^{-3/2}$. In the absence of relativistic beaming, i.e., $\beta=0$, $f(\theta)= \sin \theta$ as expected from a randomly oriented source in the plane of the sky. 

\begin{figure*}[h!]
\centerline{
\includegraphics[width=0.5\textwidth,trim=0 0 0 0]{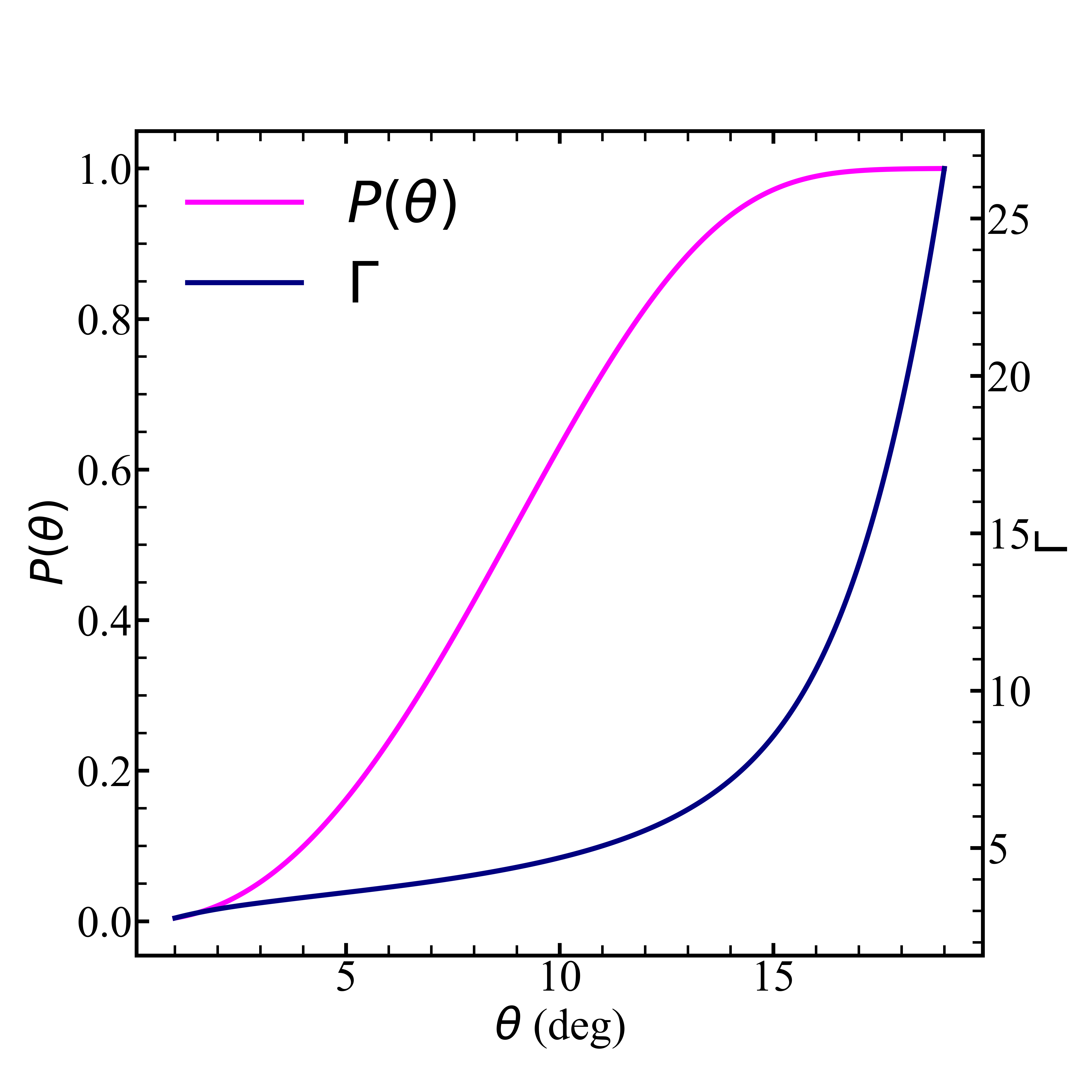}
\hspace{0.5cm}
\includegraphics[width=0.5\textwidth,trim=0 0 0 0]{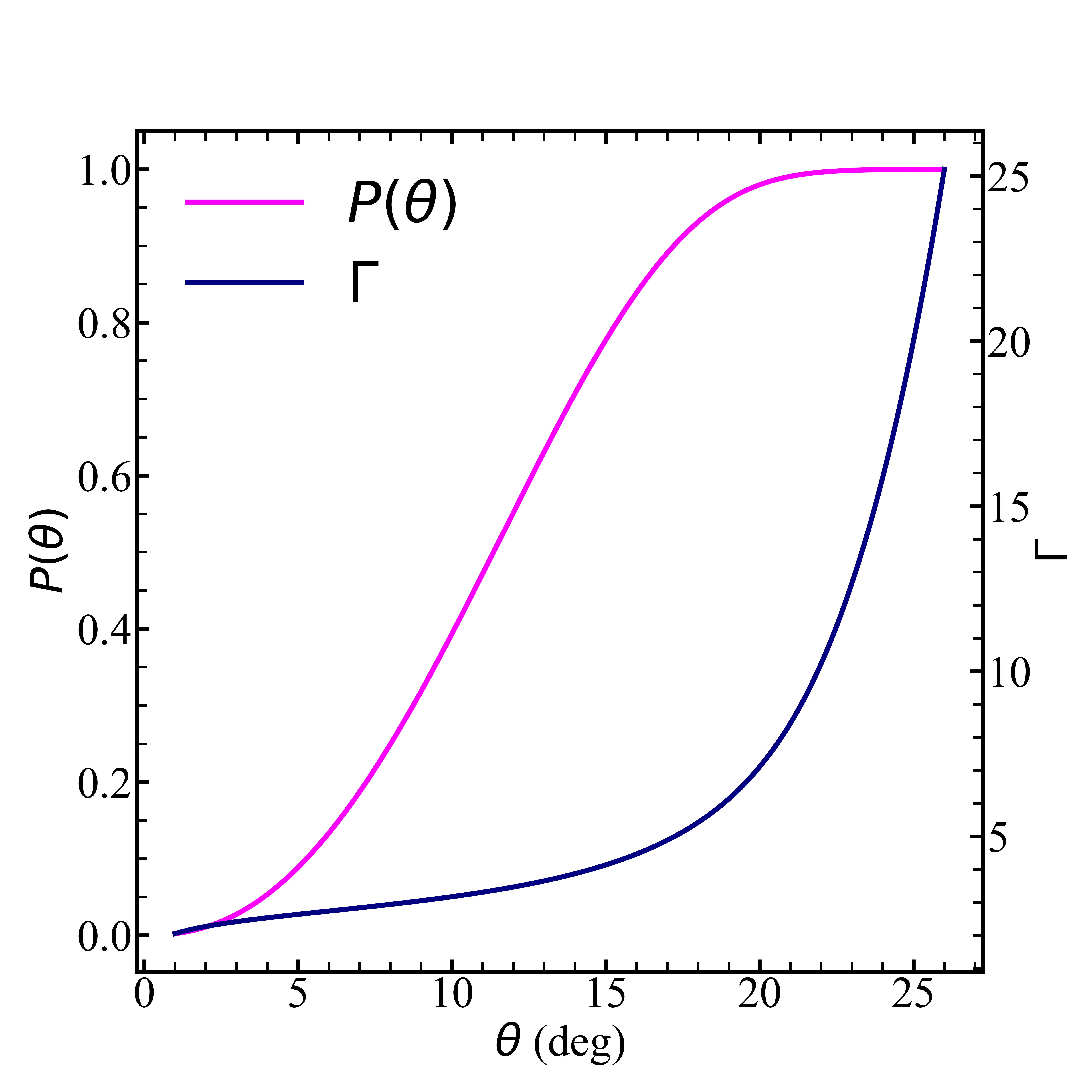}}
\caption{\small Probability and bulk Lorentz factor as a function of line of sight angle for J1405+0415 (left) and J1610+1811 (right). }
\label{ProbGamma}
\end{figure*}

Integrating the probability density function and substituting $\beta$ as a function of $\Gamma$, we get the cumulative distribution function as
\begin{equation}
   P(\theta) = \int_{0}^{\theta} f(\theta) \, d\theta = \frac{\left(-1 + \sqrt{1 - \frac{1}{\Gamma^2}}\right)^2}{\left(-2 + \sqrt{1 - \frac{1}{\Gamma^2}}\right)  \sqrt{1 - \frac{1}{\Gamma^2}}  \left(-1 +\sqrt{1 - \frac{1}{\Gamma^2}}  \cos \theta \right)^2} - \frac{1}{\left(-2 + \sqrt{1 - \frac{1}{\Gamma^2}}\right)  \sqrt{1 - \frac{1}{\Gamma^2}}}.
\end{equation}
The IC/CMB model provides the bulk Lorentz factor as a function of the given line-of-sight angle. Using the values of $\Gamma$ obtained from our model, Figure \ref{ProbGamma} shows the cumulative distribution function. $P(\theta)$ approaches unity at $\theta \approx 18 \degr$ for J1405+0415 and at $\theta \approx 25 \degr$ for J1610+1811.

Assuming that probability of line of sight angle follows a uniform distribution, i.e., $\mathcal{U}(0, 1)$, we solve for $\theta$ such that
\begin{equation}
    P(\theta) = \mathcal{U}(0, 1).
\end{equation}

\begin{figure*}[h!]
\centerline{
\includegraphics[width=0.5\textwidth,trim=0 0 0 0]{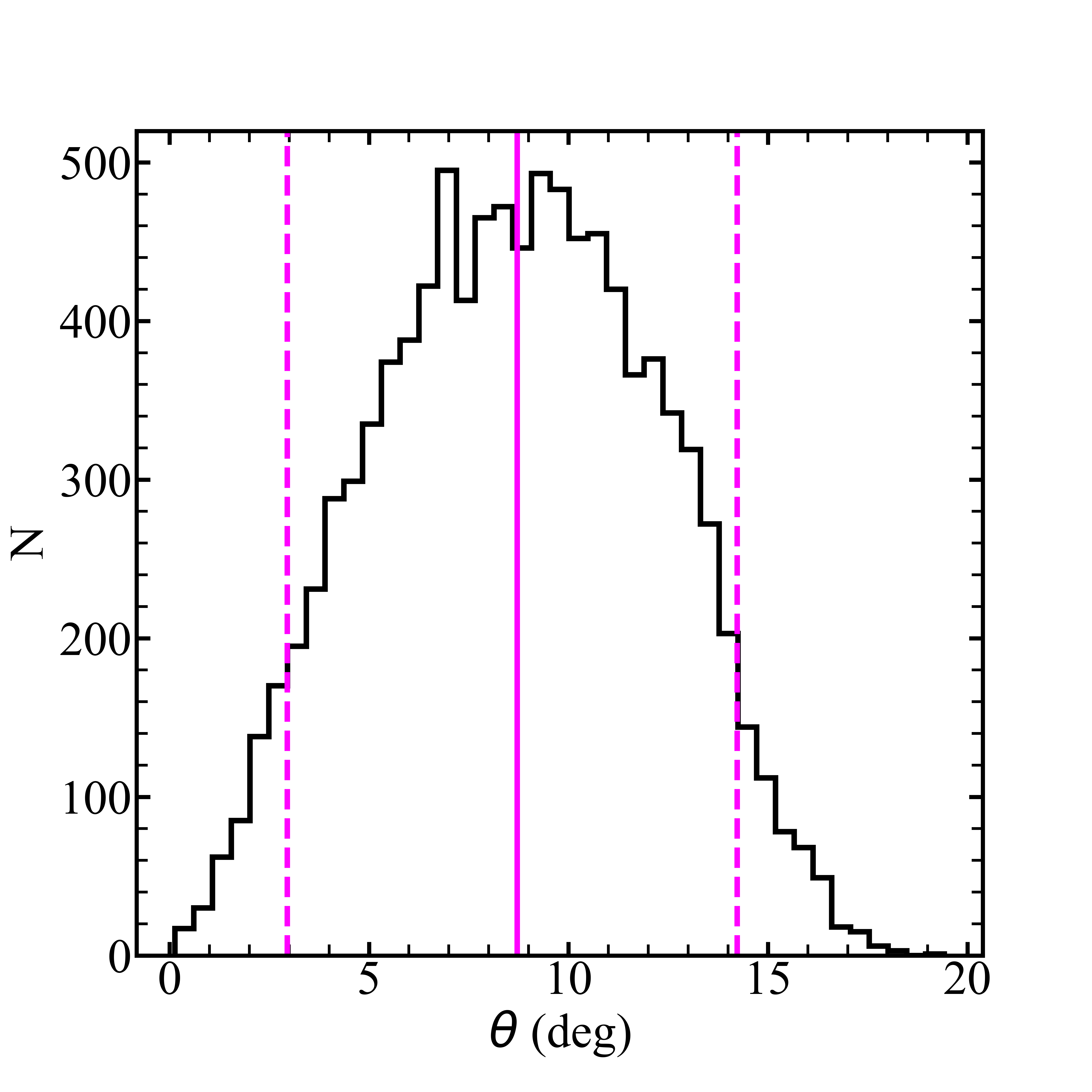}
\includegraphics[width=0.5\textwidth,trim=0 0 0 0]{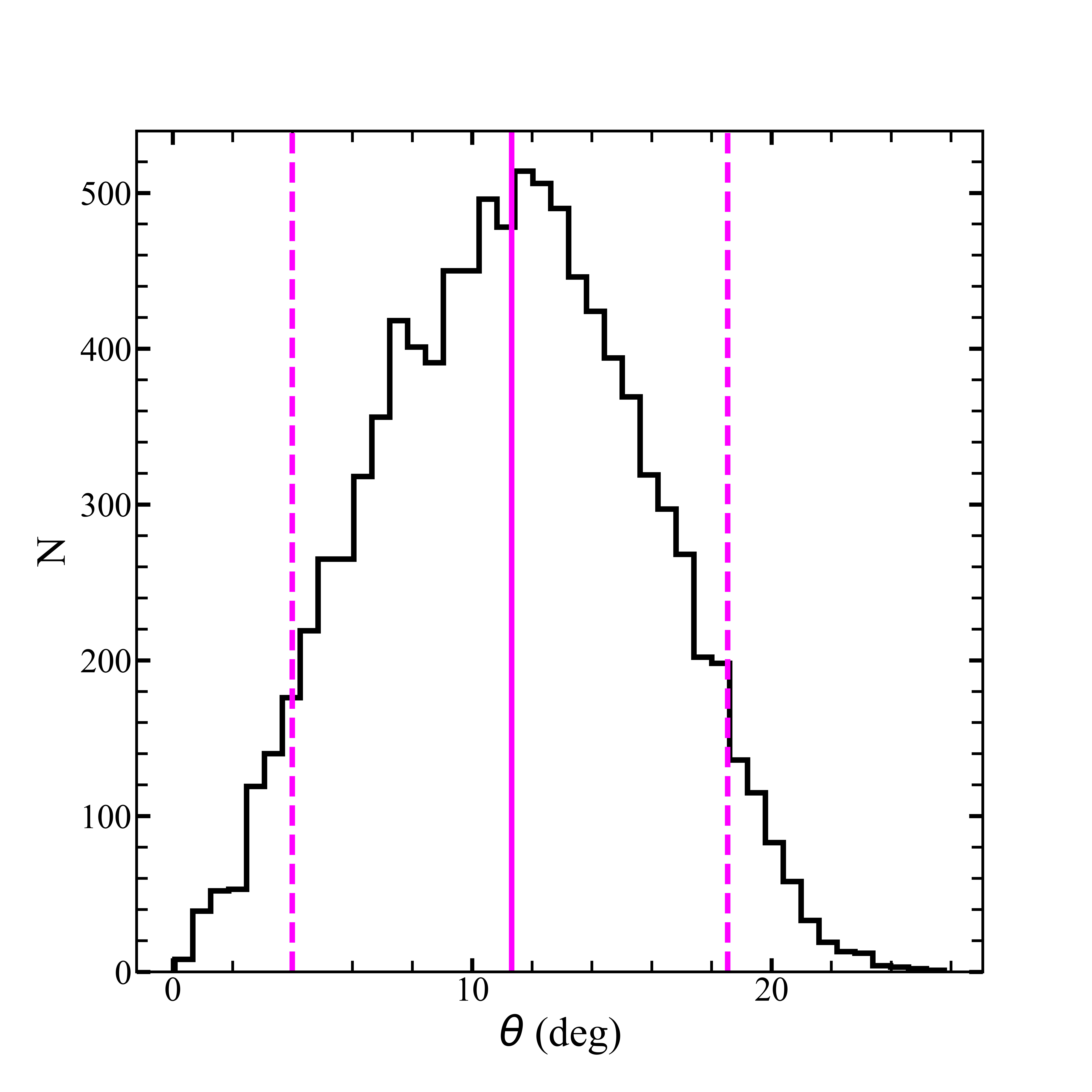}}
\caption{\small Distribution of line of sight angle $\theta$ for J1405+0415 (left) and J1610+1811 (right) based on our Bayesian method described in Section 4.2. The median $\theta$ value of 9$\degr$ and 11$\degr$ are marked by solid magenta lines for J1405+0415 and J1610+1811, respectively. Dot-dash lines represent the 90\% confidence intervals.}
\label{medianthetafig}
\end{figure*}

We repeat this 10,000 times to get the distribution of line of sight angle for the range of $\Gamma$ predicted by our IC/CMB model (shown in Figure \ref{medianthetafig}). Table~\ref{tab:table5} shows the median value and the 90\% confidence interval for $\theta$ and the corresponding values of the relativistic and jet parameters.

We denote upper limits of these parameters with downward arrows and lower limits with upward arrows. The jet parameters scale systematically with the radio-to-X-ray flux ratio. When the observed radio-to-X-ray flux ratio increases tenfold, the line-of-sight angle ($\theta$), magnetic field (B), electron number density ($n_e$), and kinetic energy (KE) of the jet increase by factors of 1.28, 2.67, 7.13, and 3.22, respectively. Conversely, the bulk Lorentz factor ($\Gamma$), Doppler factor ($\delta$), and relativistic speed ($\beta$) decrease by factors of 0.67, 0.625, and 0.98, respectively. Alternatively, if the radio-to-X-ray flux ratio decreases by a factor of 10, we find that $\theta$,  B, $n_e$, and KE decreases by a factor of 0.71, 0.33, 0.13, and 0.61, respectively. Meanwhile $\Gamma$, $\delta$, and $\beta$ increase by factors of 1.33, 1.38, and 1.01, respectively.

\begin{deluxetable}{lllll}[h!]
\tablecaption{Estimated values of relativistic and jet parameters based on our IC/CMB model. \label{tab:table5}}
\tablehead{
\colhead{Parameters} & \multicolumn{2}{c}{J1405+0415} & \multicolumn{2}{c}{J1610+1811}\\
\cline{2-3}
\cline{4-5}
\colhead{} & \colhead{Median} & \colhead{90\% C.I.} & \colhead{Median} & \colhead{90\% C.I.} \\
\cline{2-5}
\colhead{(1)} & \colhead{(2)} & \colhead{(3)} & \colhead{(4)} & \colhead{(5)}
}
\startdata
$\theta \downarrow$ & 9 & (3, 14) & 11 & (4, 18)\\
$\Gamma \uparrow$ & 4 & (3, 8) & 3 & (2, 6)\\
$\delta \uparrow$ & 6 & (6, 3) & 5 & (5, 3)\\
$\beta \uparrow$	 & 0.97 & (0.95, 0.99) & 0.95 & (0.92, 0.98)\\
B ($\mu$G) $\downarrow$	& 24 & (18, 42) & 10 & (8, 18)\\
$n_e~ (10^{-7} \rm cm^{-3}) \downarrow$ & 2.3 & (1.3, 7.2) & 0.3 & (0.2, 0.9)\\
K.E. $(10^{46} \rm erg~ s^{-1}) \downarrow$ & 4.7 & (1.3, 51.4) & 0.4 & (0.1, 3.9)\\
\enddata
\end{deluxetable}

\subsection{Benchmarking our model against other techniques and models}
Monitoring Of Jets in Active galactic nuclei with VLBA Experiments \citep[MOJAVE;][]{Lister+2018} observations of J1405+0415 reveals that the maximum apparent speeds ($\beta_{\rm app,max}$) based on five moving features is $9.2 \pm 1.8$ \citep{Lister+2019}. This is three times faster than previously determined apparent velocity upper limit of $3$ by \citep{Yang+2008}. A $\beta_{\rm app,max} = 9.2$ imply upper limit on the jet angle to the line of sight ($\theta_{\rm max} < 2 \tan^{-1}(1/\beta_{\rm app, max})$) of less than 12$\degr$. These results are consistent with our IC/CMB model.

\begin{figure}
\gridline{\fig{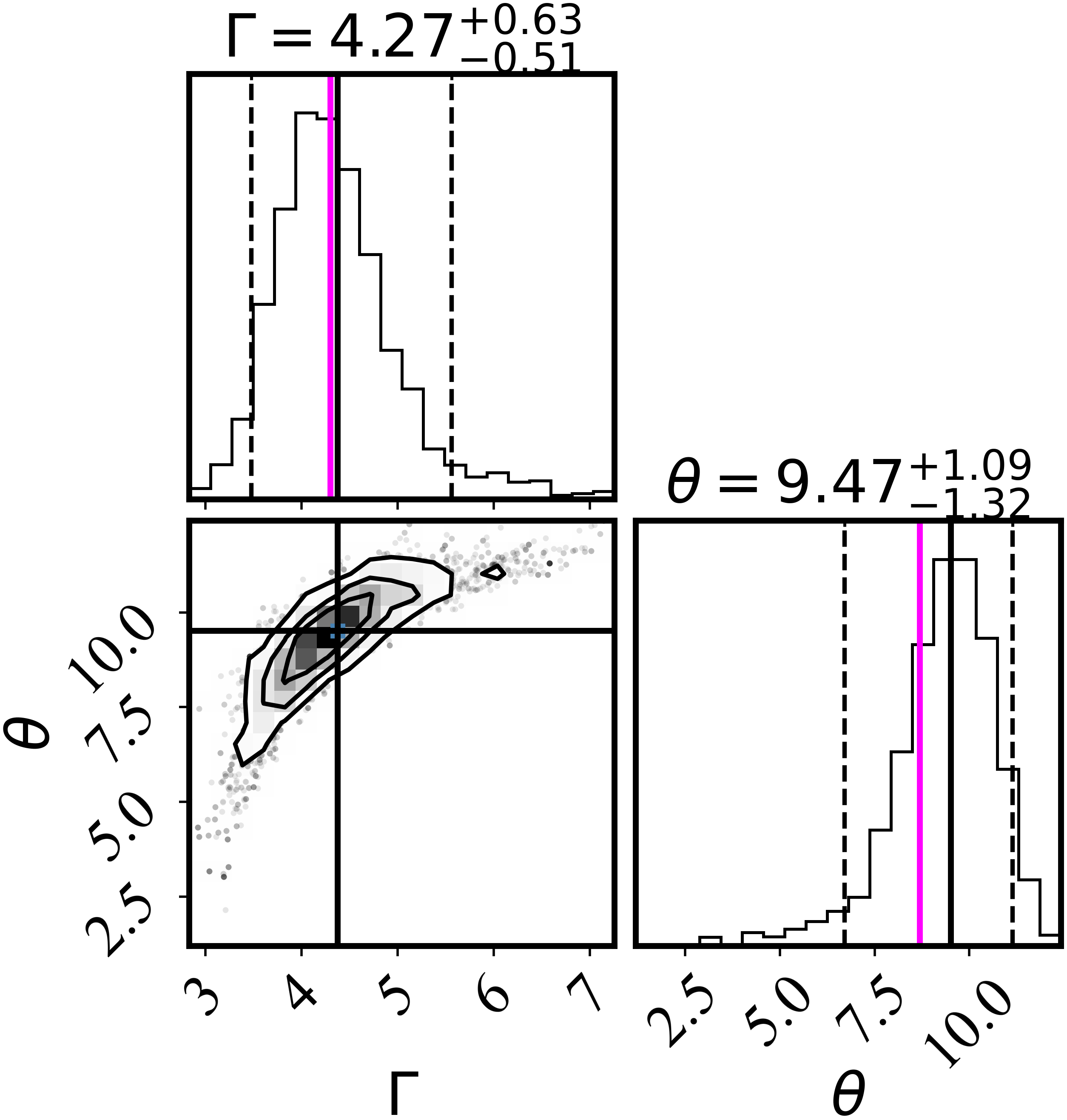}{0.4\textwidth}{}
          \fig{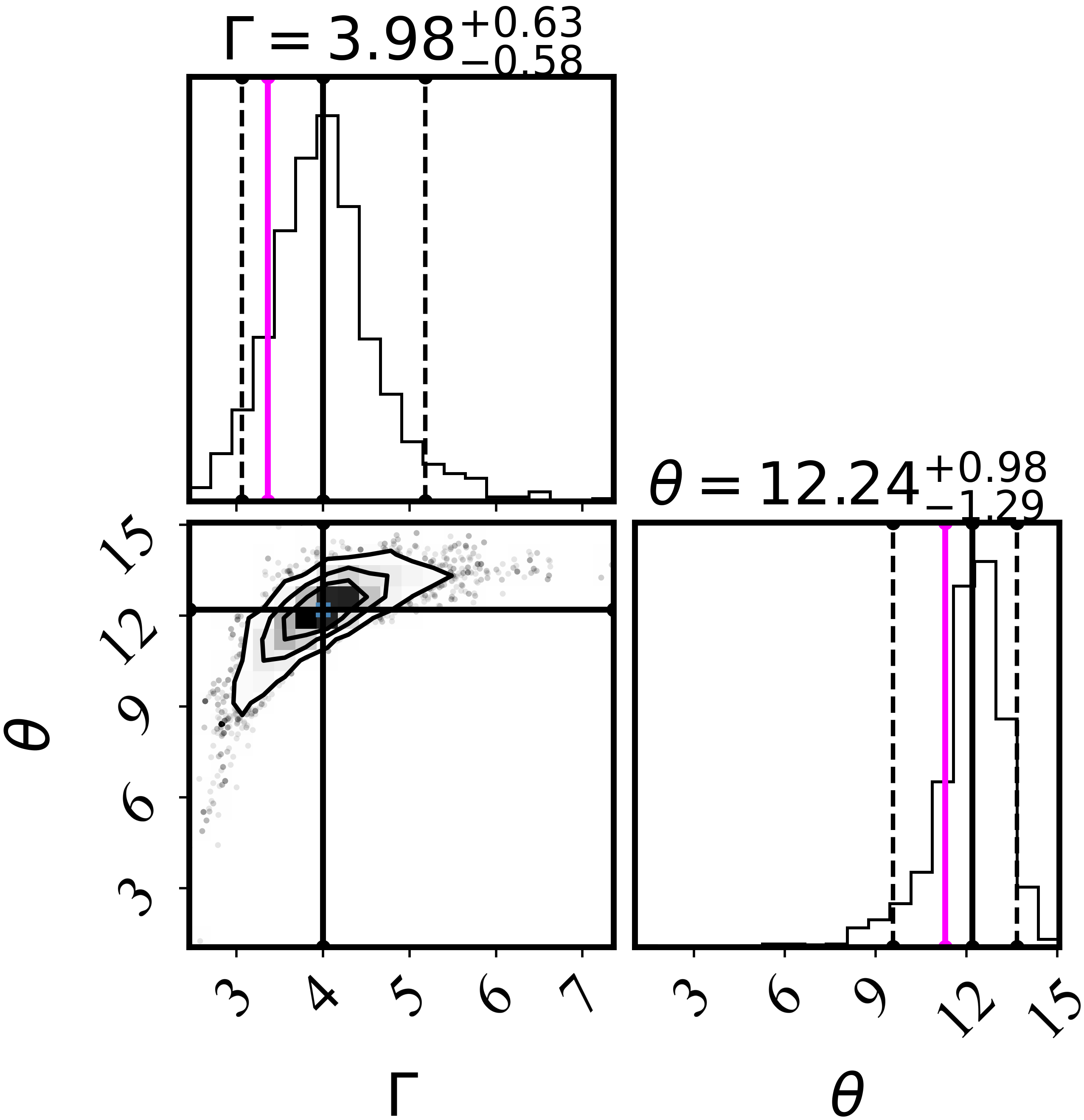}{0.4\textwidth}{}}
\caption{Two dimensional projections of the posterior probability distributions of $\theta$ and $\Gamma$ for J1405+0415 (left) and J1610+1811 (right) obtained using JetSeT. The black and magenta solid lines represent the median value of the parameter from JetSet and from our IC/CMB model, respectively. The black dashed line represents the 0.05 and 0.95 quantile from JetSet. The median values of $\theta$ and $\Gamma$ predicted by our IC/CMB model are consistent with the ones predicted by JetSet within the model uncertainty.}
\label{jetset}
\end{figure}

We further validated our results using Jet SED modeler and fitting Tool \citep[JetSeT,][]{Tramacere2020, Tramacere+2011, Tramacere+2009, Massaro+2006}. It is an open source computational tool for modeling radiative processes in astrophysical jets. We employed a simplified one-zone leptonic model, assuming a spherical emission region and a power-law energy distribution for the electrons. To maintain volumetric consistency with our cylindrical jet assumption, we calculated an effective radius for the spherical blob that produces an identical volume to our cylindrical model geometry. Synchrotron and IC/CMB processes were utilized to interpret the radio and X-ray emission. JetSeT's built-in optimization routines were employed to find the best-fit parameters for each quasar, focusing on the jet angle ($\theta$) and bulk Lorentz factor ($\Gamma$), as our data only provided two flux points. The remaining parameters were fixed to the values obtained from our IC/CMB model. Additionally, we used an Markov Chain Monte Carlo sampler to estimate the average values of $\theta$ and $\Gamma$, as illustrated in Figure \ref{jetset}. JetSeT results are consistent with our IC/CMB model within error margins. 

\section{Summary \& Conclusions}
The X-ray emission from quasar jets provides valuable insights into the physical processes of these powerful relativistic outflows. At high redshifts, the dominant X-ray emission mechanism is inverse-Compton scattering of CMB photons by synchrotron-emitting electrons. Our study focuses on two quasars, J1405+0415 and J1610+1811, at redshifts 3.215 and 3.122. These were selected from a parent sample of 123 radio-loud quasars identified by \cite{Gobeille2011} through cross-matching SDSS and FIRST surveys, with a radio flux density cutoff of 70 mJy and redshift $z\geq2.5$. 
14 sources were selected for further X-ray study by excluding triple sources, known X-ray jet sources, and raising the redshift cutoff to 3. Only J1405+0415 and J1610+1811 
showed evidence for extended X-ray jet emission in 10 ks Chandra observations by \cite{Schwartz+2019}. We present follow-up 6 GHz VLA observations in A-configuration and re-analysis of the $\sim$ 90 ks archival Chandra data to develop an IC/CMB model for these quasar jets.

We detect bright radio cores, with J1405+0415's core being ten times brighter than J1610+1811's. Quasar cores typically have a flat radio spectral index ($\alpha \sim 0$). J1405+0415 shows this expected flat spectrum, while J1610+1811 has a steeper core spectrum ($\alpha = 0.42$), indicating unresolved extended emission \citep{Maithil+2020}. In J1405+0415, we identify a compact knot that defines the jet direction and faint off-axis hotspot. J1610+1811 exhibits two equidistant hotspots, with the northwestern one ~20 times brighter than the southeastern. No continuous jet structure connecting the core and hotspot/lobe region is detected in our radio observations, which are sensitive to $\mu$Jy levels. In our IC/CMB model, we treat radio upper limits as detections, adopting a jet spectral index of $1.0$ for J1405+0415 based on detection in X-rays and $0.7$ for J1610+1811 assuming a typical value.

The 1.49 GHz archival VLA dataset with $1\arcsec$ resolution was analyzed to study radio polarization. The radio core of each quasar has a fractional polarization of $\sim2\%$. The jet edge in J1405+0415 shows $11.8\pm2.6 \%$ fractional polarization, which decreases to $2.2\pm0.8\%$
before the hotspot. J1610+1811's north-west hotspot has a fractional polarization of $3.9\pm0.7\%$. The spectral indices from the archival data are consistent with the new 6 GHz measurements.

Both quasars display X-ray cores with similar net counts and a flat/hard photon index. The co-added spectrum of J1610+1811's core shows intrinsic absorption. We simulated the X-ray cores and compared simulated and observed counts outside the core region, revealing significant excess when comparing net counts as a function of azimuth angle and surface brightness profiles. We confirm X-ray extended emission with $>5\sigma$ significance, coinciding with the direction of the radio extended emission. The X-ray jet-to-core count ratios are 0.023 for J1405+0415 and 0.004 for J1610+1811, consistent with literature values spanning 0.009 to 0.18 \citep[e.g.,][]{Siemiginowska+2003, Cheung+2012, Simionescu+2016, Worrall+2020}. \cite{Marshall+2018} report a median X-ray jet-to-core ratio of $2\%$ for high-redshift quasars.

\begin{figure*}[h!]
\gridline{
\includegraphics[width=0.52\columnwidth,trim=0cm 0.18cm 0cm 0cm,clip]{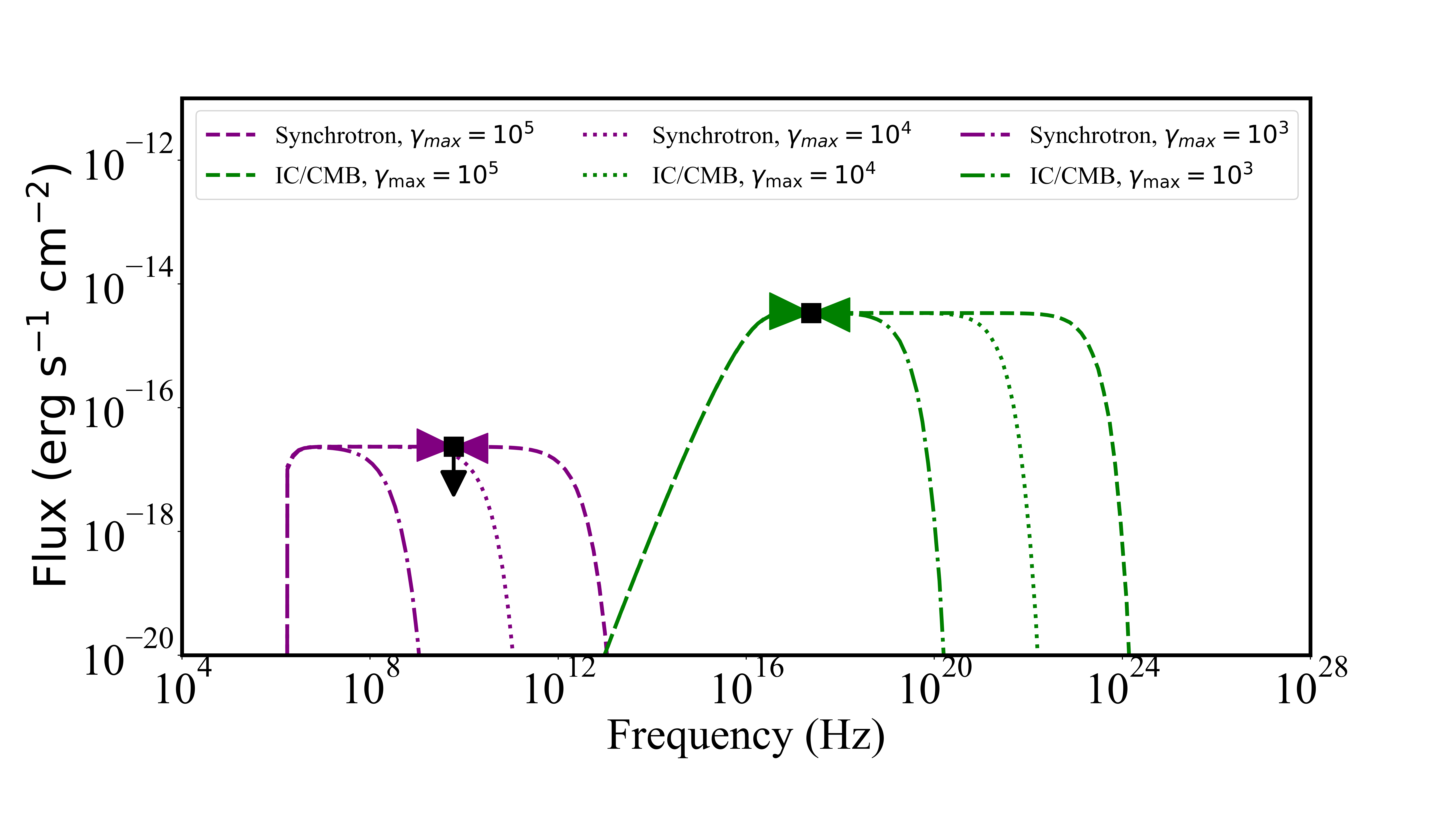}
\includegraphics[width=0.52\columnwidth,trim=0cm 0.18cm 0cm 0cm,clip]{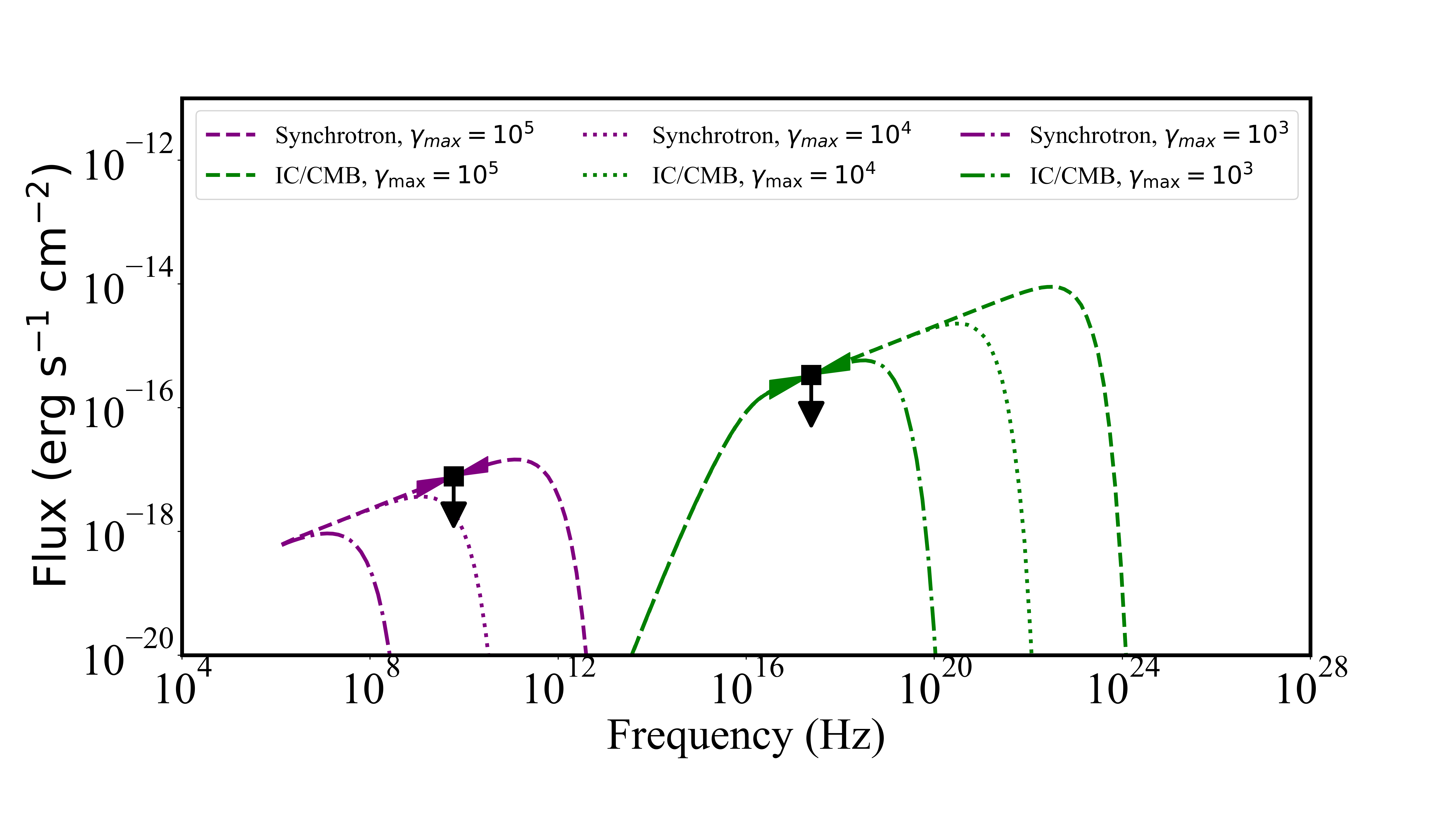}}
\caption{\small Spectral energy distribution of J1405+0415 (left) and J1610+1811 (right) obtained with JetSet using parameter values of our IC/CMB model. Black squares shows the observed fluxes. JetSet's synchrotron and IC/CMB model for J1405+0415 (J1610+1811) assuming a power-law distribution of electrons with an index of 3 (2.4) and three values of maximum Lorentz factor for electrons is shown in purple and green lines, respectively. The bow-tie shows the change in slope of the spectrum for power-law index of 2.4 and 3.8 for J1405+0415 and 2.12 and 2.8 for J1610+1811.}
\label{SEDpowerlaw}
\end{figure*}

To model the X-ray jet emission, we employed a cylindrical geometry with a radius of $3~ \rm kpc$ and a power-law distribution of electrons, with minimum and maximum Lorentz factors of 30 and $10^5$ respectively. Using the observed jet length and upper limits on radio fluxes for both targets, along with the upper limit on X-ray flux for J1610+1811, our model provides reasonable constraints on the relativistic and jet parameters. A jet with Lorentz factor of $\sim$4 (3) and magnetic field of $\sim 24\mu G$ (10$\mu G$) for J1405+0415 (1610+1811) is typical for high-redshift quasar jets \citep{Worrall+2020}. The median jet kinetic power of J1405+0415 is $5 \times 10^{46}~ \rm erg~ s^{-1}$ is $\sim 50\%$ of its bolometric luminosity ($9  \times 10^{46}~ \rm erg~ s^{-1}$, \citealt{Shen+2011}). For J1610+1811, the jet kinetic power is $\sim0.4 \times 10^{46}~ \rm erg~ s^{-1}$ is only 2\% of its bolometric luminosity ($2 \times 10^{47}~ \rm erg~ s^{-1}$, \citealt{Shen+2011}). Notably, we determined a shorter lifetime for radio-emitting electrons with $\gamma_{\rm max} = 5000$ compared to those emitting X-rays at 1 keV through IC/CMB. The spectral energy distribution (SED), illustrated in Figure \ref{SEDpowerlaw}, indicates that the synchrotron emission from electron populations with $\gamma_{\rm max} \leq 10^4$ remains below the observed 6 GHz flux limit, yet still capable of producing X-ray jet emission via the IC/CMB process. This suggests that it is the low-energy electrons that primarily produce X-rays via IC/CMB.

\begin{figure*}[h!]
\gridline{
\includegraphics[width=0.52\columnwidth,trim=0cm 0.18cm 0cm 0cm,clip]{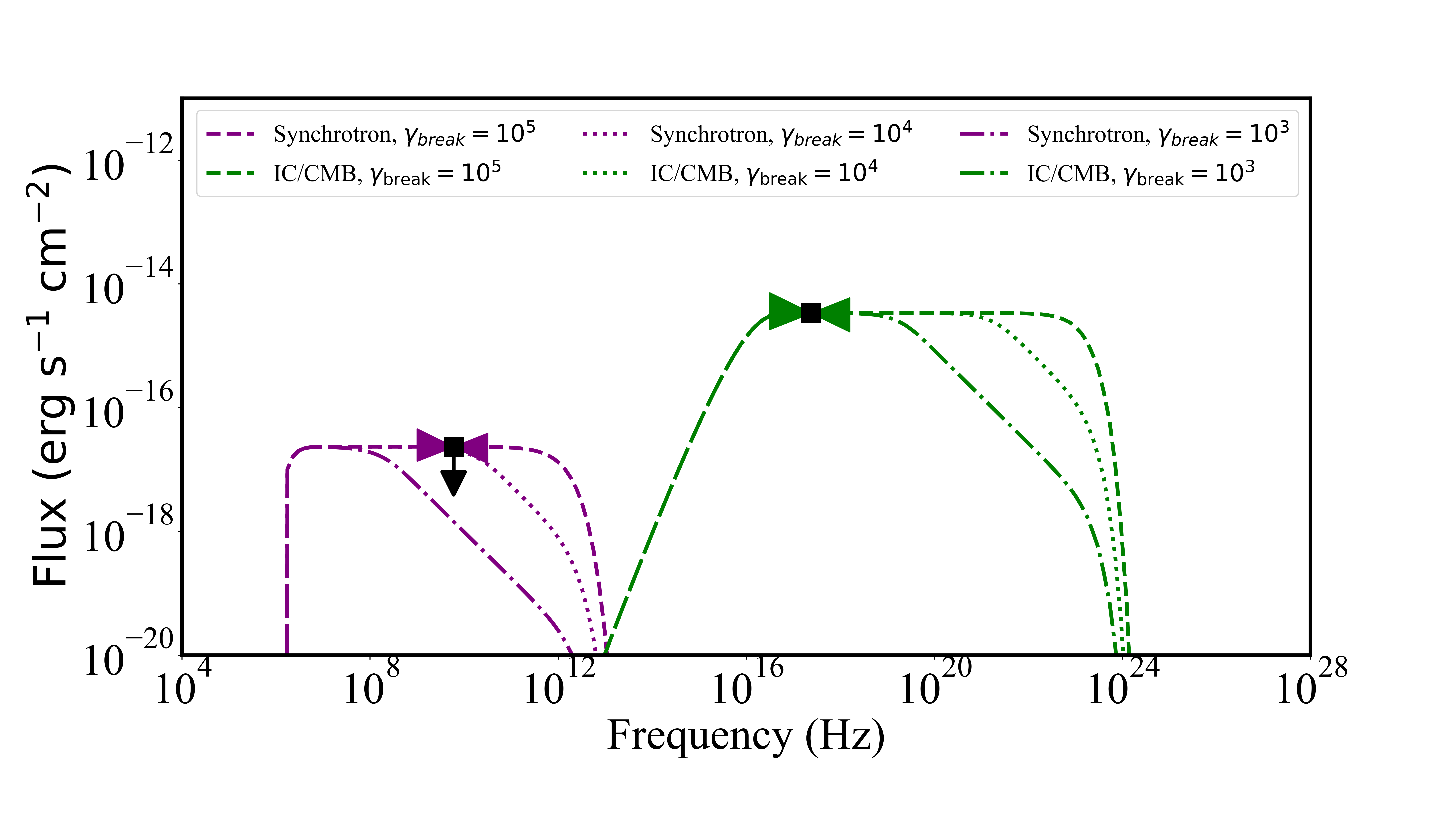}
\includegraphics[width=0.52\columnwidth,trim=0cm 0cm 0cm 0cm,clip]{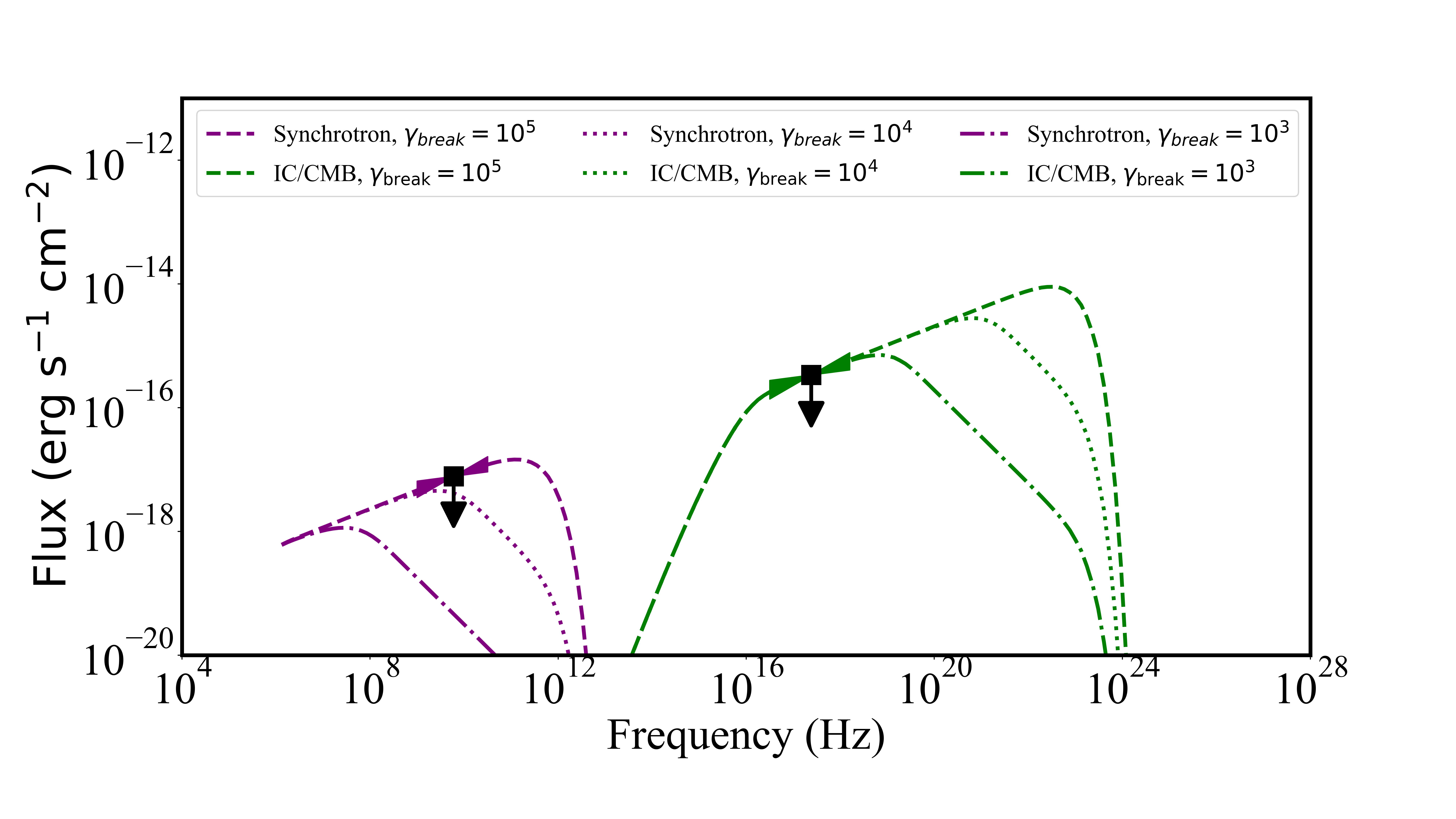}}
\caption{\small Spectral energy distribution of J1405+0415 (left) and J1610+1811 (right) obtained with JetSet using parameter values of our IC/CMB model. Black squares shows the observed fluxes. The JetSet model assumes a broken power-law distribution of electrons with low-energy $(\gamma \leq \gamma_{\rm break})$ spectral slope of 3 (2.4) and high-energy ($\gamma > \gamma_{\rm break}$) spectral slope of 4.5 for different values of $\gamma_{\rm break}$ for J1405+0415 (J1610+1811).}
\label{SEDbrokenpowerlaw}
\end{figure*}
Our findings indicate that IC/CMB is a feasible mechanism for generating the observed X-ray emission from quasar jets. However, our model relies on the availability of only radio and X-ray emissions from the jets. In our case, radio is upper limit but treated as a measurement in the IC/CMB model. Figure \ref{SEDbrokenpowerlaw} illustrates the SED for a broken power-law distribution of electrons, showcasing three different break Lorentz factors. Notably, when the break occurs below $10^4$, the 6 GHz radio emission dips below the observed flux limit while still generating X-ray emissions via IC/CMB. This behavior is seen in \cite{Worrall+2020} for the quasar PKS J1421-0643 at $z=3.69$, based on deep radio, optical and X-ray observations. Nevertheless, the SED shape varies significantly, necessitating multi-wavelength observations to accurately constrain the underlying electron distribution. For both power-law and broken power-law electron distribution, the predicted gamma-ray flux remains below the sensitivity of Fermi $\gamma$-ray space telescope for the low-energy electrons.
\newpage

\section{Acknowledgments}

 Support for this work was provided by the National Aeronautics and Space Administration through Chandra Award Numbers GO8-19077X, GO8-19093X, GO1-22086X
issued by the Chandra X-ray Observatory Center, which is operated by the Smithsonian Astrophysical Observatory for and on behalf of the National Aeronautics Space Administration under contract NAS8-03060.
D.A.S. and A.S. acknowledge support from NASA Contract NAS8-03060 to the {\sl Chandra X-ray Center}.
C.C.C. at the Naval Research Laboratory is supported by NASA DPR S-15633-Y. 

This research has made use of data obtained from the Chandra Data Archive, and software 
provided by the Chandra X-ray Center (CXC) in the application packages CIAO and Sherpa.

This paper employs a list of Chandra datasets, obtained by the Chandra X-ray Observatory, contained in the Chandra Data Collection (CDC) 293~{\href{https://doi.org/10.25574/cdc.293}{doi:1025574/cdc.293}}

The National Radio Astronomy Observatory is a facility of the National Science Foundation operated under a cooperative agreement by Associated Universities, Inc. This research has made use of data from the MOJAVE database that is maintained by the MOJAVE team \citep{Lister+2018}.


\vspace{5mm}
\facilities{Chandra, VLA}


\software{CIAO \citep{Fruscione2006}, Sherpa \citep{Freeman2001,Sherpa2024}, BEHR \citep{Park2006}, CASA \citep{CASA2022}, JetSet \citep{Tramacere+2009, Tramacere+2011, Tramacere2020}}



\newpage

\appendix

\section{Historical VLA Data} \label{Appendix_Historic}
In addition to the newly acquired VLA data we also reduced a couple of historical VLA datasets to obtain radio polarization information. We were looking for datasets that had multiple scans (and extensive parallactic angle coverage) of well-known polarization calibrators like 3C286 and others.\footnote{Table 7.2.6, \url{https://science.nrao.edu/facilities/vla/docs/manuals/obsguide/modes/pol}} 

We found only one dataset at ~1" resolution: VLA A-array at 1.49 GHz. Here, we report the polarization results from project AB449, dated August 16, 1987.
While the dataset had two scans of 1328+307 (3C286), the lack of good parallactic angle coverage did not allow a robust instrumental leakage calibration. We therefore used the unpolarized calibrator 1358+624\footnote{Table 7.2.4, \url{https://science.nrao.edu/facilities/vla/docs/manuals/obsguide/modes/pol}} to calibrate the instrumental leakages. The basic initial calibration was carried out by following the step-by-step procedure outlined in the AIPS Cookbook.\footnote{ \url{http://www.aips.nrao.edu/CookHTML/CookBookse124.html\#x172-375000A.2}} After the initial calibration, the instrumental leakages were estimated using 1358+624 with the AIPS task PCAL. The antenna leakages turned out to be between $2-5$\%. The absolute polarization angle calibration was completed using 1328+307 (3C286) and AIPS tasks RLDIF and CLCOR. The final Stokes I, Q, and U images were made after several rounds of phase-only and amplitude+phase self-calibration using the AIPS tasks IMAGR and CALIB. The final r.m.s. noise in the total intensity (I) image at 1.49 GHz was $\sim$0.17~mJy~beam$^{-1}$ for J1405+0415 and $\sim$0.13~mJy~beam$^{-1}$ for J1610+1811. 

Using the AIPS task COMB, and the Stokes Q and U images, we created images of the polarized intensity (P), polarization angle ($\chi$) and fractional polarization (F=P/I). The polarized intensity images were corrected for Ricean bias while using COMB. The P image was made after blanking pixels with signal-to-noise ratios (SNR) less than 3, while the $\chi$ image was made after blanking pixels with angle errors greater than 10 degrees. Finally, the F image was made after blanking pixels with greater than 10\% error in fractional polarization. The polarization images for J1405+0415 and J1610+1811 are presented in Figure~\ref{fig1402pol}. 

The $1.5-4.9$ GHz spectral index image for J1405+0415 was created using L-band data described above and an archival 4.86~GHz dataset (Project ID AW122) from December 10, 1984. Similarly, the spectral index image for J1610+1811 was created using the L-band data and archival 4.86~GHz dataset from project AB750 from September 3, 1995. Images at 4.9 GHz were made with a cell-size, image-size and beam-size identical to the 1.5 GHz data, and the image centres were made to be coincident using the AIPS task OGEOM, before creating the spectral index images with task COMB. Total intensity pixels with SNR values below 3 were blanked. The spectral index images are presented in Figure~\ref{fig1402spix}. All estimates of flux densities and mean spectral indices reported in the paper were obtained using the AIPS verbs TVWIN and IMSTAT. 

\begin{deluxetable}{crcccc}[h!]
\tablecaption{ Observation log for the Chandra data. \label{tab:table2}}
\tablehead{
\colhead{Object name} & \colhead{$z$} & \colhead{$N_{\rm H}$} & \colhead{ObsID} & \colhead{Date} & \colhead{Live time} \\\colhead{} & \colhead{} & \colhead{ ($10^{20} \rm cm^{-2}$)} & \colhead{} & \colhead{} & \colhead{(ks)}\\
\colhead{(1)} & \colhead{(2)} & \colhead{(3)} & \colhead{(4)} & \colhead{(5)} & \colhead{(6)}
}
\startdata
J1405+0415	& 3.215  & 2.17     & 23649    & 2021 Apr 29  &  15.28  \\
{ }	& { } & { }                 & 24316    & 2021 Apr 18  &  14.33  \\
{ }	& { } & { }                 & 24317    & 2021 Apr 21  &  24.79  \\
{ }	& { } & { }                 & 25011    & 2021 Apr 19  &  20.03  \\
{ }	& { } & { }                 & 25014    & 2021 May 01  &  13.37  \\
\hline
J1610+1811	& 3.122  &  3.62    &  23648   & 2021 May 04  &  26.69  \\
{ }	& { } & { }                 &  24486   & 2021 May 12  &  14.33  \\
{ }	& { } & { }                 &  24487   & 2021 May 29  &  28.59  \\
{ }	& { } & { }                 &  25035   & 2021 May 12  &  14.71  \\
\enddata
\tablecomments{(1) Source name. (2) Redshift of the sources from \cite{Sowards+2005, Paris+2018}. (3) Galactic neutral hydrogen column density ($N_{\rm H}$) adopted from \cite{DickeyLockman1990}. (4) Observation Id. (5) Date of observation. (6) Exposure time.}
\end{deluxetable}

\subsection{Radio Polarization \& Spectral Indices from Archival VLA data}

\begin{figure*}[h!]
\includegraphics[width=9.0cm,trim=50 170 0 100]{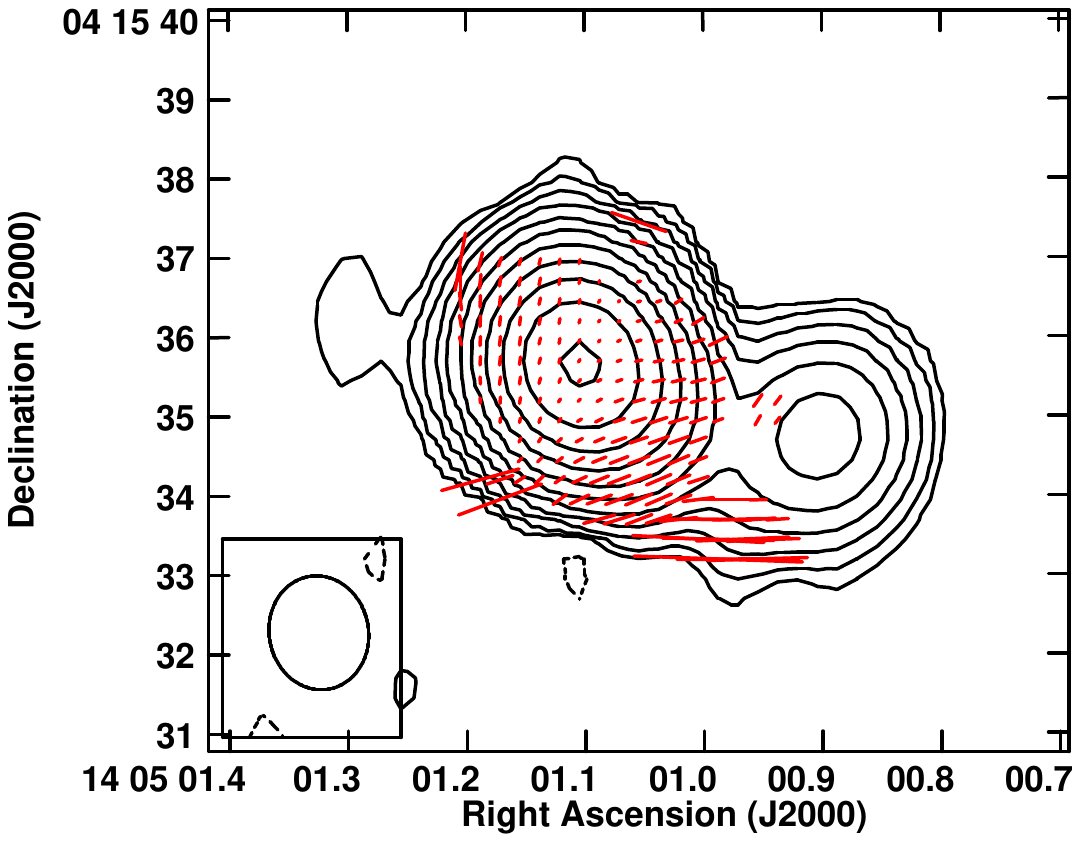}
\includegraphics[width=8.2cm,trim=50 150 0 140]{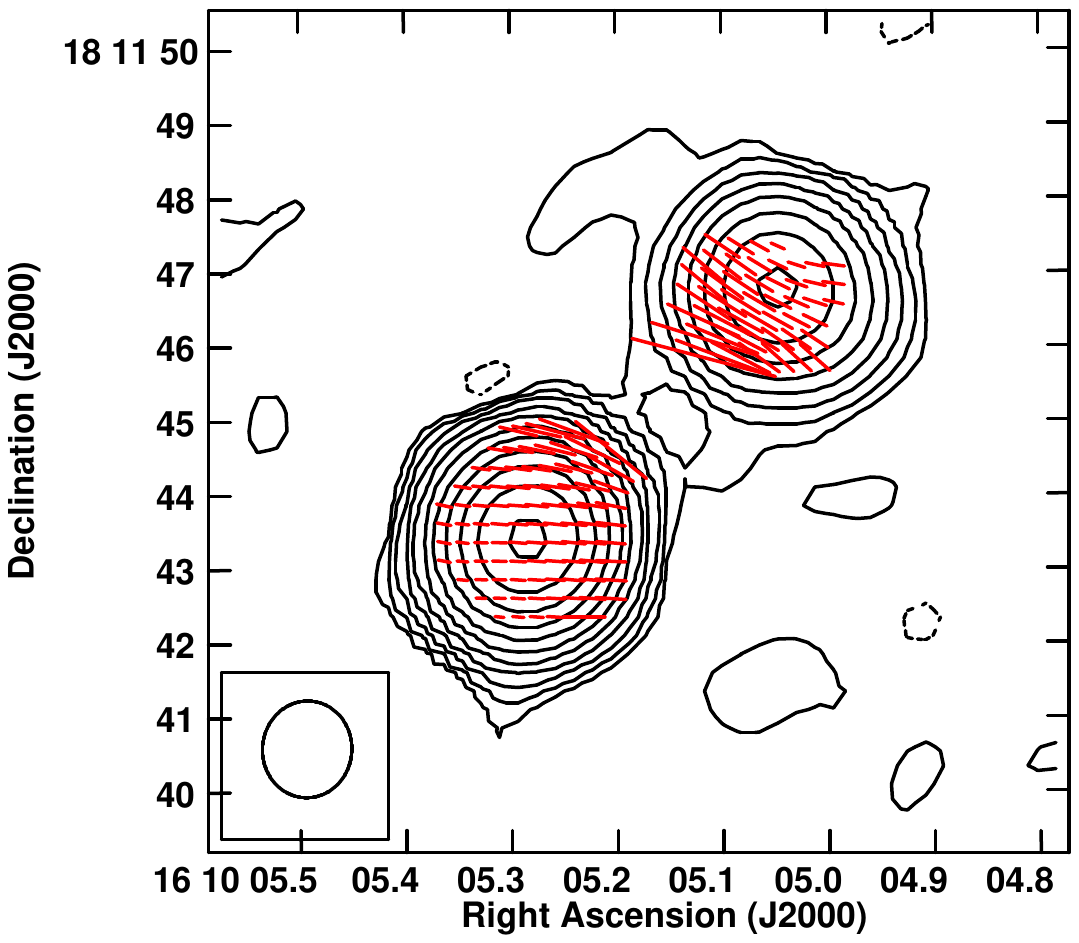}
\caption{\small VLA 1.49~GHz contour image of (left) J1405+0415 and (right) J1610+1811 with the inferred magnetic field vectors superimposed in red. These vectors are obtained by rotating the polarization vectors by 90 degrees assuming optically thin emission. The contour levels are (left) $6.414\times10^{-3}\times(-0.085, 0.085, 0.17, 0.35, 0.70, 1.40, 2.80, 5.60, 11.25, 22.50, 45, 90)$~Jy~beam$^{-1}$; 1 arcsec of polarization vector corresponds to 15.4\% fractional polarization and the synthesized beam (shown in the bottom left corner) is of size $1.44\arcsec \times 1.25\arcsec$ at a PA of $10.2\degr$. The contour levels are (right) $2.162\times10^{-3}\times(-0.085, 0.085, 0.17, 0.35, 0.70, 1.40, 2.80, 5.60, 11.25, 22.50, 45, 90)$~Jy~beam$^{-1}$; 1 arcsec of polarization vector corresponds to 7\% fractional polarization and the synthesized beam (shown in the bottom left corner) is of size $1.31\arcsec \times 1.20\arcsec$ at a PA of $-4.7\degr$.}
\label{fig1402pol}
\end{figure*}

Historical VLA data reveals the core fractional polarization in J1405+0415 to be $1.9\pm0.1$\%, the jet edge to be $11.8\pm2.6$\% and the jet region before the hotspot to be $2.2\pm0.8$\% (see Figure~\ref{fig1402pol}). There is a sudden change in the polarization vectors inside the core region suggesting that there is an unresolved jet component there along with the core and the optical depth is changing between the two components going from optically thick in the core to optically thin in the inner jet. Assuming optically thin emission, the magnetic (B-) field direction can be inferred to be perpendicular to the $\chi$ vectors \citep{Pacholczyk1970}. The B-fields appear to be tangential to the jet/lobe axis towards the south in J1405+0415. Such a structure could arise due to B-field shearing as the jet bends and interacts with the medium \citep[e.g.,][]{Kharb2008, Baghel2023, Ghosh2023}. The B-field in the region just before the hotspot to the west is transverse to the jet direction, consistent with compressed magnetic fields as are often observed in jet knots or hotspots \citep[e.g.,][]{Baghel2023,Ghosh2023}.

For J1610+1811, the fractional polarization in the core was $2.2\pm0.2$\% and the hotspot region was $3.9\pm0.7$\% (see Figure~\ref{fig1402pol}). The fractional polarization increases at the edge of the core region to $3.7\pm0.5$\% consistent with the presence of a jet or lobe region that is not fully resolved from the core. Similarly, the fractional polarization reaches its highest at $4.9\pm0.8$\% in the region before the hotspot. The inferred B-field is oblique to the jet direction in the core but becomes largely transverse in the jet/lobe region close to the core and stays transverse to the hotspot including the region just before it. Transverse B-fields in the hotspot region is consistent with B-field compression in the terminal shock as mentioned above.
\begin{figure*}[h!]
\centerline{
\includegraphics[width=10cm,trim=20 140 0 120]{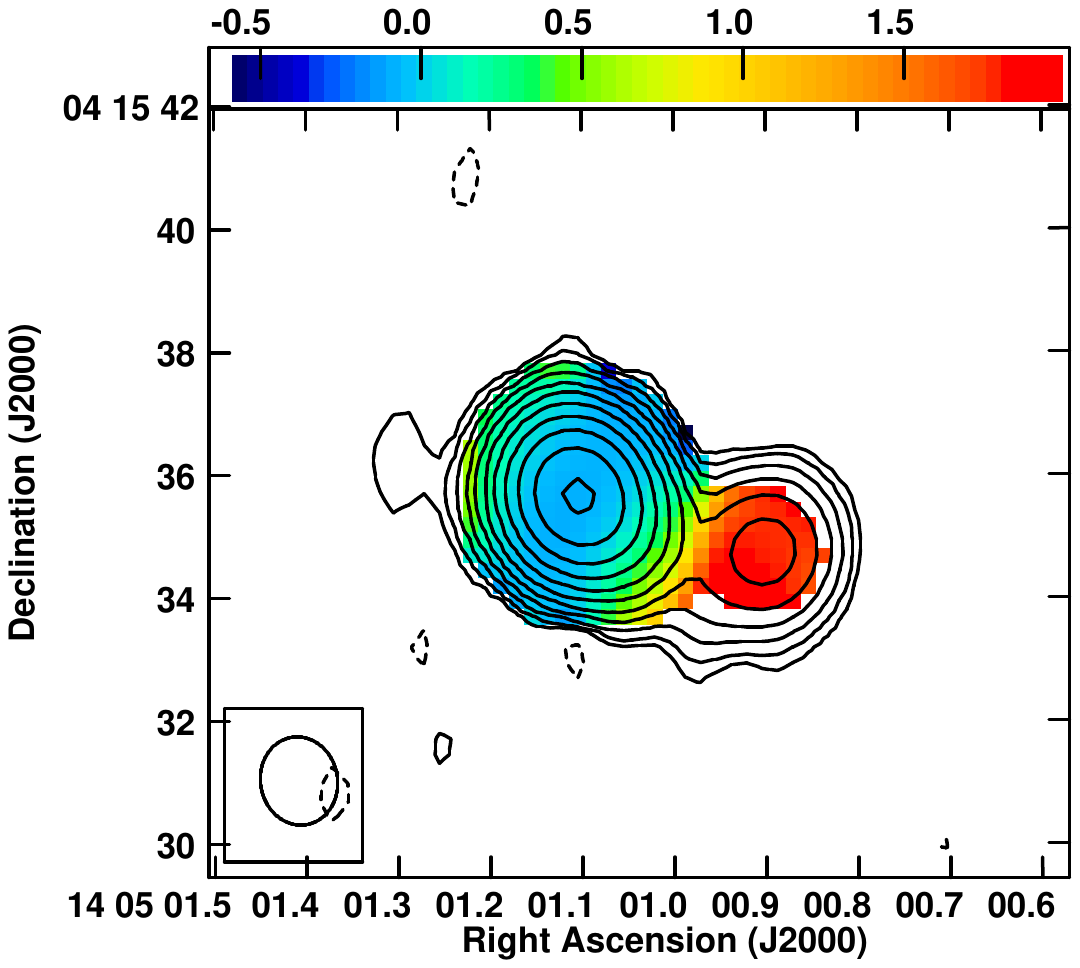}
\includegraphics[width=8.6cm,trim=75 130 0 135]{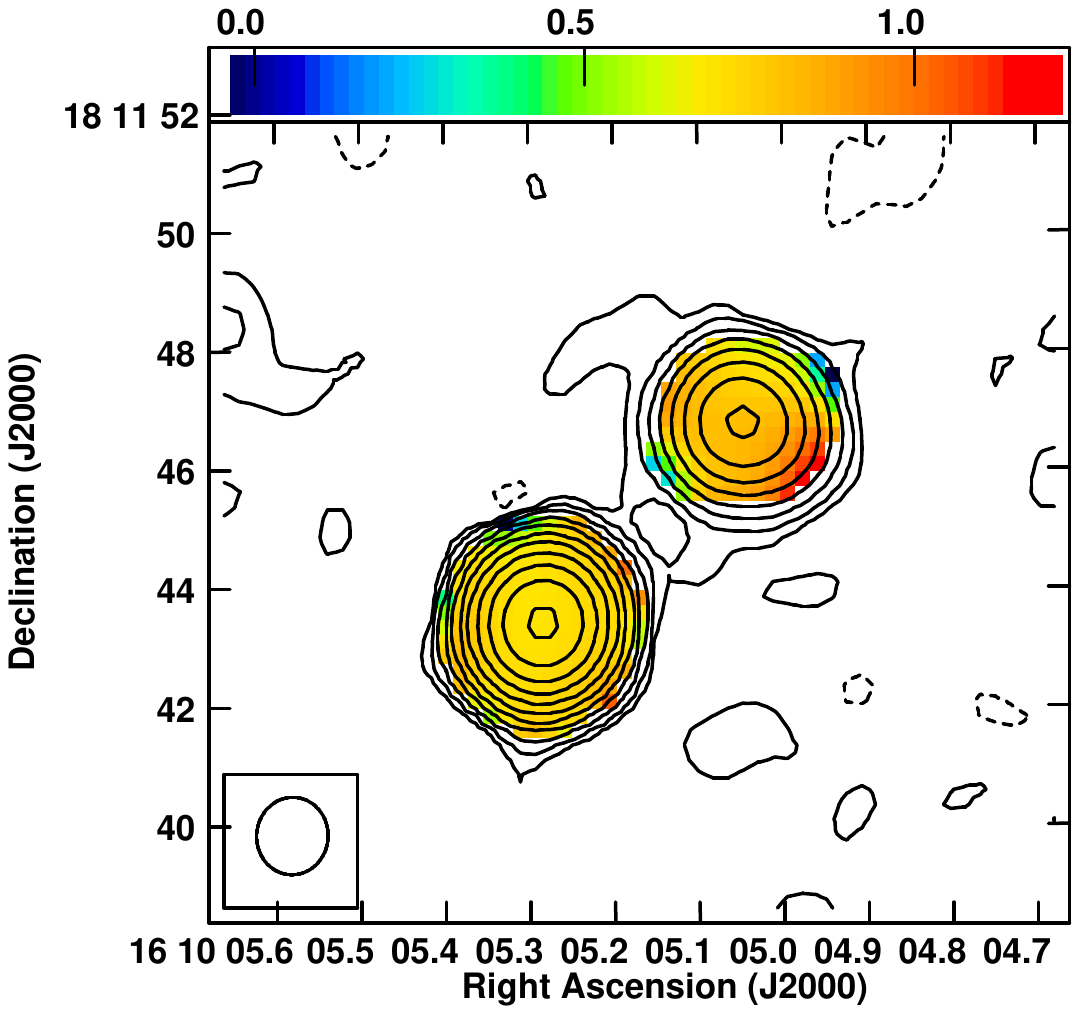}}
\caption{\small (Left) VLA contour image of J1405+0415  at 1.49 GHz with $1.5 - 4.9$~GHz spectral index values in color. The contour levels are $6.414\times10^{-3}\times(-0.085, 0.085, 0.17, 0.35, 0.70, 1.40, 2.80, 5.60, 11.25, 22.50, 45, 90)$~Jy~beam$^{-1}$. The synthesized beam shown in the bottom left corner is of size $1.44\arcsec \times 1.25\arcsec$ at a PA of $10.2\degr$. (Right) VLA contour image of J1610+1811 at 1.49 GHz with $1.5 - 4.9$~GHz spectral index values in color. The contour levels are $2.162\times10^{-3}\times(-0.085, 0.085, 0.17, 0.35, 0.70, 1.40, 2.80, 5.60, 11.25, 22.50, 45, 90)$~Jy~beam$^{-1}$. The synthesized beam shown in the bottom left corner is of size $1.31\arcsec \times 1.20\arcsec$ at a PA of $-4.7\degr$.}
\label{fig1402spix}
\end{figure*}

The mean 1.5 and 4.9 GHz spectral index value for the core and hotspot region for J1405+0415 are $0.15\pm0.04$ and $1.69\pm0.15$, respectively. Similarly, for J1610+1811, the core has a $\alpha$ value of $0.71\pm0.06$ while the hotspot region has a spectral index value of $0.76\pm0.01$. The spectral index value for the hotspot in J1405+0415 and the spectral index value for the core in J1610+1811 appear to be steeper than those observed in nearby radio galaxies and quasars \citep[e.g.,][]{Kharb2008,Baghel2023}.

\counterwithin{figure}{section}

\section{Determining the appropriate \textit{AspectBlur} Parameter}\label{Appendix1}
\begin{figure*}[h!]
\gridline{
\includegraphics[width=0.52\columnwidth,trim=0cm 0cm 0cm 0cm,clip]{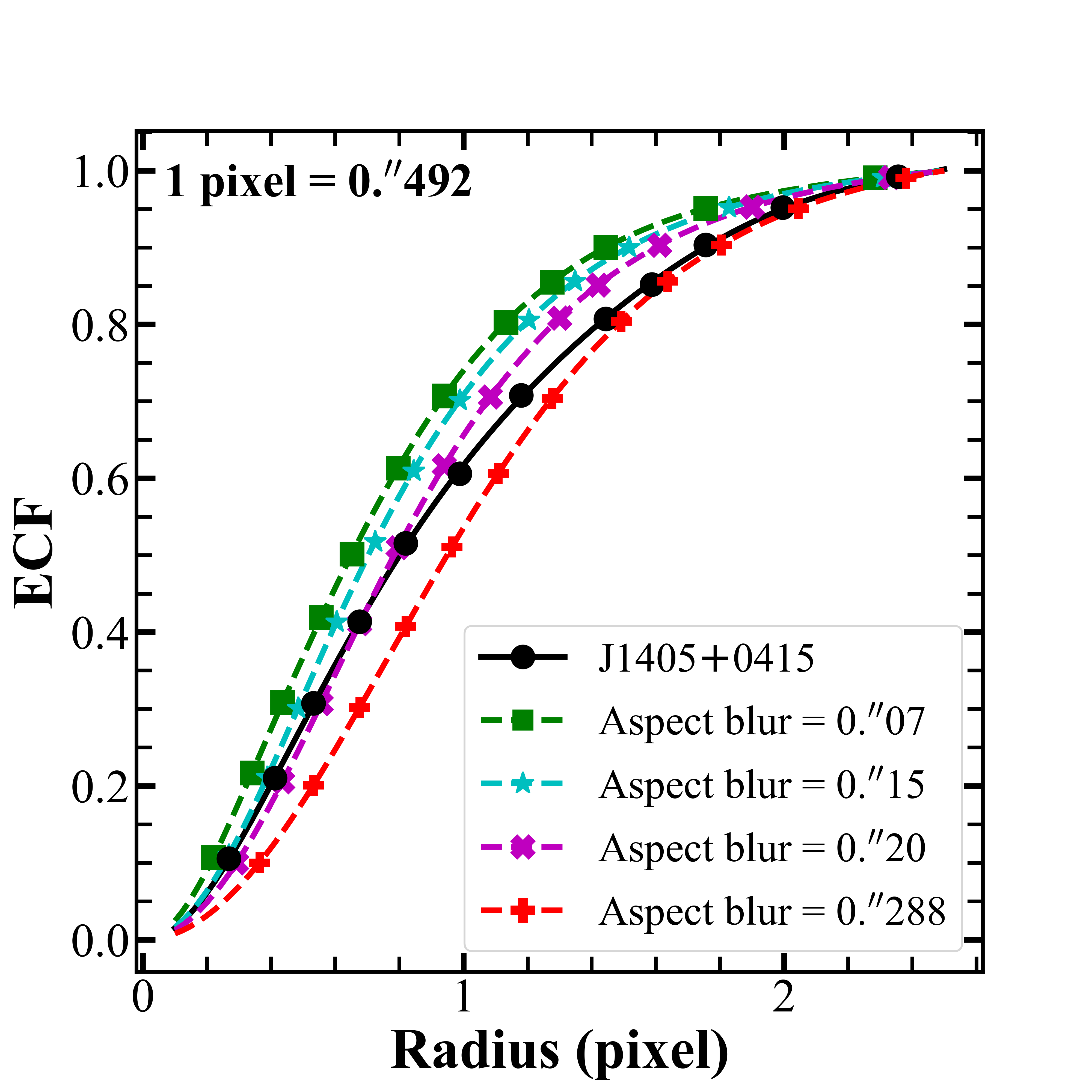}
\includegraphics[width=0.52\columnwidth,trim=0cm 0cm 0cm 0cm,clip]{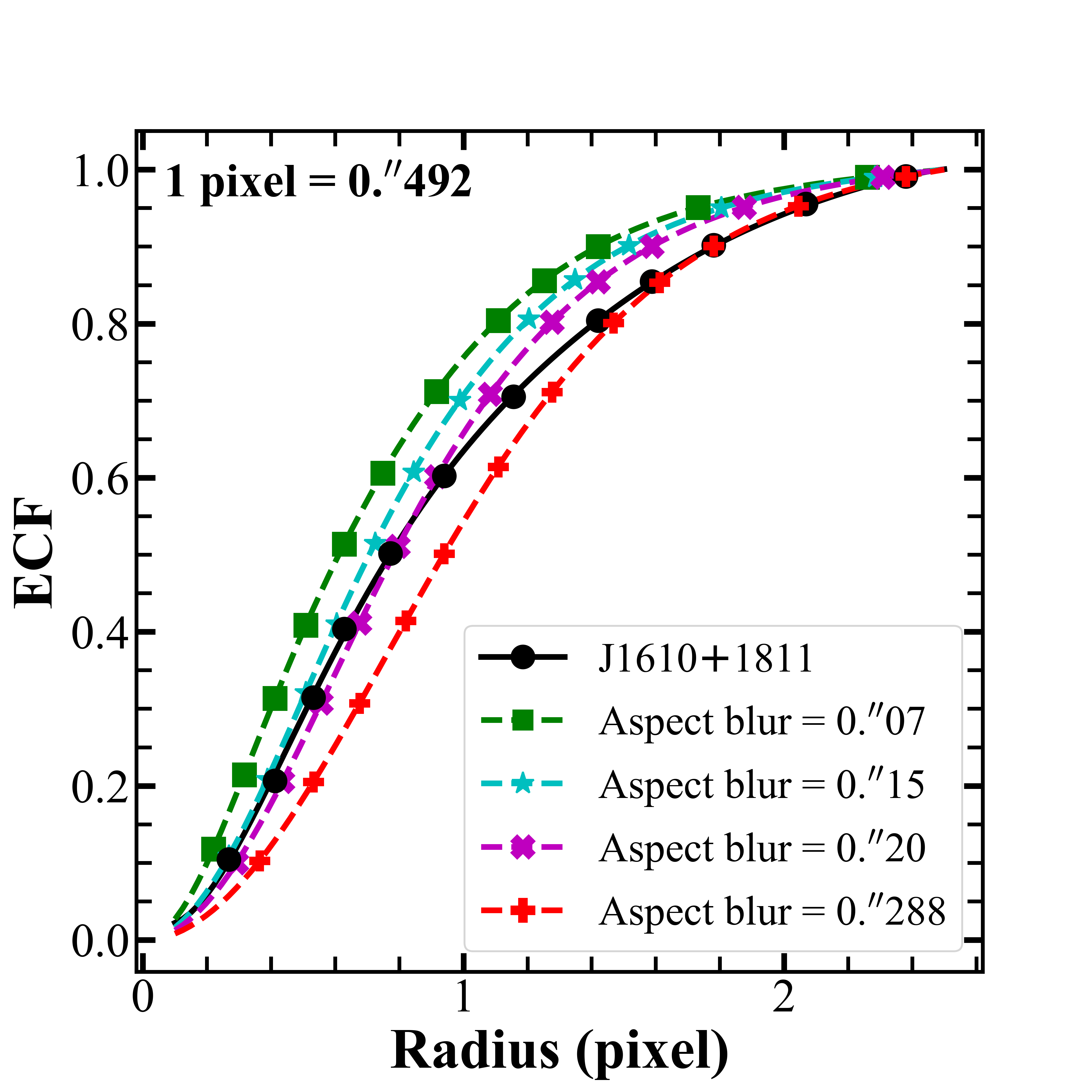}}
\caption{\small Enclosed count fraction for J1405+0415 (left) and J1610+1811 (right) as a function of radius compared with the simulated PSF with different aspect blur parameter values. ECF of the observed data is shown in black. $AspectBlur=0 \farcs 20$ gives the best match with our data below 1.5 pixel radius, while $AspectBlur=0 \farcs 288$ is a better match with the observed profile for larger pixel radii.}
\label{ECF}
\end{figure*}

The Chandra PSF documentation\footnote{https://cxc.cfa.harvard.edu/ciao/PSFs/chart2/caveats.html} recommends using an AspectBlur parameter value of $0 \farcs 25$ for ACIS-S. However, \cite{Ma+2023} suggests $0 \farcs 07$ as suitable for quasars with extended X-ray jets. Previous studies by \cite{Schwartz2020} and \cite{Snios2021} used $0 \farcs 288$ to simulate the psf of our targets. To determine the optimal aspect solution for simulating the core, we tested values $0 \farcs 07$, $0 \farcs 15$, $0 \farcs 20$, and $0 \farcs 288$ for the AspectBlur parameter. Figure \ref{ECF} compares the encircled counts fraction profiles of the observed and simulated PSFs, indicating none of the aspect solutions fully matches the observed profile. An aspect blur of $0 \farcs 07$ \& $0 \farcs 15$ provides a narrower PSF, whereas, aspect blur of $0 \farcs 288$ provides a broader PSF in the core.  The simulated PSF with an AspectBlur of $0 \farcs 20$ provides a good fit with the observed profile up to 1.5 pixel radius beyond which an AspectBlur of $0 \farcs 288$ matches the observed profile. 
We simulated each observation ID with aspect blurs of $0 \farcs 20$ \& $0 \farcs 288$. After aligning the images with the radio core position, we merged them to create a co-added simulated image for each aspect blur value. These merged PSF images are then used for our comprehensive comparison. For radius below 1.5 pixel i.e.,  $0 \farcs 74$, we compare the observed counts with the simulated counts using $0 \farcs 20$ aspect blur and beyond 1.5 pixel radius with simulated counts using $0 \farcs 288$ blur.

\section{Supplementary figures}\label{Appendix2}
\begin{figure*}[h!]
\gridline{
\includegraphics[width=0.47\columnwidth,trim=0.cm 0.cm 0.cm 0.cm]{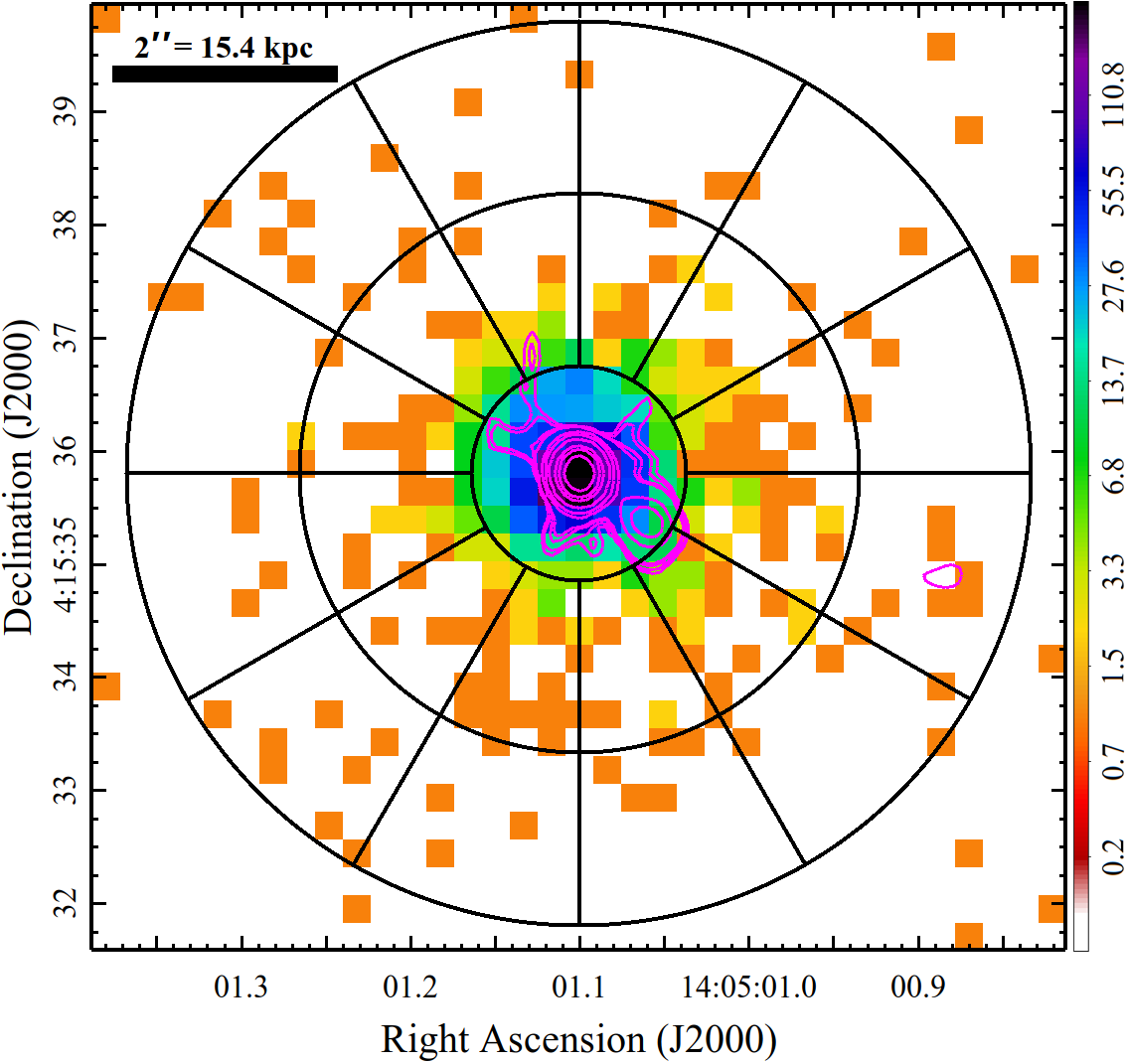}
\includegraphics[width=0.47\columnwidth,trim=0cm 0.5cm 0.5cm 1.5cm,clip]{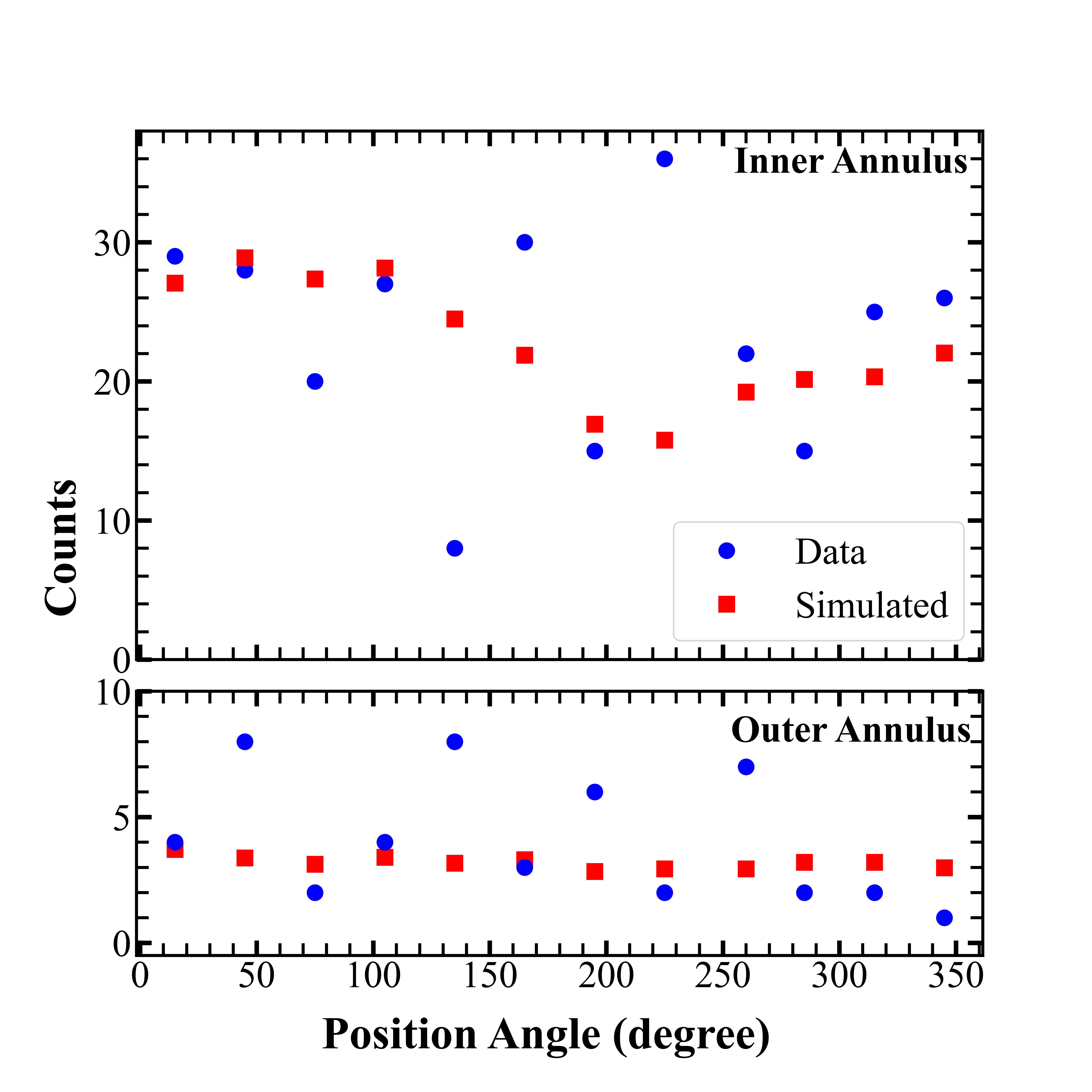}}
\caption{\small Left: Chandra image of J1405+0415 at 0.5-7 keV range is shown with $0 \farcs 246$ pixels.  Contours represent the radio emission at 6 GHz. Color scheme and scaling is the same as Figure \ref{ChandraImage}. The inner annulus has a radius ranging from $0 \farcs 95$ to $2 \farcs 475$, while the outer annulus extends to $4 \farcs 0.$ Each annulus is further divided into twelve sectors spanning 30 degrees each, increasing counterclockwise from the zero position angle pointing North. Right: Counts in each sector of the inner and outer annulus as a function of position angle. The simulated psf with AspectBlur=$0 \farcs 288$. The error in counts (N) is $\sqrt{N}$. }
\label{J1405PA}
\end{figure*}

\begin{figure*}[h!]
\gridline{
\includegraphics[width=0.47\columnwidth,trim=0.cm 0.cm 0cm 0.0cm]{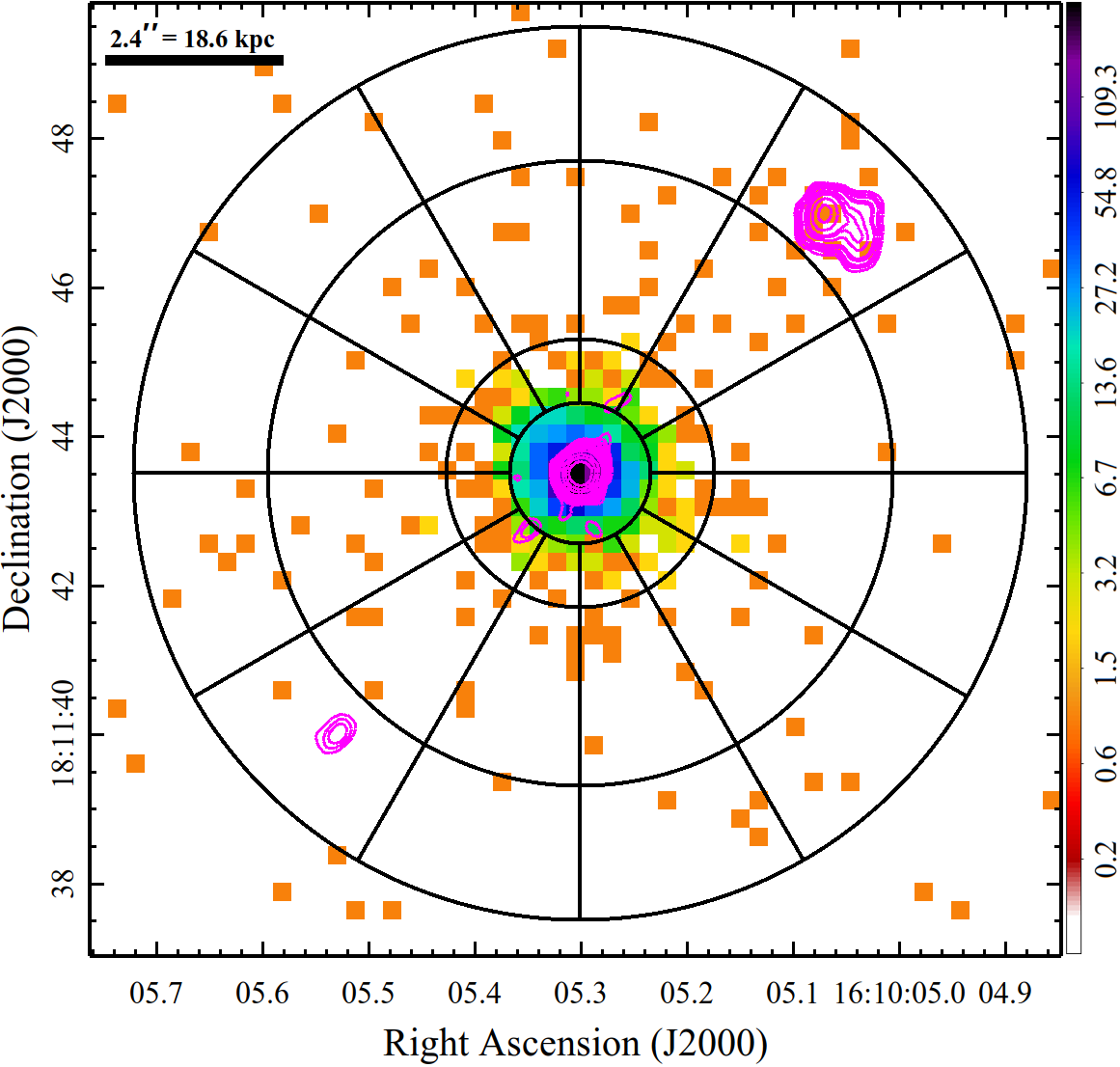}
\includegraphics[width=0.47\columnwidth,trim=0cm 0.5cm 0.5cm 1.5cm,clip]{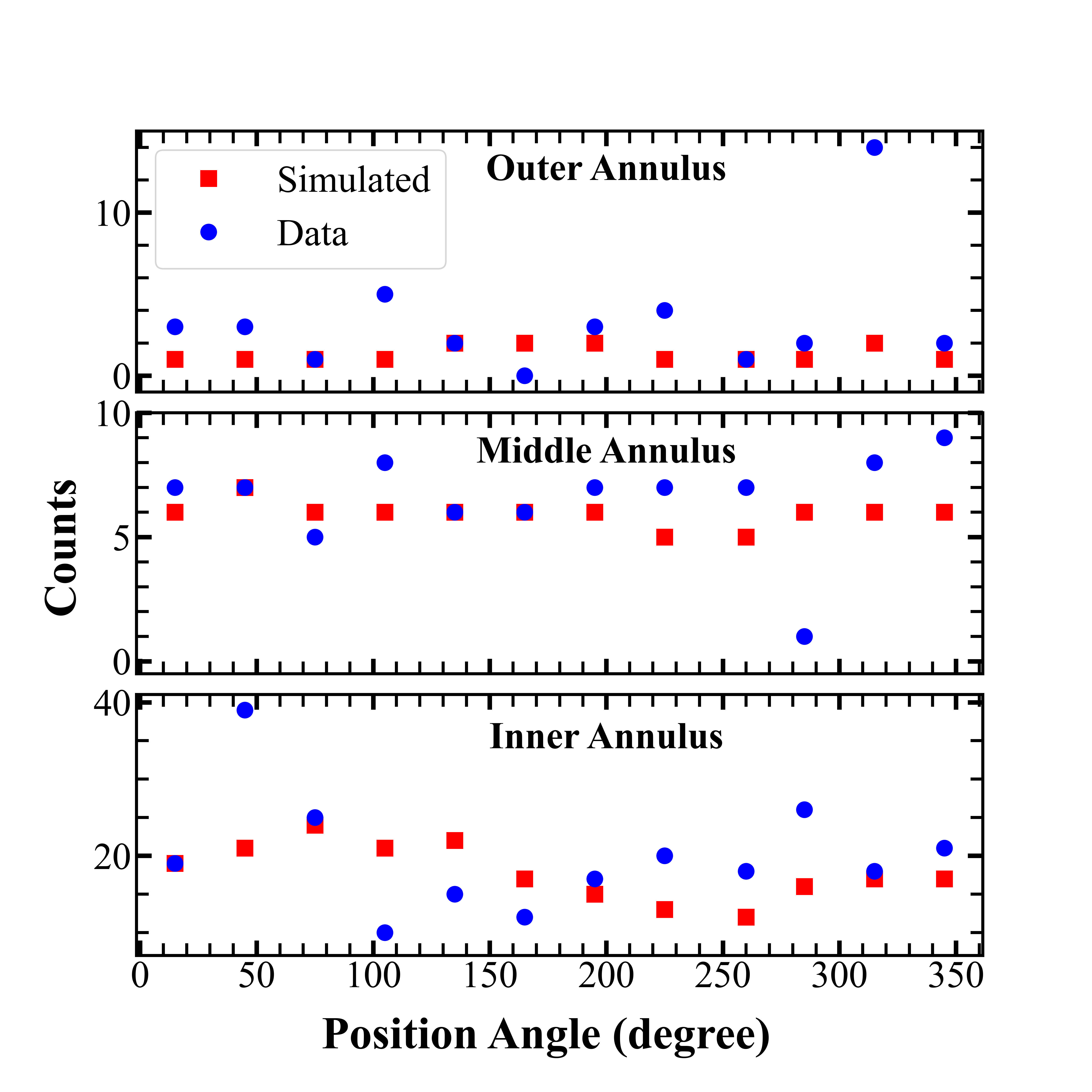}}
\caption{\small Left: Chandra image of J1610+1811 at 0.5-7 keV range is shown with $0 \farcs 246$ pixels. Contours represent the radio emission at 6 GHz. Color scheme and scaling is the same as Figure \ref{ChandraImage}. The inner annulus has a radius ranging from $0 \farcs 95$ to $1 \farcs 8$, the middle annulus goes to $4 \farcs 2$, whereas the outer annulus extends to $6 \farcs 0.$ Each sector spans 30 degrees, increasing counterclockwise from the zero position angle pointing North. Right: Counts in each sector of the inner, middle, and outer annulus as a function of position angle. Simulated psf with AspectBlur=$0 \farcs 288$. The error in counts (N) is $\sqrt{N}$.}
\label{J1610PA}
\end{figure*}

To assess the significance of extended X-ray emission, we examine the azimuthal distribution of counts in both observed and simulated images. 
The simulated counts were normalized by the ratio of the observed-to-simulated core counts in the $0 \farcs 5$ circular aperture. Assuming that the distribution of X-ray counts per sector follows Poisson statistics, we determined a projected probability distribution for each annular sector. The number of trials was not explicitly considered because the annular sectors were chosen based on their consistency with the radio emission. This selection criterion significantly reduces the likelihood that the detected excess is merely a statistical fluctuation. Afterward, we compared observed counts with projected sector probability distributions, setting a detection threshold at probability, $p$, less than 0.001. Figure \ref{J1405PA} (left) depicts two annular regions utilized for J1405+0415: the inner one with a radius of $0 \farcs 95$  to $2 \farcs 475$  and the outer region from $2 \farcs 475$ to $4 \farcs 0$, each divided into 12 sectors of 30 degrees. Figure \ref{J1405PA} (right) indicates that only the inner sector at a position angle of 225$\degr$ exhibits a significant detection of extended emission with a significance of $p<5\times 10^{-6}$. We observed 36 counts in this sector, while the simulation predicts 16 counts scattered from the core. For all other sectors, the observed counts fall below our detection threshold. For J1610+1811, we utilized three annular regions: an inner annulus ranging from $0 \farcs 95$ to $1 \farcs 8$, a middle annulus from $1 \farcs 8$ to $4 \farcs 2$, and an outer annulus from $4 \farcs 2$ to $6 \farcs 0$. In Figure \ref{J1610PA}, a significant detection of extended emission is observed in the outer annulus sector at a position angle of 315$\degr$ with a significance of $p< 4\times 10^{-9}$. This sector coincides with the radio hotspot and contains 14 X-ray counts, while the predicted counts scattered from the core are only two. The sector coincident with the radio counter-hotspot shows no X-ray excess. Sectors in the middle annulus also do not show a significant excess. However, we detect an excess in the inner annulus sector at a position angle of 45$\degr$, with observed counts of 39 and predicted counts of 21, resulting in a significance of $p=0.00014$. This is in the direction of the known HRMA artifact\citep{Kashyap2010}. Our detection of extended X-ray emission is consistent with that determined by \cite{Snios+2022}, \cite{Snios2021} \& \cite{Schwartz2020}.


\begin{figure*}[]
\gridline{
\includegraphics[width=0.5\columnwidth,trim=0.cm 0cm 0cm 0.cm]{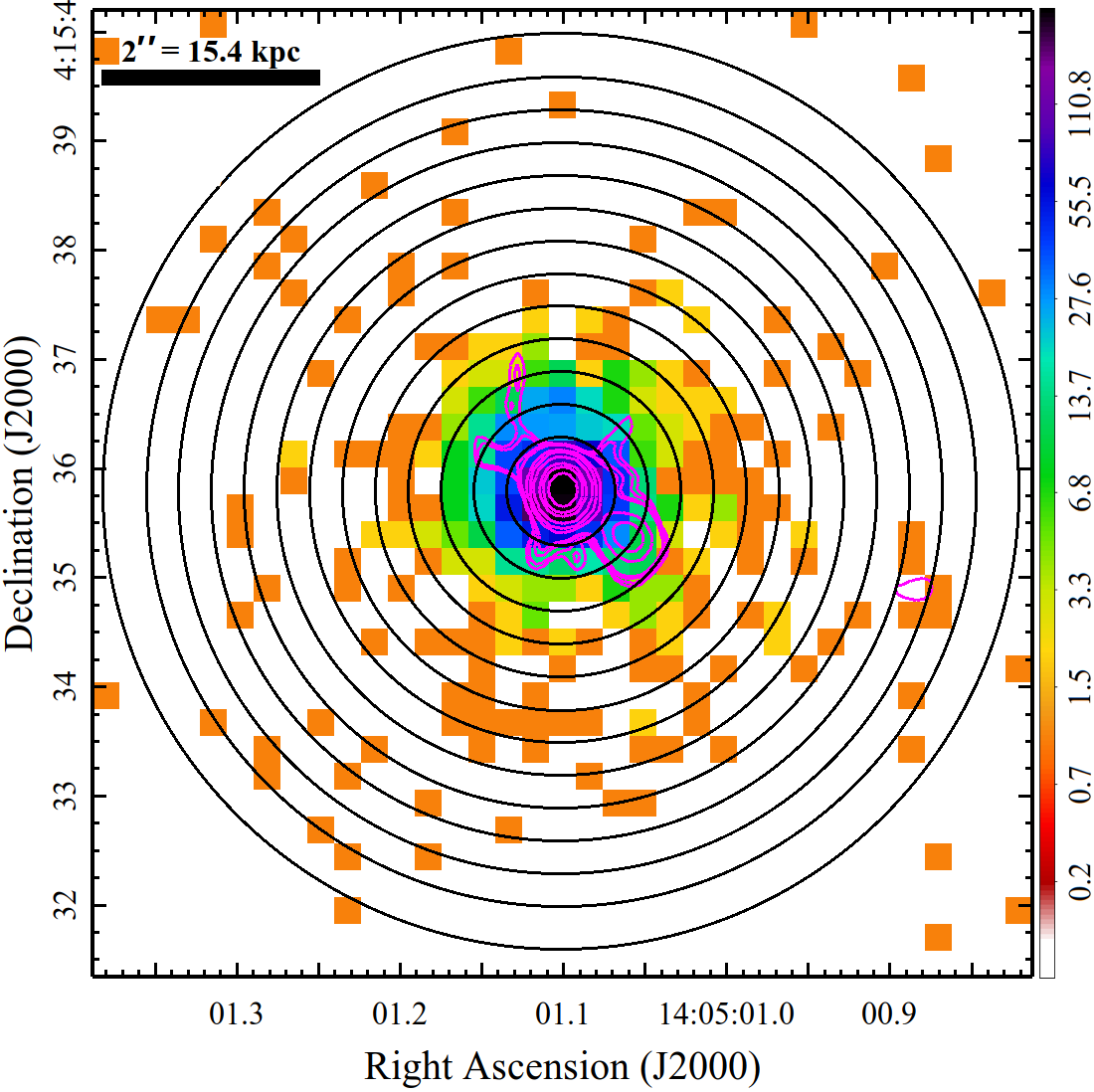}
\includegraphics[width=0.51\columnwidth,trim=0cm 0.5cm 1.4cm 3.cm,clip]{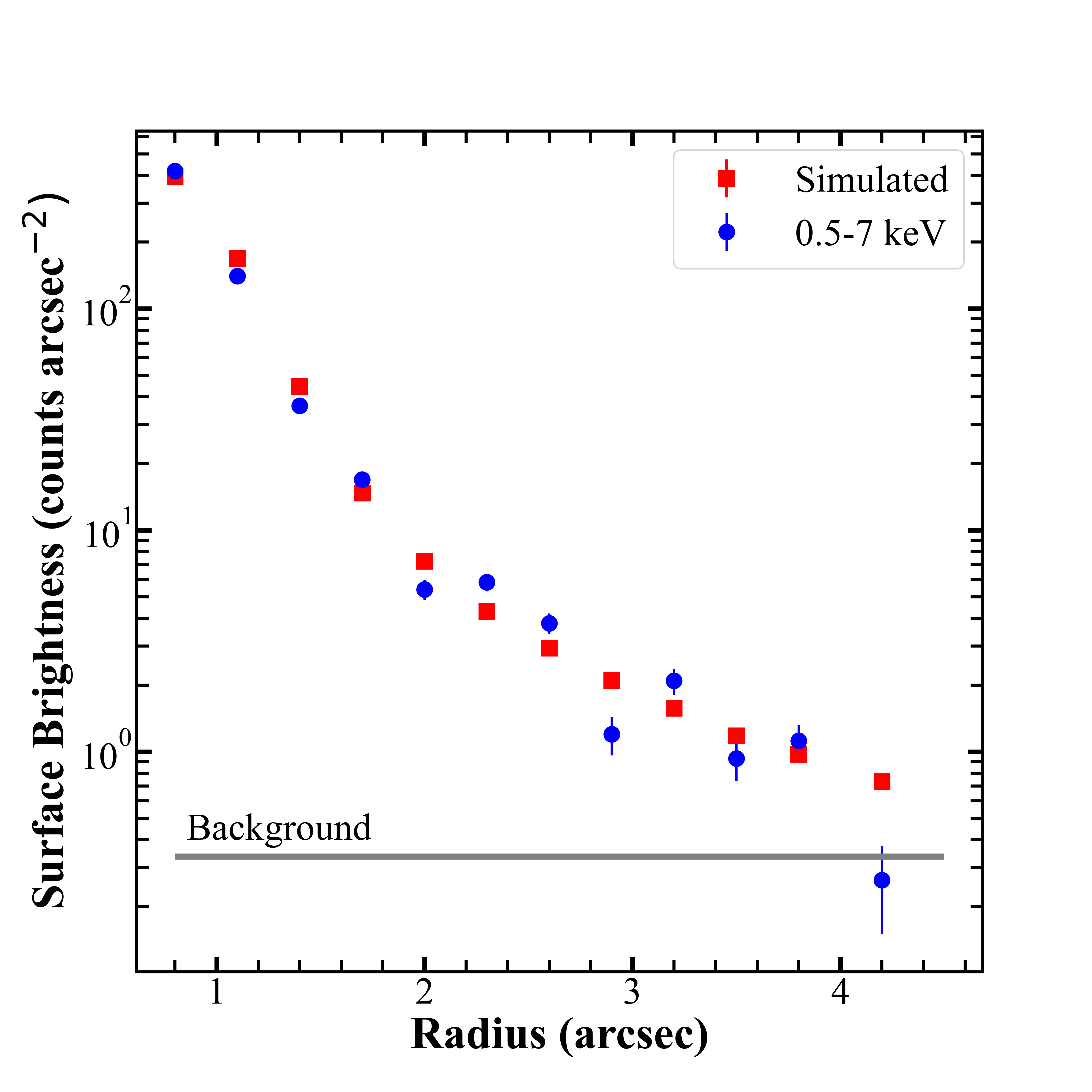}}
\gridline{
\includegraphics[width=0.51\columnwidth,trim=0.1cm 0cm 1.4cm 3.cm,clip]{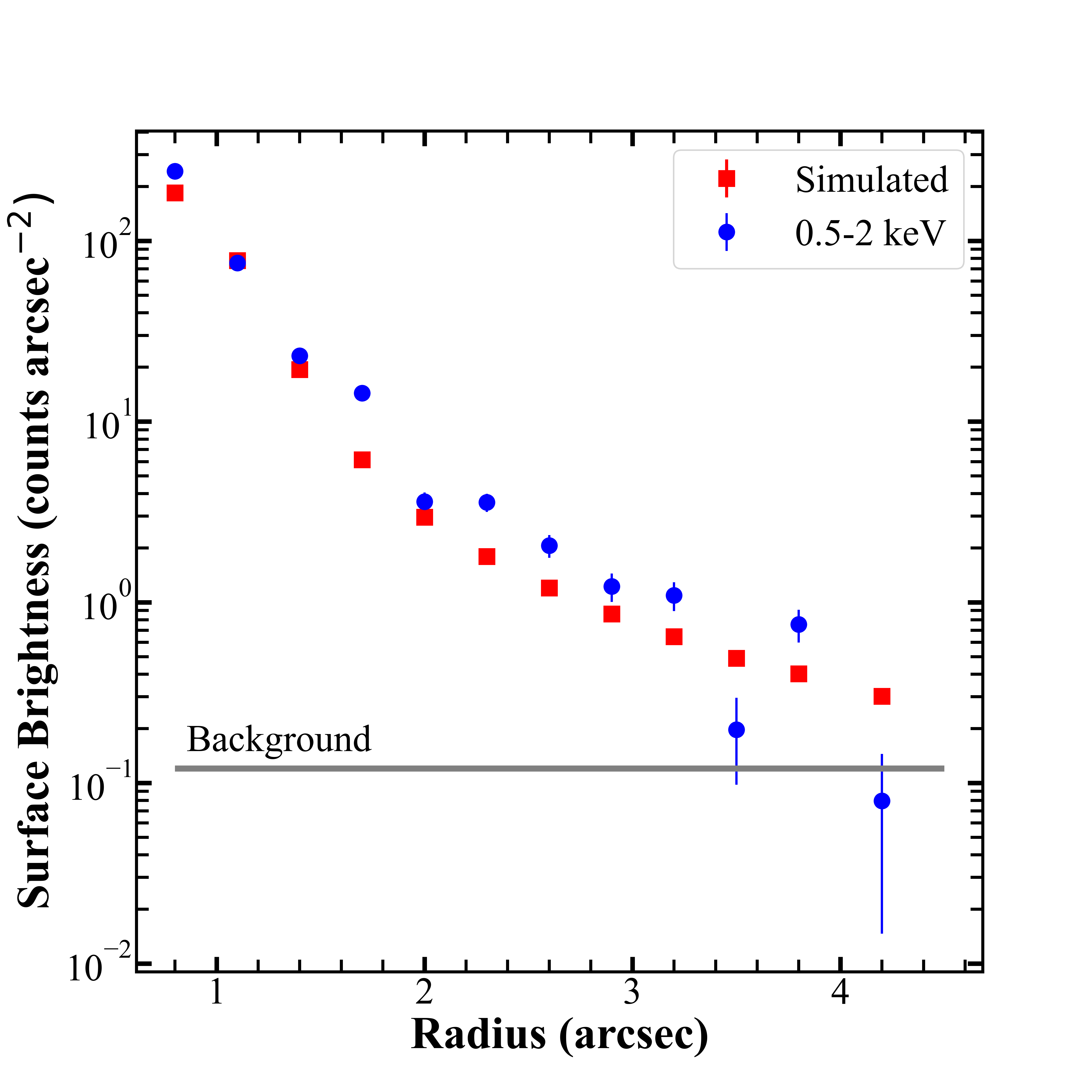}
\includegraphics[width=0.51\columnwidth,trim=0.4cm 0cm 1.4cm 3.cm,clip]{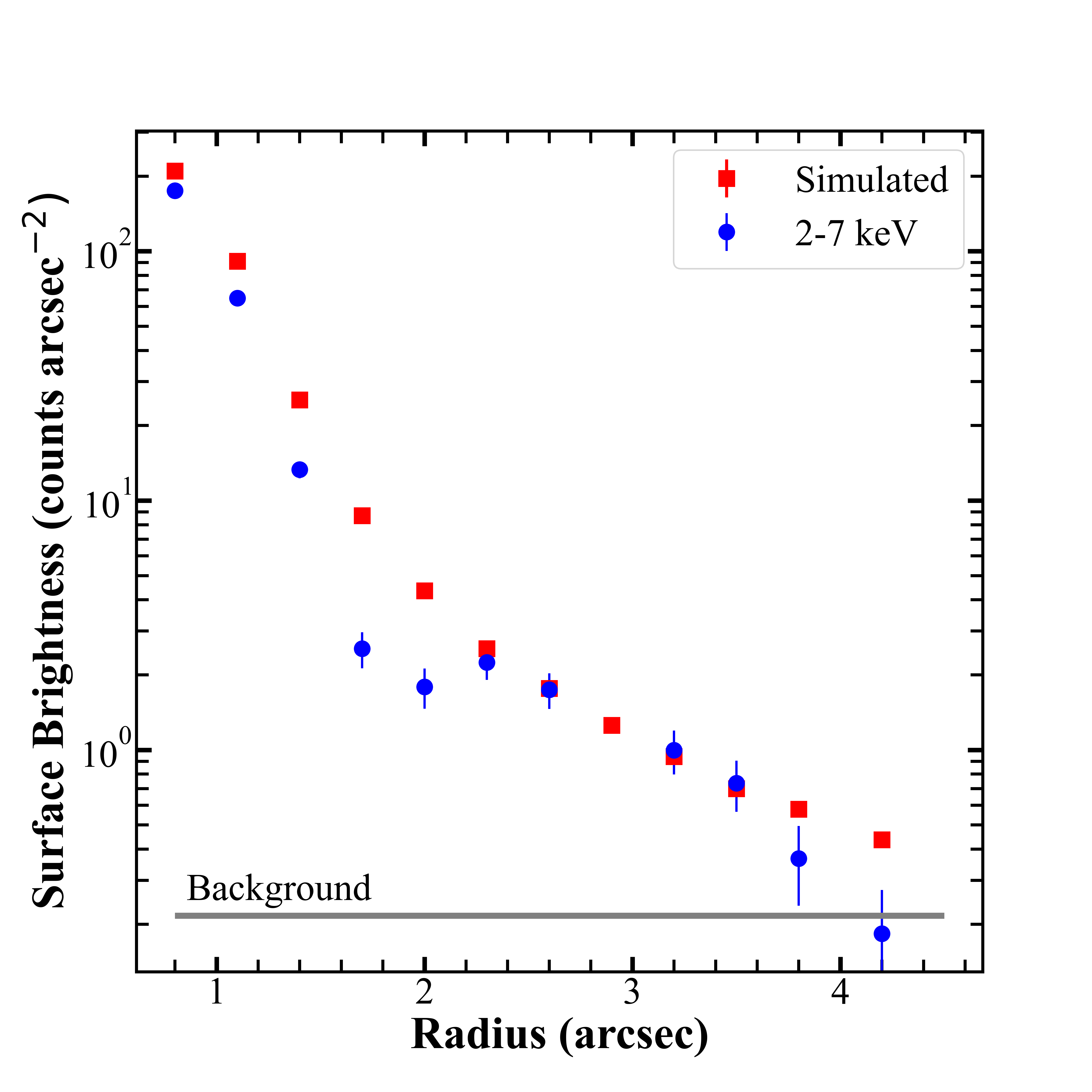}}
\caption{\small Chandra image of J1405+0415 (top left) same as Figure C.1. Contours represent the radio emission at 6 GHz. Color scheme and scaling is the same as Figure \ref{ChandraImage}. Twelve annuli with inner radius $0 \farcs 5$ and outer radius $4 \farcs 2$ are used to extract the surface brightness. Except for the twelfth annulus which has a width of $0 \farcs 4$, all annuli have same width of $0 \farcs 3.$ Surface brightness as a function of radius over the 0.5-7 keV (top right), 0.5-2 keV (bottom left), and 2-7 keV (bottom right) energy band. Gray line shows the background level in each energy band. Note -- The counts in the inner annulus 0.5-0.8 arcsec is simulated using AspectBlur = $0 \farcs 20$. Counts in all other annuli are simulated using AspectBlur = $0 \farcs 288$. }
\label{J1405radialprofile}
\end{figure*}

\begin{figure*}[]
\gridline{
\includegraphics[width=0.5\columnwidth,trim=0.cm 0.cm 0cm 0.cm]{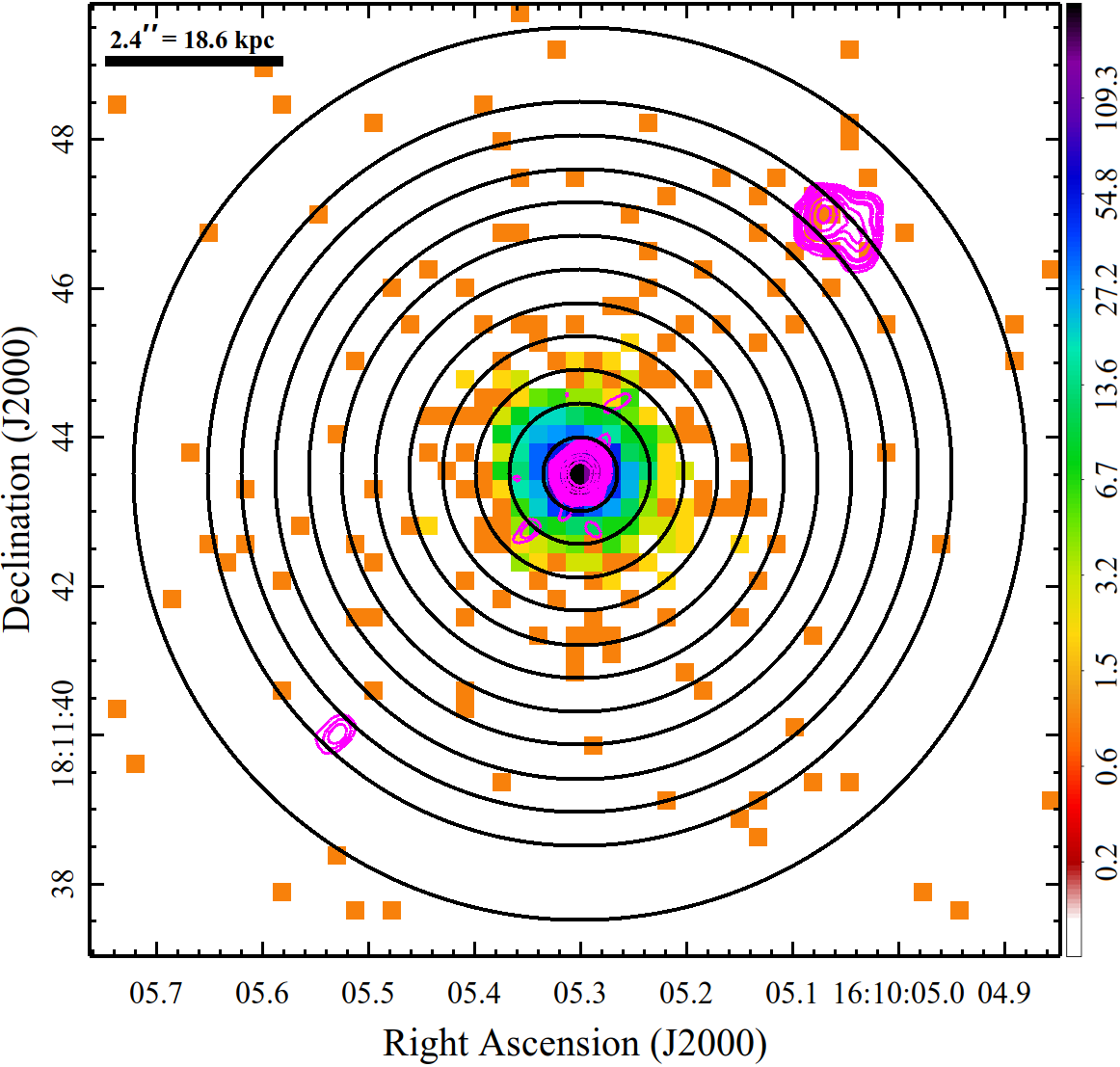}
\includegraphics[width=0.5\columnwidth,trim=0cm 0.5cm 1.4cm 3.cm,clip]{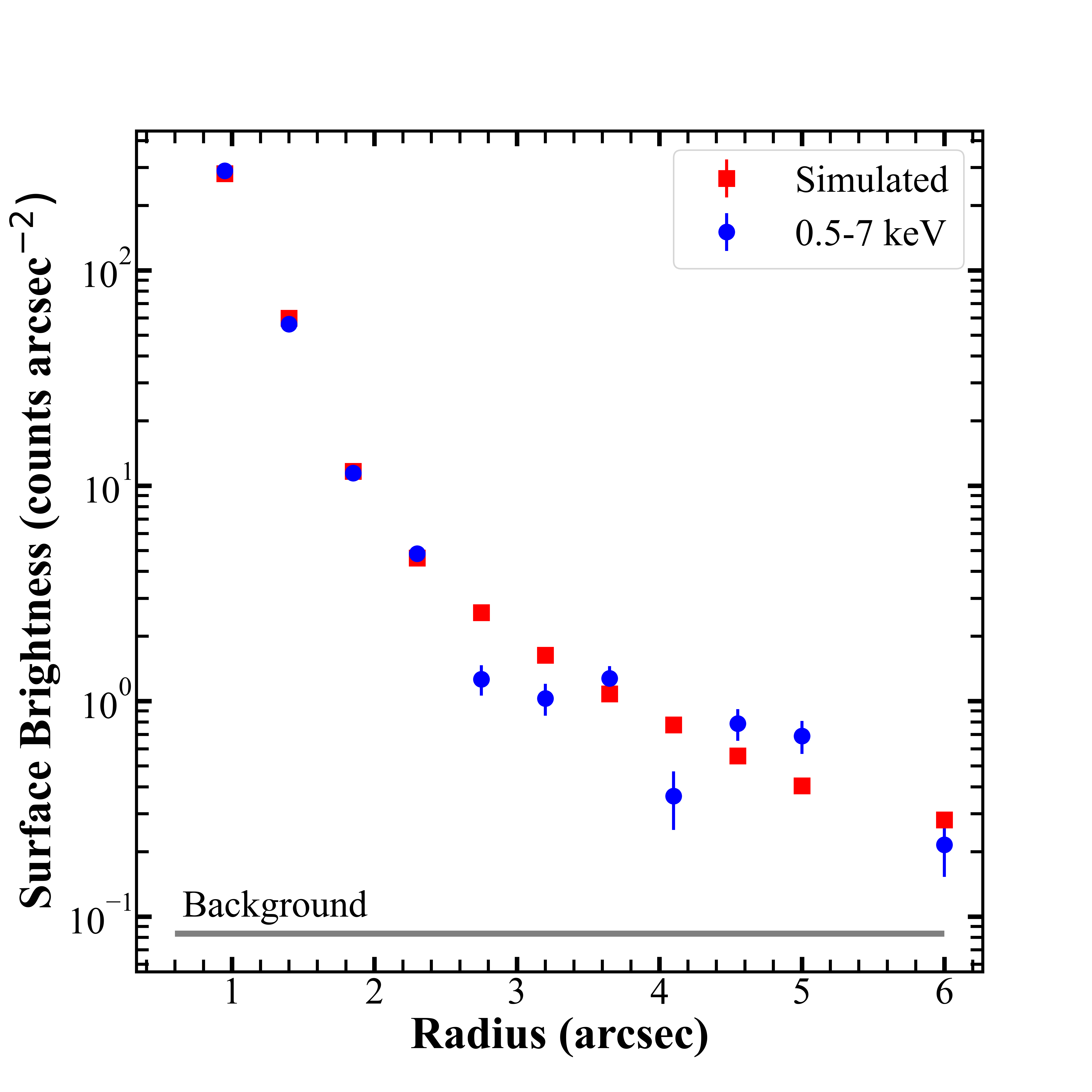}}
\gridline{
\includegraphics[width=0.5\columnwidth,trim=0.cm 0cm 0cm 0.cm,clip]{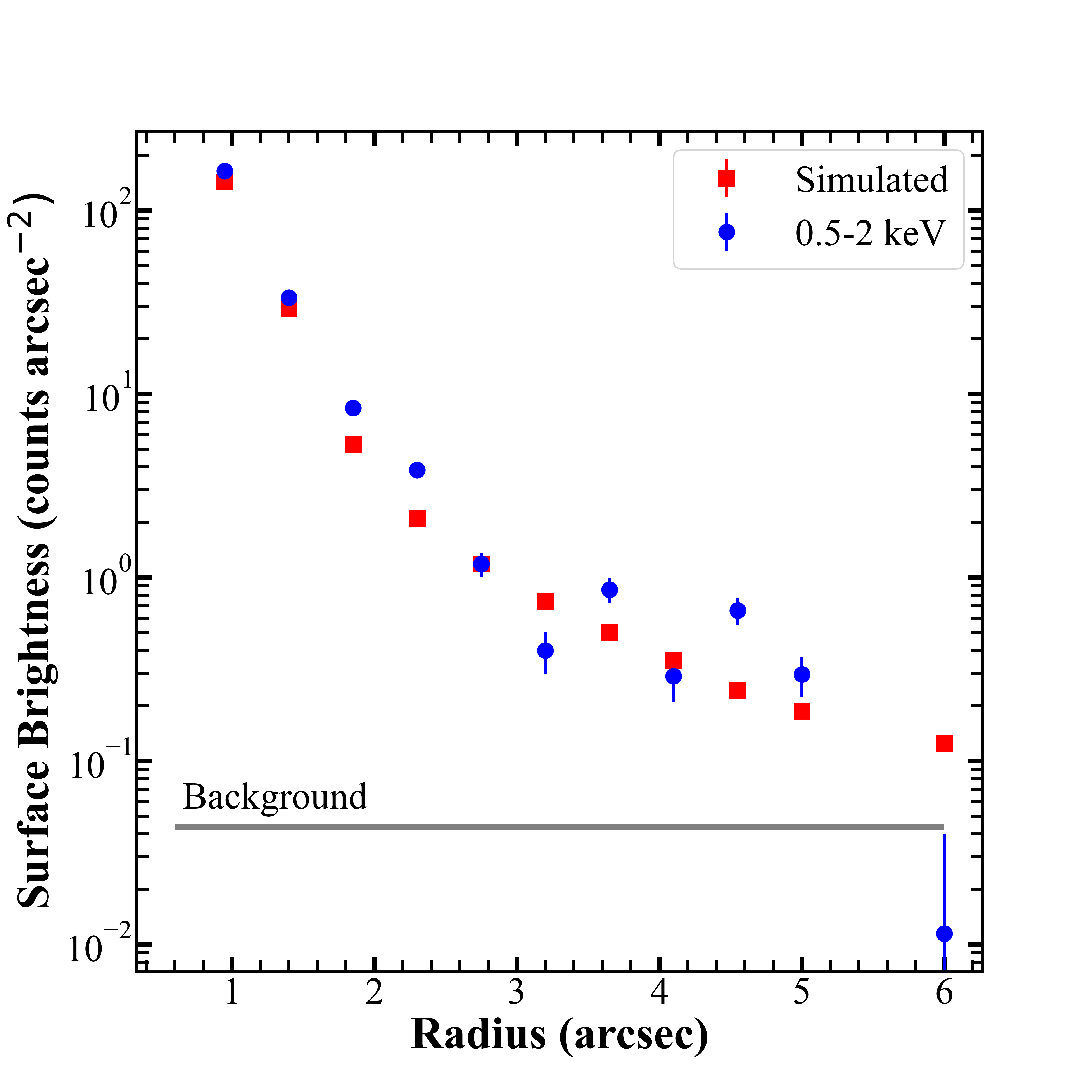}
\includegraphics[width=0.5\columnwidth,trim=0cm 0.5cm 1.4cm 3.cm,clip]{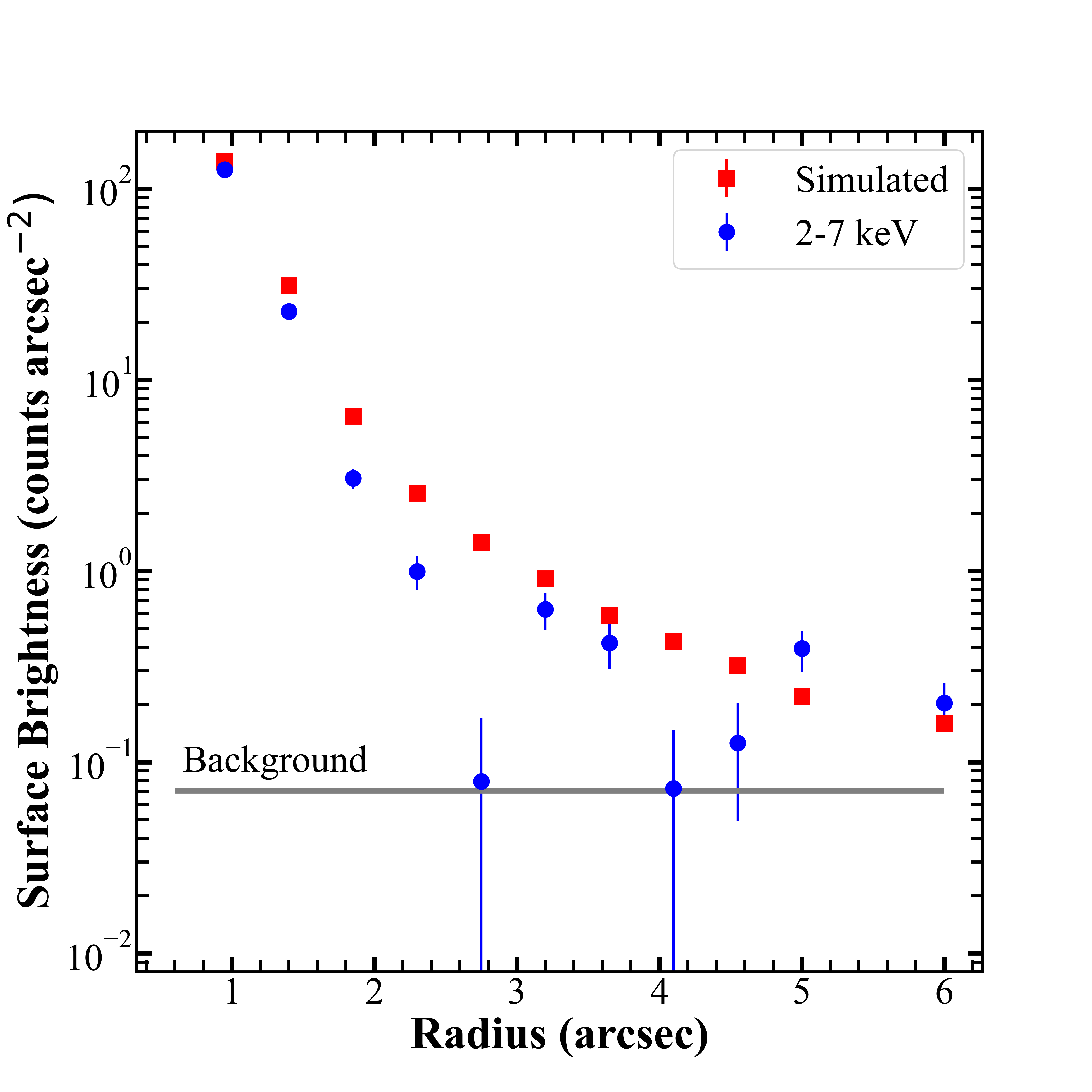}}
\caption{\small Chandra image of J1610+1811 (top left) same as Figure C.2. Contours represent the radio emission at 6 GHz. Color scheme and scaling is the same as Figure \ref{ChandraImage}. Eleven annuli with inner radius $0 \farcs 5$ and outer radius $6 \farcs 0$ are used to extract the surface brightness. Except the twelfth annulus which has a width of $1 \farcs 0$, all annuli have same width of $0 \farcs 45.$ Surface brightness as a function of radius over the 0.5-7 keV (top right), 0.5-2 keV (bottom left), and 2-7 keV (bottom right) energy band. Gray line shows the background level in each energy band. Note -- The counts in the inner annulus 0.5-0.95 arcsec is simulated using AspectBlur = $0 \farcs 20$. Counts in all other annuli are simulated using AspectBlur = $0 \farcs 288$. }
\label{J610radialprofile}
\end{figure*}

\begin{figure*}[]
\gridline{
\includegraphics[width=0.5\columnwidth,trim=0.cm 0.cm 0cm 0.cm,clip]{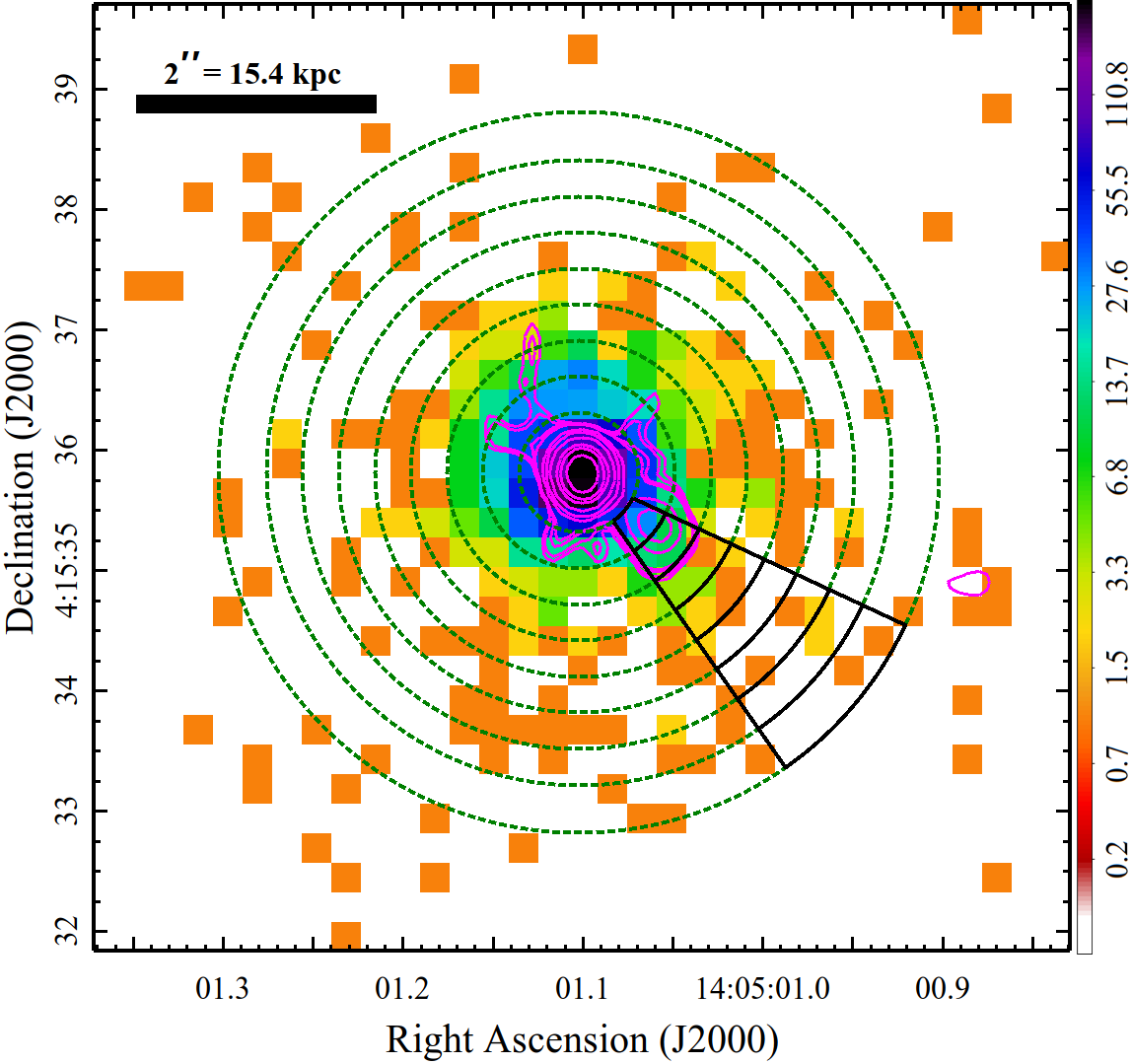}
\includegraphics[width=0.51\columnwidth,trim=0cm 0.cm 1.4cm 2.cm,clip]{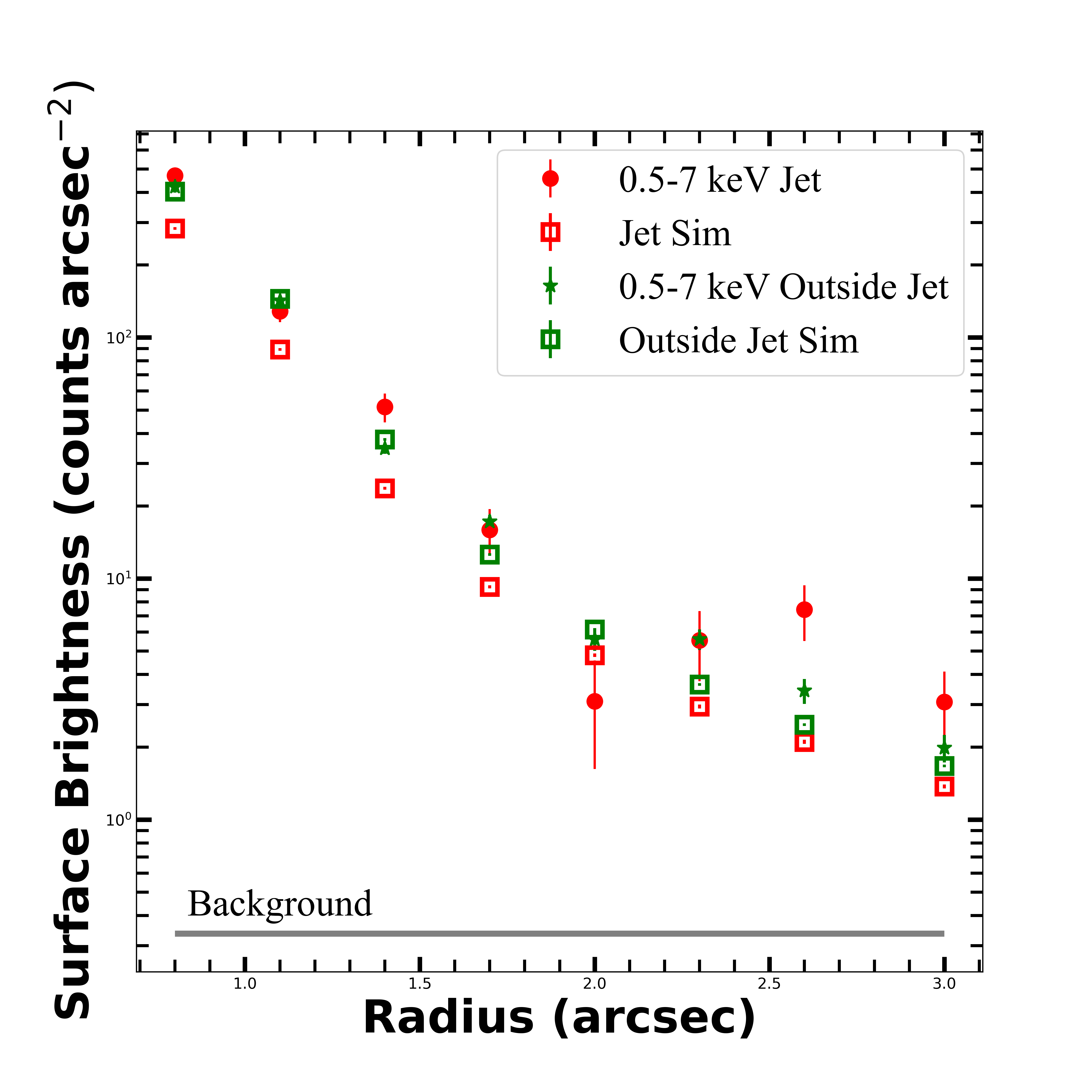}}
\gridline{
\includegraphics[width=0.51\columnwidth,trim=0.1cm 0cm 1.4cm 3.cm,clip]{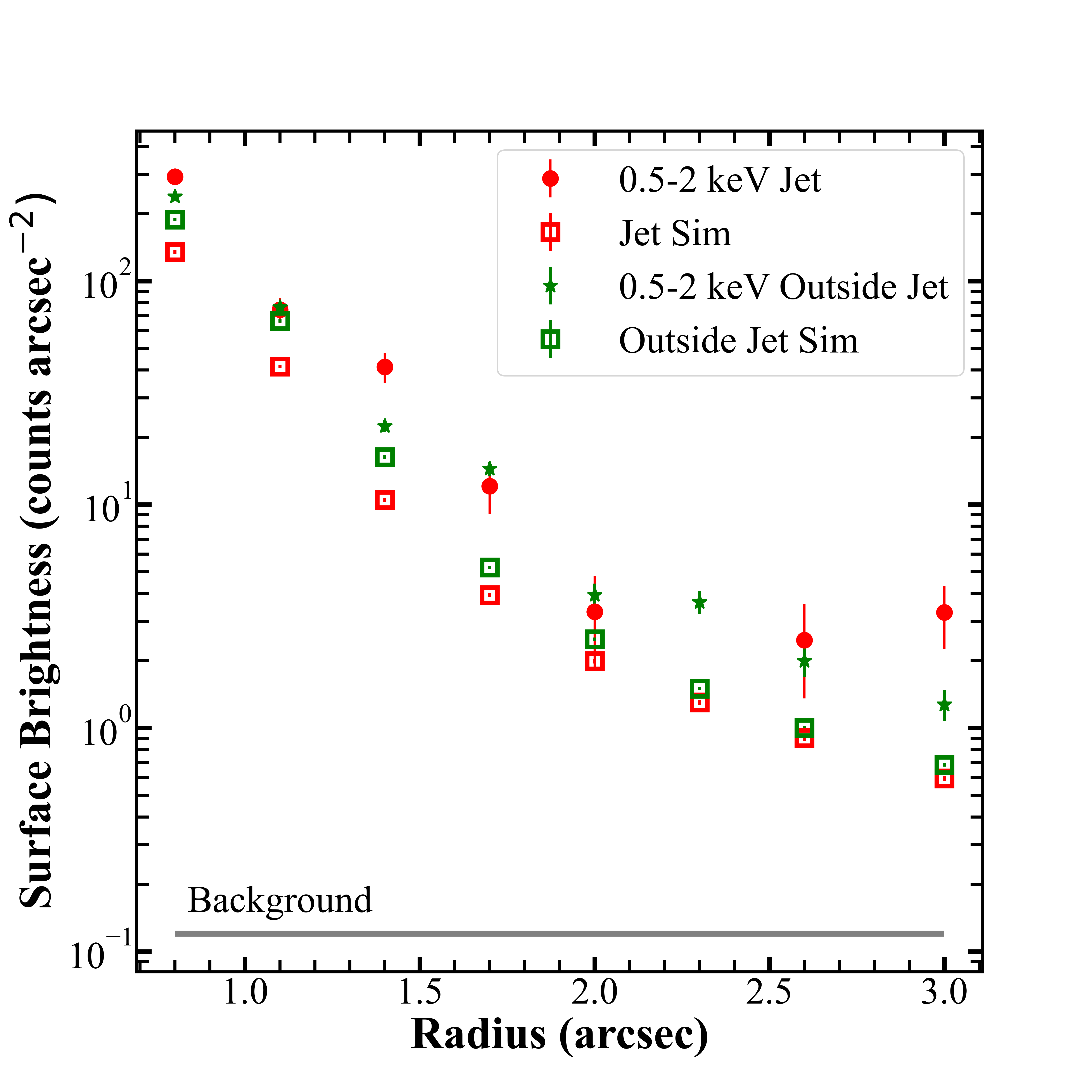}
\includegraphics[width=0.51\columnwidth,trim=0.4cm 0cm 1.4cm 3.cm,clip]{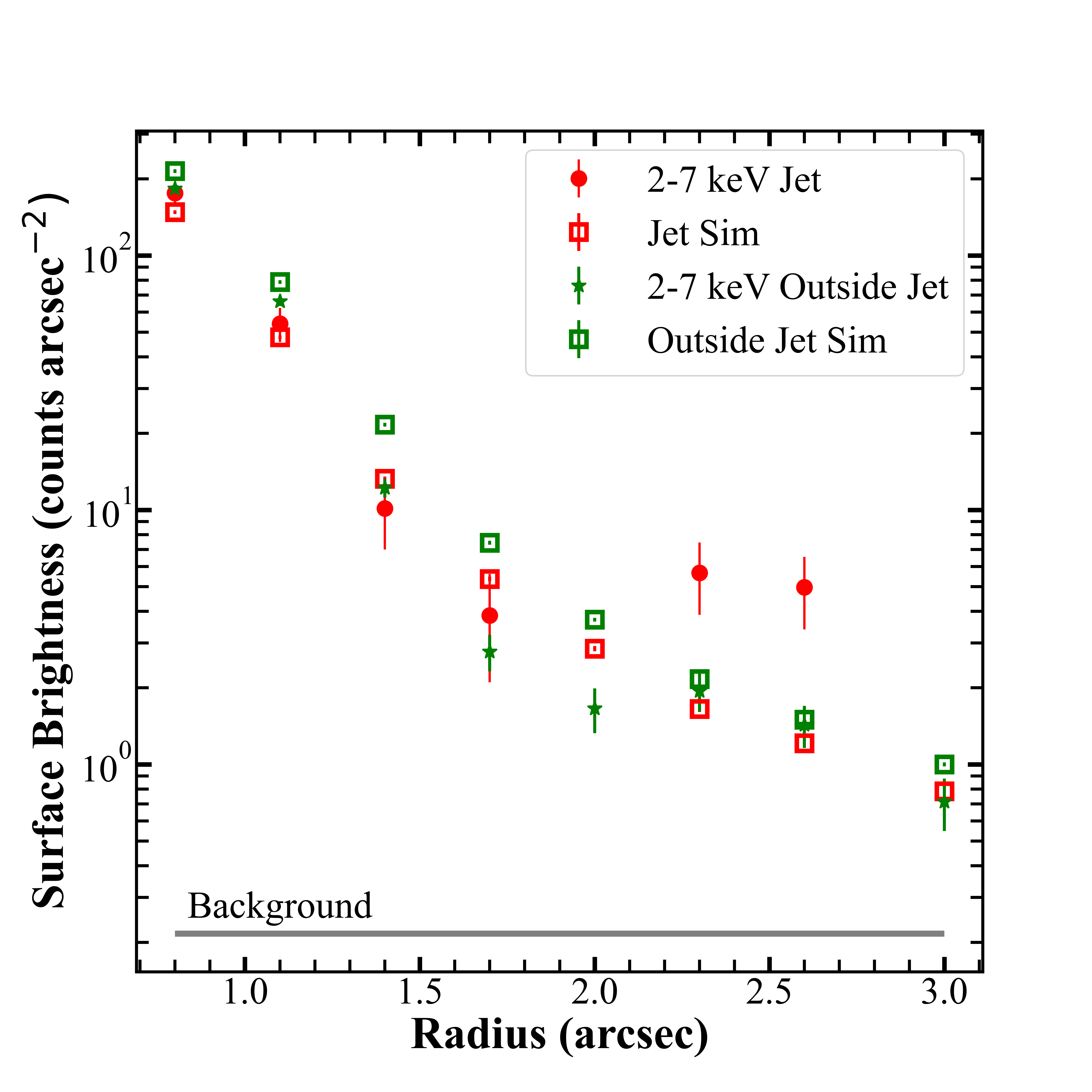}}
\caption{\small Chandra image of J1405+0415 (top left) same as Figure C.1. Contours represents the radio emission at 6 GHz. Color scheme and scaling is the same as Figure \ref{ChandraImage}. Eight annuli with inner radius $0 \farcs 5$ and outer radius $3 \farcs 0$ are used to extract the surface brightness. Except the twelfth annulus which has a width of $0 \farcs 4$, all annuli have same width of $0 \farcs 3.$ Surface brightness as a function of radius over the 0.5-7 keV (top right), 0.5-2 keV (bottom left), and 2-7 keV (bottom right) energy band. Gray line shows the background level in each energy band. Note -- The counts in the inner annulus 0.5-0.95 arcsec is simulated using AspectBlur = $0 \farcs 20$. Counts in all other annuli are simulated using AspectBlur = $0 \farcs 288$. }
\label{J1405jetprofile}
\end{figure*}

\begin{figure*}[]
\centerline{(a) Jet and counter-jet sectors.}
\centering
\includegraphics[width=0.4\columnwidth,trim=0cm 0.cm 0cm 0.cm]{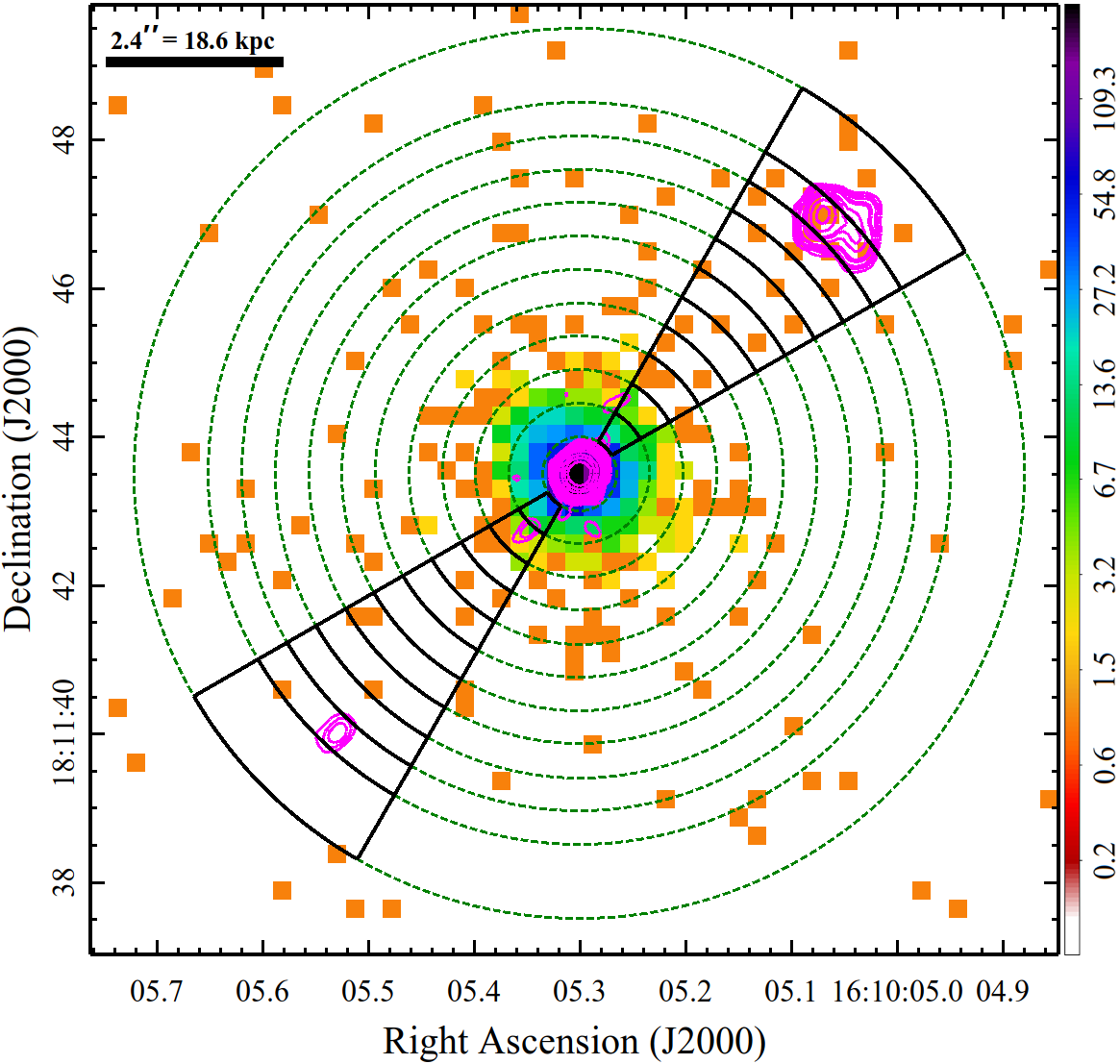}
\centerline{(b) Radial profile of the jet.}
\gridline{
\includegraphics[width=0.35\columnwidth,trim=0cm 0.5cm 1.4cm 3.cm,clip]{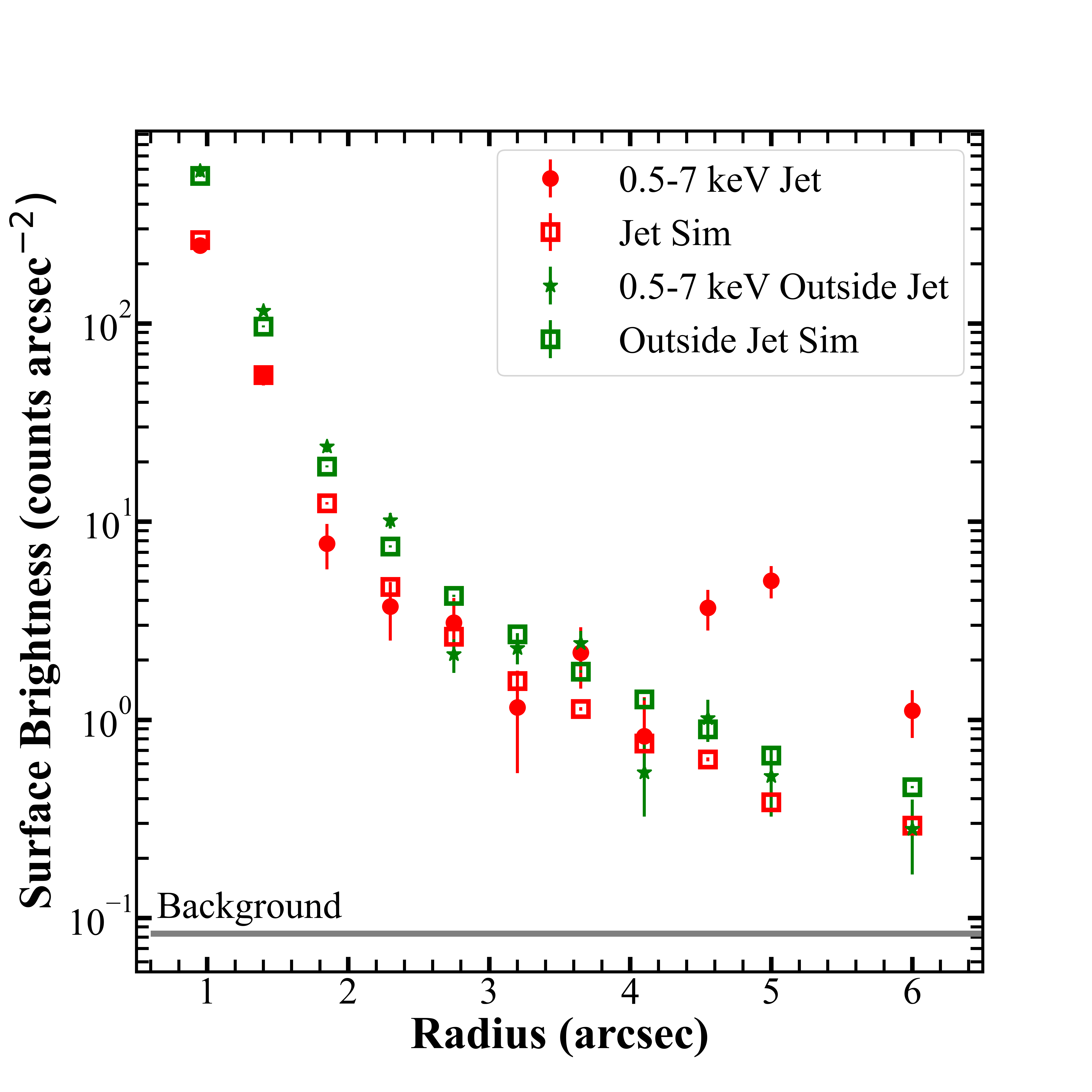}
\includegraphics[width=0.35\columnwidth,trim=0.cm 0.5cm 1.4cm 3.cm,clip]{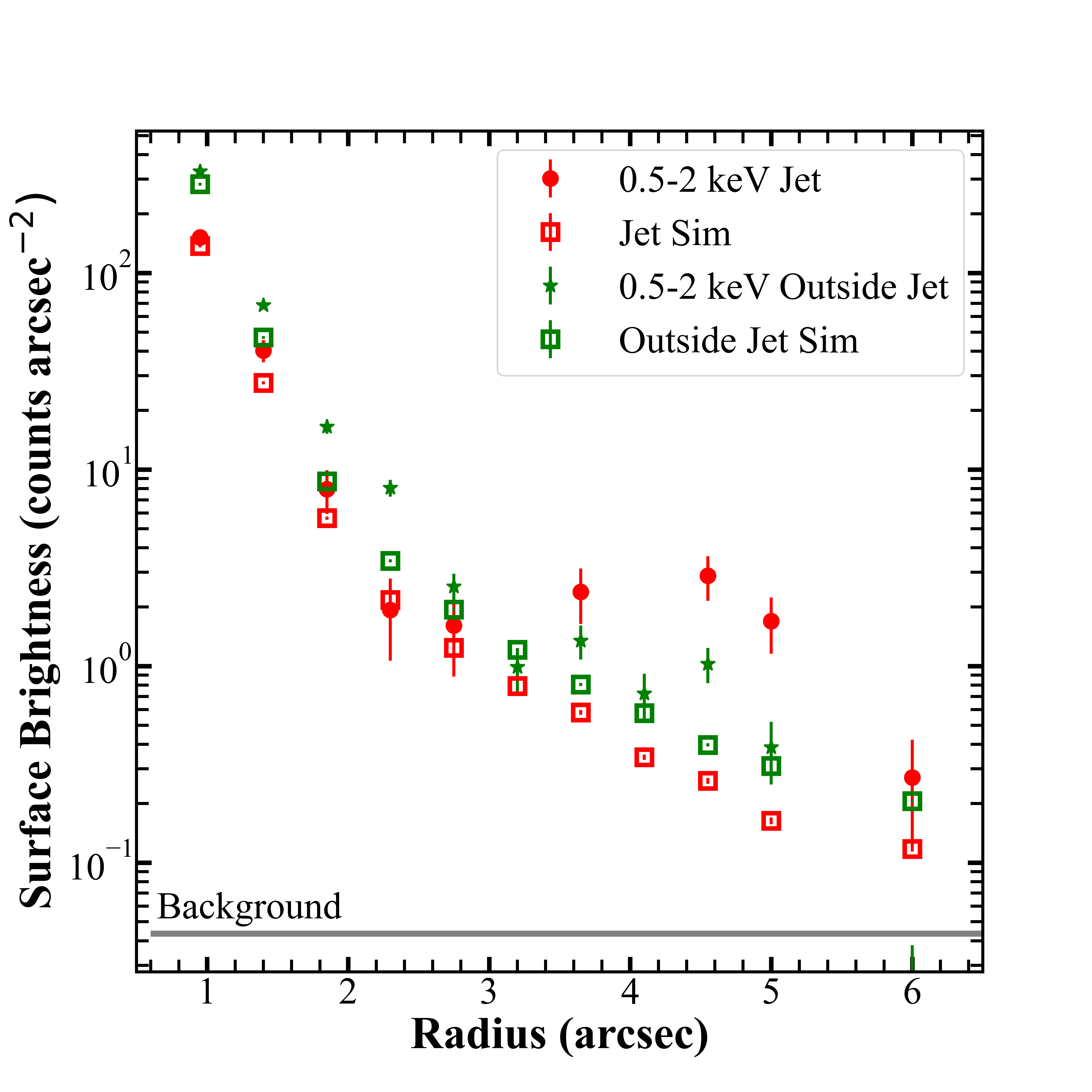}
\includegraphics[width=0.35\columnwidth,trim=0cm 0.5cm 1.4cm 3.cm,clip]{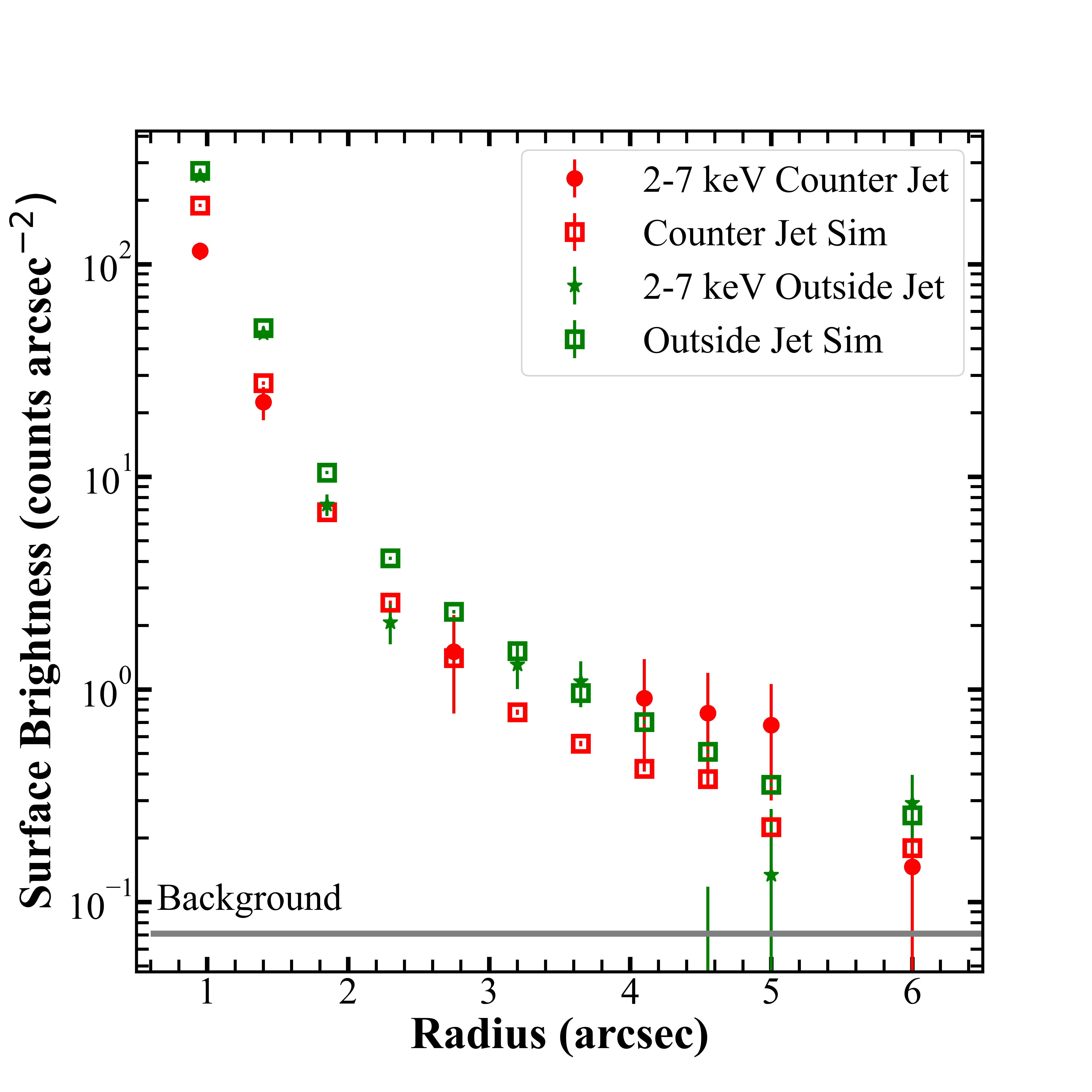}}
\centerline{(c) Radial profile of the counter-jet.}
\gridline{
\includegraphics[width=0.35\columnwidth,trim=0cm 0.5cm 1.4cm 3.cm,clip]{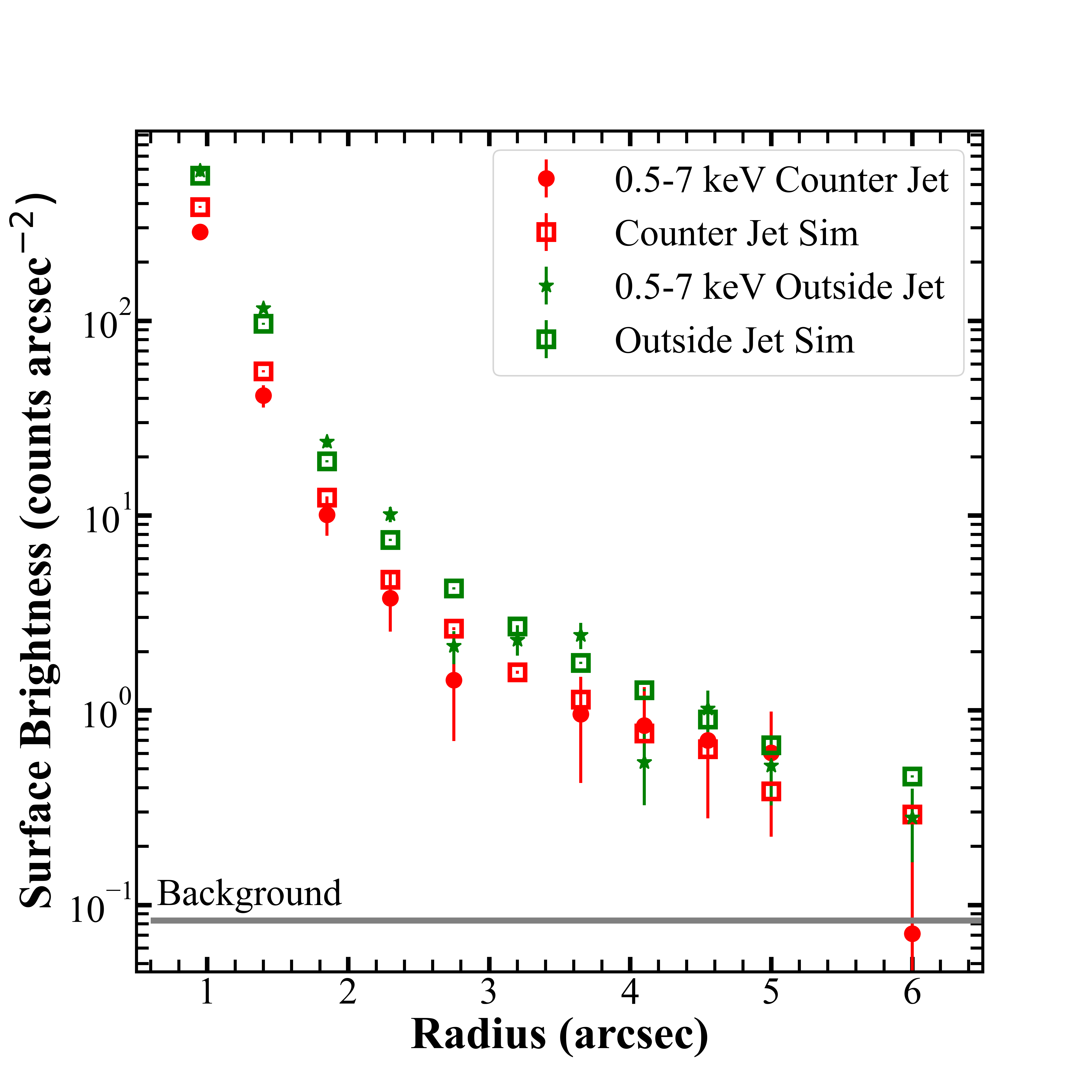}
\includegraphics[width=0.35\columnwidth,trim=0.cm 0.5cm 1.4cm 3.cm,clip]{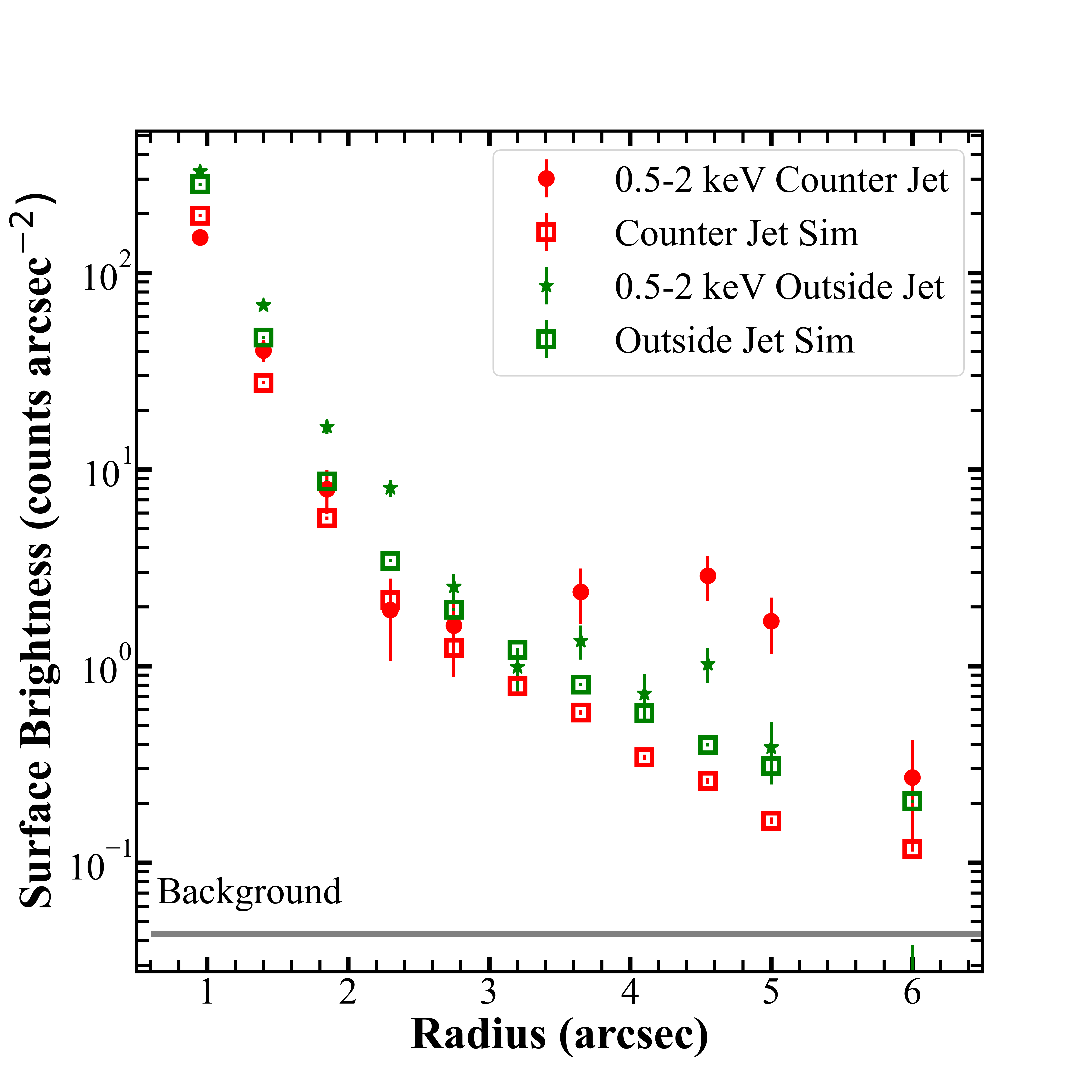}
\includegraphics[width=0.35\columnwidth,trim=0cm 0.5cm 1.4cm 3.cm,clip]{J1610RadialProfileHardCounterJetmixedoutsidefinal.png}}
\caption{\small (a) Chandra image of J1610+1811 same as Figure C.2. Contours represents the radio emission at 6 GHz. Color scheme and scaling is same as Figure \ref{ChandraImage}. Eleven sectors with inner radius $0 \farcs 5$ and outer radius $6 \farcs 0$ are used to extract the surface brightness of the jet and counter-jet. Except the outermost sector which has a width of $1 \farcs 0$, all sectors have same width of $0 \farcs 45.$ (b) Surface brightness as a function of radius for the jet over the 0.5-7 keV (left), 0.5-2 keV (middle), and 2-7 keV (right) energy band. Gray line shows the background level in each energy band. (c) Surface brightness as a function of radius for the counter-jet in the same energy bands as above. Note -- The counts in the inner annulus 0.5-0.95 arcsec is simulated using AspectBlur = $0 \farcs 20$. Counts in all other annulii are simulated using AspectBlur= $0 \farcs 288$. }
\label{J610jetprofile}
\end{figure*}

Next, we analyze the radial surface brightness profiles of our quasars across soft, hard, and broad energy bands. Figure \ref{J1405radialprofile} displays the radial profile for J1405+0415 using twelve concentric annuli from $0 \farcs 5$ to $4 \farcs 2$. The annular bin size is $0 \farcs 3$, except for the last bin, which has an increased size of $0 \farcs 4$ to ensure a minimum of 5 counts. It appears that the surface brightness in the soft energy band exhibits excess compared to the simulated PSF. We infer a similar trend in the radial surface brightness profile for J1610+1811 using concentric annuli extending up to $6 \farcs 0$ shown in Figure \ref{J610radialprofile}. We explore the radial profile in the pie sector at the position angle indicated by our analysis of the azimuthal distribution of X-ray counts (Figures \ref{J1405PA} \& \ref{J1610PA}). We compare the observed counts with simulated counts both within the pie sector and outside it. Figure \ref{J1405jetprofile} presents the radial profile of the jet for J1405+0415, revealing excess emission beyond 2\arcsec in all bands. Similarly, Figure \ref{J610jetprofile} illustrates the radial profiles for both the jet and counterjet of J1610+1811. While excess X-ray emission is observed in the broad and soft bands beyond 4\arcsec in the jet, the counterjet exhibits excess emission only in the soft band.


\newpage

\section{Sumplementry results for IC/CMB model for J1610+1811}\label{Appendix3}
This appendix presents comprehensive supplementary analyses of results for J1610+1811 using an extended extraction region ($3 \farcs 7 \times 1 \farcs 5$) that encompasses both the jet and the radio lobe/hotspot complex, including the corresponding IC/CMB model parameters.

We extracted and modeled the X-ray spectrum using the same methodology described in Section 3.4 but this time using a $3 \farcs 7 \times 1 \farcs 5$ jet box. The best-fit parameters are presented in Table 6. We derived the radio jet upper limit with a jet length of $3 \farcs 7$, following our standard procedure outlined in Section 3.4. Table 6 also documents the corresponding radio flux density. The median jet parameters are shown in Table 7 and the median $\theta$ and the probability distribution of $\theta$ is shown in Figure D1. We find that a longer jet length do not give a different median $\theta$. It results in slighly higher median values of relativistic parameters like bulk Lorentz factor, Doppler factor and bulk speed. The main difference lies in the median electron number density that decreases by a factor of ten when using a longer jet. 


\begin{table}[h!]
    \centering
\caption{X-ray and radio properties of the jet.}
\label{tab:Jetpropsv1}
    \begin{tabular}{|l|c|} \hline 
         Jet Properties & J1610+1811\\ \hline 
         Observed Counts (0.5-7 keV) & 17\\
         Background Counts (0.5-7 keV) &  \\ 
         ~~   Observed X-ray background &  2 \\ 
         ~~   Scattered from core & 8 \\ 
         Net Counts (0.5-7 keV) & 15.3$\pm$2.3\\ 
         Hardness Ratio & $-0.06_{-0.32}^{0.16}$\\ 
         $\Gamma_{\rm 0.5-7~  keV}$ & $1.53^{+0.53}_{-0.52}$\\ 
         $f_{\rm 0.5-7~  keV}~ \rm (10^{-14}~ erg~ s^{-1}~ cm^{-2})$ & $0.25\pm0.07$\\ 
         $f_{\rm 1~  keV}~ (\rm nJy)$ &  $0.28\pm0.12$\\ 
         6 GHz Flux density upper-limit (mJy) & $0.28\pm0.12$\\\hline
    \end{tabular}
\end{table}

\begin{deluxetable}{lll}[h!]
\tablecaption{Estimated values of relativistic and jet parameters based on our IC/CMB model with radio jet upper-limit for J1610. \label{tab:table5app}}
\tablehead{
\colhead{Parameters}& \multicolumn{2}{c}{J1610+1811}\\
\cline{2-3}
\colhead{} & \colhead{Median} & \colhead{90\% C.I.} \\
\cline{2-3}
\colhead{(1)} & \colhead{(2)} & \colhead{(3)}
}
\startdata
$\theta \downarrow$ & 11 & (4, 17) \\
$\Gamma \uparrow$ & 4 & (3, 6) \\
$\delta \uparrow$ & 5 & (5, 3)\\
$\beta \uparrow$ & 0.96 & (0.93, 0.99) \\
B ($\mu$G) $\downarrow$	& 10 & (7, 17) \\
$n_e~ (10^{-7} \rm cm^{-3}) \downarrow$ & 0.26 & (0.14, 0.81)\\
K.E. $(10^{46} \rm erg~ s^{-1}) \downarrow$ & 0.39 & (0.11, 4.26)\\
\enddata
\end{deluxetable}

\begin{figure*}[h!]
\centerline{
\includegraphics[width=0.5\textwidth,trim=0 0 0 0]{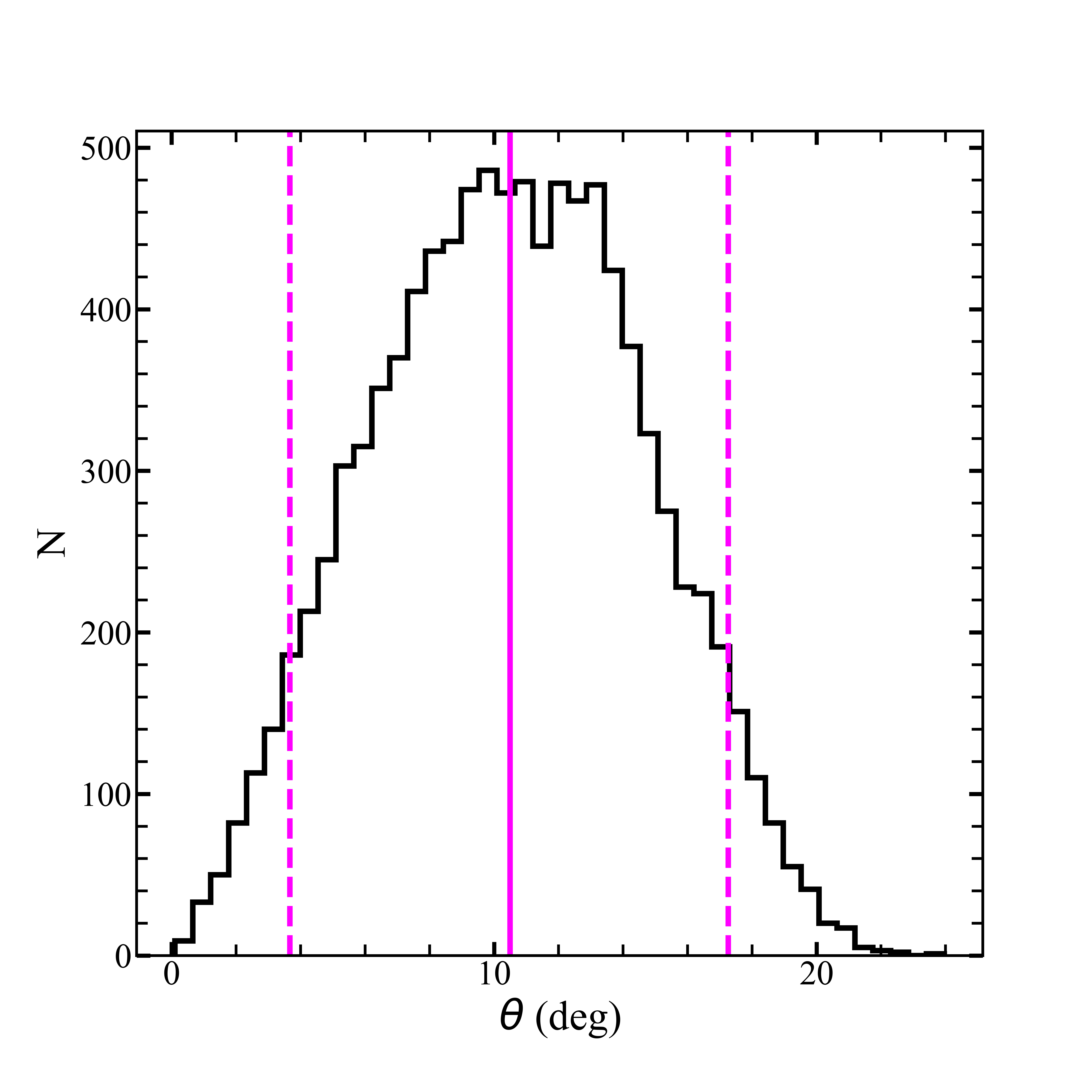}
\includegraphics[width=0.5\textwidth,trim=0 0 0 0]{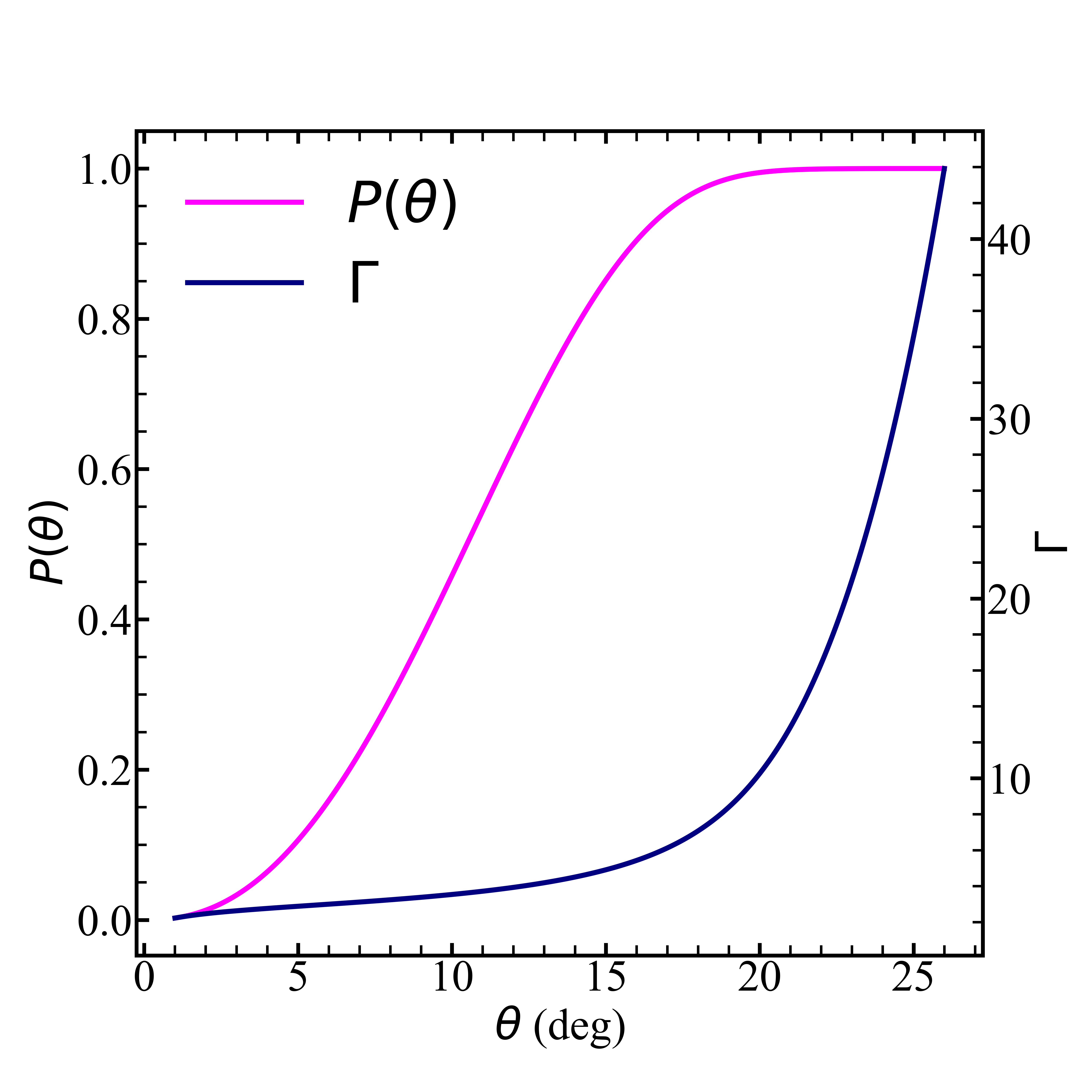}}
\caption{\small Left Distribution of line of sight angle $\theta$ for J1610+1811 based on our Bayesian method described in Section 4.2 using a $3\farcs7$ jet length covering the X-ray detection.  We use the radio flux upperlimit from a $3\farcs7 \times 0\farcs 75$ jet box region and X-ray flux measurement from a $3\farcs7 \times 1\farcs 5$ jet box. The median $\theta$ value of $\approx 11 \degr$ is marked by solid magenta lines. Dot-dash lines represent the 90\% confidence intervals. Right: Probability and bulk Lorentz factor as a function of line of sight angle shows that for steeper spectral index the probability normalizes at larger line of sight angle and results is a smaller bulk Lorentz factor.}
\label{medianthetaappen2}
\end{figure*}
\newpage

\bibliography{Highz}{}

\begin{thebibliography}{}
\expandafter\ifx\csname natexlab\endcsname\relax\def\natexlab#1{#1}\fi
\providecommand{\url}[1]{\href{#1}{#1}}
\providecommand{\dodoi}[1]{doi:~\href{http://doi.org/#1}{\nolinkurl{#1}}}
\providecommand{\doeprint}[1]{\href{http://ascl.net/#1}{\nolinkurl{http://ascl.net/#1}}}
\providecommand{\doarXiv}[1]{\href{https://arxiv.org/abs/#1}{\nolinkurl{https://arxiv.org/abs/#1}}}

\bibitem[{{Baghel} {et~al.}(2023){Baghel}, {Kharb}, {Silpa}, {Ho}, \& {Harrison}}]{Baghel2023}
{Baghel}, J., {Kharb}, P., {Silpa}, {Ho}, L.~C., \& {Harrison}, C.~M. 2023, \mnras, 519, 2773, \dodoi{10.1093/mnras/stac3691}

\bibitem[{{Bicknell}(1994)}]{Bicknell1994}
{Bicknell}, G.~V. 1994, \apj, 422, 542, \dodoi{10.1086/173748}

\bibitem[{{Breiding} {et~al.}(2017){Breiding}, {Meyer}, {Georganopoulos}, {Keenan}, {DeNigris}, \& {Hewitt}}]{Breiding+2017}
{Breiding}, P., {Meyer}, E.~T., {Georganopoulos}, M., {et~al.} 2017, \apj, 849, 95, \dodoi{10.3847/1538-4357/aa907a}

\bibitem[{{Briggs}(1995)}]{Briggs1995}
{Briggs}, D.~S. 1995, PhD thesis, New Mexico Institute of Mining and Technology

\bibitem[{{Briggs} {et~al.}(1999){Briggs}, {Schwab}, \& {Sramek}}]{Briggs+1999}
{Briggs}, D.~S., {Schwab}, F.~R., \& {Sramek}, R.~A. 1999, in Astronomical Society of the Pacific Conference Series, Vol. 180, Synthesis Imaging in Radio Astronomy II, ed. G.~B. {Taylor}, C.~L. {Carilli}, \& R.~A. {Perley}, 127

\bibitem[{{CASA Team} {et~al.}(2022){CASA Team}, {Bean}, {Bhatnagar}, {Castro}, {Donovan Meyer}, {Emonts}, {Garcia}, {Garwood}, {Golap}, {Gonzalez Villalba}, {Harris}, {Hayashi}, {Hoskins}, {Hsieh}, {Jagannathan}, {Kawasaki}, {Keimpema}, {Kettenis}, {Lopez}, {Marvil}, {Masters}, {McNichols}, {Mehringer}, {Miel}, {Moellenbrock}, {Montesino}, {Nakazato}, {Ott}, {Petry}, {Pokorny}, {Raba}, {Rau}, {Schiebel}, {Schweighart}, {Sekhar}, {Shimada}, {Small}, {Steeb}, {Sugimoto}, {Suoranta}, {Tsutsumi}, {van Bemmel}, {Verkouter}, {Wells}, {Xiong}, {Szomoru}, {Griffith}, {Glendenning}, \& {Kern}}]{CASA2022}
{CASA Team}, {Bean}, B., {Bhatnagar}, S., {et~al.} 2022, \pasp, 134, 114501, \dodoi{10.1088/1538-3873/ac9642}

\bibitem[{{Celotti} {et~al.}(2001){Celotti}, {Ghisellini}, \& {Chiaberge}}]{Celotti2001}
{Celotti}, A., {Ghisellini}, G., \& {Chiaberge}, M. 2001, \mnras, 321, L1, \dodoi{10.1046/j.1365-8711.2001.04160.x}

\bibitem[{{Cheung} {et~al.}(2012){Cheung}, {Stawarz}, {Siemiginowska}, {Gobeille}, {Wardle}, {Harris}, \& {Schwartz}}]{Cheung+2012}
{Cheung}, C.~C., {Stawarz}, {\L}., {Siemiginowska}, A., {et~al.} 2012, \apjl, 756, L20, \dodoi{10.1088/2041-8205/756/1/L20}

\bibitem[{{Comastri}(2004)}]{Comastri2004}
{Comastri}, A. 2004, in Astrophysics and Space Science Library, Vol. 308, Supermassive Black Holes in the Distant Universe, ed. A.~J. {Barger}, 245, \dodoi{10.1007/978-1-4020-2471-9_8}

\bibitem[{{Condon}(1997)}]{Condon1997}
{Condon}, J.~J. 1997, \pasp, 109, 166, \dodoi{10.1086/133871}

\bibitem[{{Daly} {et~al.}(2012){Daly}, {Sprinkle}, {O'Dea}, {Kharb}, \& {Baum}}]{Daly2012}
{Daly}, R.~A., {Sprinkle}, T.~B., {O'Dea}, C.~P., {Kharb}, P., \& {Baum}, S.~A. 2012, \mnras, 423, 2498, \dodoi{10.1111/j.1365-2966.2012.21060.x}

\bibitem[{{Dickey} \& {Lockman}(1990)}]{DickeyLockman1990}
{Dickey}, J.~M., \& {Lockman}, F.~J. 1990, \araa, 28, 215, \dodoi{10.1146/annurev.aa.28.090190.001243}

\bibitem[{{Fabian}(2012)}]{Fabian2012}
{Fabian}, A.~C. 2012, \araa, 50, 455, \dodoi{10.1146/annurev-astro-081811-125521}

\bibitem[{{Felten} \& {Morrison}(1966)}]{FeltenMorrison1966}
{Felten}, J.~E., \& {Morrison}, P. 1966, \apj, 146, 686, \dodoi{10.1086/148946}

\bibitem[{{Freeman} {et~al.}(2001){Freeman}, {Doe}, \& {Siemiginowska}}]{Freeman2001}
{Freeman}, P., {Doe}, S., \& {Siemiginowska}, A. 2001, in Society of Photo-Optical Instrumentation Engineers (SPIE) Conference Series, Vol. 4477, Astronomical Data Analysis, ed. J.-L. {Starck} \& F.~D. {Murtagh}, 76--87, \dodoi{10.1117/12.447161}

\bibitem[{{Fruscione} {et~al.}(2006){Fruscione}, {McDowell}, {Allen}, {Brickhouse}, {Burke}, {Davis}, {Durham}, {Elvis}, {Galle}, {Harris}, {Huenemoerder}, {Houck}, {Ishibashi}, {Karovska}, {Nicastro}, {Noble}, {Nowak}, {Primini}, {Siemiginowska}, {Smith}, \& {Wise}}]{Fruscione2006}
{Fruscione}, A., {McDowell}, J.~C., {Allen}, G.~E., {et~al.} 2006, in Society of Photo-Optical Instrumentation Engineers (SPIE) Conference Series, Vol. 6270, Society of Photo-Optical Instrumentation Engineers (SPIE) Conference Series, ed. D.~R. {Silva} \& R.~E. {Doxsey}, 62701V, \dodoi{10.1117/12.671760}

\bibitem[{{Ghisellini} {et~al.}(2014){Ghisellini}, {Tavecchio}, {Maraschi}, {Celotti}, \& {Sbarrato}}]{Ghisellini+2014}
{Ghisellini}, G., {Tavecchio}, F., {Maraschi}, L., {Celotti}, A., \& {Sbarrato}, T. 2014, \nat, 515, 376, \dodoi{10.1038/nature13856}

\bibitem[{{Ghosh} {et~al.}(2023){Ghosh}, {Kharb}, {Baghel}, \& {Silpa}}]{Ghosh2023}
{Ghosh}, S., {Kharb}, P., {Baghel}, J., \& {Silpa}, S. 2023, \apj, 958, 71, \dodoi{10.3847/1538-4357/acfa00}

\bibitem[{{Gobeille} {et~al.}(2014){Gobeille}, {Wardle}, \& {Cheung}}]{Gobeille2014}
{Gobeille}, D.~B., {Wardle}, J. F.~C., \& {Cheung}, C.~C. 2014, arXiv e-prints, arXiv:1406.4797, \dodoi{10.48550/arXiv.1406.4797}

\bibitem[{{Gobeille}(2011)}]{Gobeille2011}
{Gobeille}, D. B.~P. 2011, PhD thesis, Brandeis University, Massachusetts

\bibitem[{{Gofford} {et~al.}(2015){Gofford}, {Reeves}, {McLaughlin}, {Braito}, {Turner}, {Tombesi}, \& {Cappi}}]{Gofford+2015}
{Gofford}, J., {Reeves}, J.~N., {McLaughlin}, D.~E., {et~al.} 2015, \mnras, 451, 4169, \dodoi{10.1093/mnras/stv1207}

\bibitem[{{Hardcastle} \& {Croston}(2020)}]{Hardcastle+Croston2020}
{Hardcastle}, M.~J., \& {Croston}, J.~H. 2020, \nar, 88, 101539, \dodoi{10.1016/j.newar.2020.101539}

\bibitem[{{Heckman} \& {Best}(2014)}]{Heckman+2014}
{Heckman}, T.~M., \& {Best}, P.~N. 2014, \araa, 52, 589, \dodoi{10.1146/annurev-astro-081913-035722}

\bibitem[{{Ighina} {et~al.}(2022){Ighina}, {Moretti}, {Tavecchio}, {Caccianiga}, {Belladitta}, {Dallacasa}, {Della Ceca}, {Sbarrato}, \& {Spingola}}]{Ighina2022}
{Ighina}, L., {Moretti}, A., {Tavecchio}, F., {et~al.} 2022, \aap, 659, A93, \dodoi{10.1051/0004-6361/202142676}

\bibitem[{{Kashyap}(2010)}]{Kashyap2010}
{Kashyap}, V. 2010, {Analysis of Chandra PSF feature using ACIS data}, https://cxc.cfa.harvard.edu/cal/Hrc/PSF/acis\_psf\_2010oct.html

\bibitem[{{Kharb} {et~al.}(2008){Kharb}, {O'Dea}, {Baum}, {Daly}, {Mory}, {Donahue}, \& {Guerra}}]{Kharb2008}
{Kharb}, P., {O'Dea}, C.~P., {Baum}, S.~A., {et~al.} 2008, \apjs, 174, 74, \dodoi{10.1086/520840}

\bibitem[{{Kormendy} \& {Richstone}(1995)}]{Kormendy1995}
{Kormendy}, J., \& {Richstone}, D. 1995, \araa, 33, 581, \dodoi{10.1146/annurev.aa.33.090195.003053}

\bibitem[{{Lister} {et~al.}(2018){Lister}, {Aller}, {Aller}, {Hodge}, {Homan}, {Kovalev}, {Pushkarev}, \& {Savolainen}}]{Lister+2018}
{Lister}, M.~L., {Aller}, M.~F., {Aller}, H.~D., {et~al.} 2018, \apjs, 234, 12, \dodoi{10.3847/1538-4365/aa9c44}

\bibitem[{{Lister} {et~al.}(2019){Lister}, {Homan}, {Hovatta}, {Kellermann}, {Kiehlmann}, {Kovalev}, {Max-Moerbeck}, {Pushkarev}, {Readhead}, {Ros}, \& {Savolainen}}]{Lister+2019}
{Lister}, M.~L., {Homan}, D.~C., {Hovatta}, T., {et~al.} 2019, \apj, 874, 43, \dodoi{10.3847/1538-4357/ab08ee}

\bibitem[{{Ma} {et~al.}(2023){Ma}, {Elvis}, {Fabbiano}, {Balokovi{\'c}}, {Maksym}, \& {Risaliti}}]{Ma+2023}
{Ma}, J., {Elvis}, M., {Fabbiano}, G., {et~al.} 2023, \apj, 948, 61, \dodoi{10.3847/1538-4357/acba8d}

\bibitem[{{Maithil} {et~al.}(2020){Maithil}, {Runnoe}, {Brotherton}, {Wardle}, {Wills}, {DiPompeo}, \& {De Breuck}}]{Maithil+2020}
{Maithil}, J., {Runnoe}, J.~C., {Brotherton}, M.~S., {et~al.} 2020, \apj, 904, 179, \dodoi{10.3847/1538-4357/abc257}

\bibitem[{{Marscher}(1988)}]{Marscher+1988}
{Marscher}, A.~P. 1988, \apj, 334, 552, \dodoi{10.1086/166859}

\bibitem[{{Marshall} {et~al.}(2018{\natexlab{a}}){Marshall}, Gelbord, Worrall, Birkinshaw, Schwartz, Jauncey, Griffiths, Murphy, Lovell, Perlman, \& Godfrey}]{Marshall2018}
{Marshall}, H.~L., Gelbord, J.~M., Worrall, D.~M., {et~al.} 2018{\natexlab{a}}, ApJ, 856, 66, \dodoi{10.3847/1538-4357/aaaf66}

\bibitem[{{Marshall} {et~al.}(2018{\natexlab{b}}){Marshall}, {Gelbord}, {Worrall}, {Birkinshaw}, {Schwartz}, {Jauncey}, {Griffiths}, {Murphy}, {Lovell}, {Perlman}, \& {Godfrey}}]{Marshall+2018}
{Marshall}, H.~L., {Gelbord}, J.~M., {Worrall}, D.~M., {et~al.} 2018{\natexlab{b}}, \apj, 856, 66, \dodoi{10.3847/1538-4357/aaaf66}

\bibitem[{{Massaro} {et~al.}(2006){Massaro}, {Tramacere}, {Perri}, {Giommi}, \& {Tosti}}]{Massaro+2006}
{Massaro}, E., {Tramacere}, A., {Perri}, M., {Giommi}, P., \& {Tosti}, G. 2006, \aap, 448, 861, \dodoi{10.1051/0004-6361:20053644}

\bibitem[{{McKeough} {et~al.}(2016){McKeough}, {Siemiginowska}, {Cheung}, {Stawarz}, {Kashyap}, {Stein}, {Stampoulis}, {van Dyk}, {Wardle}, {Lee}, {Harris}, {Schwartz}, {Donato}, {Maraschi}, \& {Tavecchio}}]{McKeough2016}
{McKeough}, K., {Siemiginowska}, A., {Cheung}, C.~C., {et~al.} 2016, \apj, 833, 123, \dodoi{10.3847/1538-4357/833/1/123}

\bibitem[{{Meyer} {et~al.}(2017){Meyer}, {Breiding}, {Georganopoulos}, {Oteo}, {Zwaan}, {Laing}, {Godfrey}, \& {Ivison}}]{Meyer2017}
{Meyer}, E.~T., {Breiding}, P., {Georganopoulos}, M., {et~al.} 2017, \apjl, 835, L35, \dodoi{10.3847/2041-8213/835/2/L35}

\bibitem[{{Meyer} {et~al.}(2015){Meyer}, {Georganopoulos}, {Sparks}, {Godfrey}, {Lovell}, \& {Perlman}}]{Meyer2015}
{Meyer}, E.~T., {Georganopoulos}, M., {Sparks}, W.~B., {et~al.} 2015, \apj, 805, 154, \dodoi{10.1088/0004-637X/805/2/154}

\bibitem[{{Meyer} {et~al.}(2019){Meyer}, {Iyer}, {Reddy}, {Georganopoulos}, {Breiding}, \& {Keenan}}]{Meyer+2019}
{Meyer}, E.~T., {Iyer}, A.~R., {Reddy}, K., {et~al.} 2019, \apjl, 883, L2, \dodoi{10.3847/2041-8213/ab3db3}

\bibitem[{{Meyer} {et~al.}(2023){Meyer}, {Shaik}, {Reddy}, \& {Georganopoulos}}]{Meyer2023}
{Meyer}, E.~T., {Shaik}, A., {Reddy}, K., \& {Georganopoulos}, M. 2023, arXiv e-prints, arXiv:2303.08897, \dodoi{10.48550/arXiv.2303.08897}

\bibitem[{{Migliori} {et~al.}(2023){Migliori}, {Siemiginowska}, {Sobolewska}, {Cheung}, {Stawarz}, {Schwartz}, {Snios}, {Saxena}, \& {Kashyap}}]{Migliori2023}
{Migliori}, G., {Siemiginowska}, A., {Sobolewska}, M., {et~al.} 2023, \mnras, 524, 1087, \dodoi{10.1093/mnras/stad1959}

\bibitem[{{Murphy}(1988)}]{Murphy1988}
{Murphy}, D.~W. 1988, PhD thesis, University of Manchester, UK

\bibitem[{{O'Dea} {et~al.}(2009){O'Dea}, {Daly}, {Kharb}, {Freeman}, \& {Baum}}]{Odea2009}
{O'Dea}, C.~P., {Daly}, R.~A., {Kharb}, P., {Freeman}, K.~A., \& {Baum}, S.~A. 2009, \aap, 494, 471, \dodoi{10.1051/0004-6361:200809416}

\bibitem[{{Pacholczyk}(1970)}]{Pacholczyk1970}
{Pacholczyk}, A.~G. 1970, {Radio astrophysics. Nonthermal processes in galactic and extragalactic sources} (IOP ebooks. Bristol, UK: IOP Publishing)

\bibitem[{{P{\^a}ris} {et~al.}(2018){P{\^a}ris}, {Petitjean}, {Aubourg}, {Myers}, {Streblyanska}, {Lyke}, {Anderson}, {Armengaud}, {Bautista}, {Blanton}, {Blomqvist}, {Brinkmann}, {Brownstein}, {Brandt}, {Burtin}, {Dawson}, {de la Torre}, {Georgakakis}, {Gil-Mar{\'\i}n}, {Green}, {Hall}, {Kneib}, {LaMassa}, {Le Goff}, {MacLeod}, {Mariappan}, {McGreer}, {Merloni}, {Noterdaeme}, {Palanque-Delabrouille}, {Percival}, {Ross}, {Rossi}, {Schneider}, {Seo}, {Tojeiro}, {Weaver}, {Weijmans}, {Y{\`e}che}, {Zarrouk}, \& {Zhao}}]{Paris+2018}
{P{\^a}ris}, I., {Petitjean}, P., {Aubourg}, {\'E}., {et~al.} 2018, \aap, 613, A51, \dodoi{10.1051/0004-6361/201732445}

\bibitem[{{Park} {et~al.}(2006){Park}, {Kashyap}, {Siemiginowska}, {van Dyk}, {Zezas}, {Heinke}, \& {Wargelin}}]{Park2006}
{Park}, T., {Kashyap}, V.~L., {Siemiginowska}, A., {et~al.} 2006, \apj, 652, 610, \dodoi{10.1086/507406}

\bibitem[{{Planck Collaboration} {et~al.}(2016){Planck Collaboration}, {Ade, P. A. R.}, {Aghanim, N.}, {Arnaud, M.}, {Ashdown, M.}, {Aumont, J.}, {Baccigalupi, C.}, {Banday, A. J.}, {Barreiro, R. B.}, {Bartlett, J. G.}, {Bartolo, N.}, {Battaner, E.}, {Battye, R.}, {Benabed, K.}, {Beno\^{\i}t, A.}, {Benoit-L\'evy, A.}, {Bernard, J.-P.}, {Bersanelli, M.}, {Bielewicz, P.}, {Bock, J. J.}, {Bonaldi, A.}, {Bonavera, L.}, {Bond, J. R.}, {Borrill, J.}, {Bouchet, F. R.}, {Boulanger, F.}, {Bucher, M.}, {Burigana, C.}, {Butler, R. C.}, {Calabrese, E.}, {Cardoso, J.-F.}, {Catalano, A.}, {Challinor, A.}, {Chamballu, A.}, {Chary, R.-R.}, {Chiang, H. C.}, {Chluba, J.}, {Christensen, P. R.}, {Church, S.}, {Clements, D. L.}, {Colombi, S.}, {Colombo, L. P. L.}, {Combet, C.}, {Coulais, A.}, {Crill, B. P.}, {Curto, A.}, {Cuttaia, F.}, {Danese, L.}, {Davies, R. D.}, {Davis, R. J.}, {de Bernardis, P.}, {de Rosa, A.}, {de Zotti, G.}, {Delabrouille, J.}, {D\'esert, F.-X.}, {Di Valentino, E.}, {Dickinson, C.}, {Diego, J. M.}, {Dolag,
  K.}, {Dole, H.}, {Donzelli, S.}, {Dor\'e, O.}, {Douspis, M.}, {Ducout, A.}, {Dunkley, J.}, {Dupac, X.}, {Efstathiou, G.}, {Elsner, F.}, {En\ss{}lin, T. A.}, {Eriksen, H. K.}, {Farhang, M.}, {Fergusson, J.}, {Finelli, F.}, {Forni, O.}, {Frailis, M.}, {Fraisse, A. A.}, {Franceschi, E.}, {Frejsel, A.}, {Galeotta, S.}, {Galli, S.}, {Ganga, K.}, {Gauthier, C.}, {Gerbino, M.}, {Ghosh, T.}, {Giard, M.}, {Giraud-H\'eraud, Y.}, {Giusarma, E.}, {Gjerl\o{}w, E.}, {Gonz\'alez-Nuevo, J.}, {G\'orski, K. M.}, {Gratton, S.}, {Gregorio, A.}, {Gruppuso, A.}, {Gudmundsson, J. E.}, {Hamann, J.}, {Hansen, F. K.}, {Hanson, D.}, {Harrison, D. L.}, {Helou, G.}, {Henrot-Versill\'e, S.}, {Hern\'andez-Monteagudo, C.}, {Herranz, D.}, {Hildebrandt, S. R.}, {Hivon, E.}, {Hobson, M.}, {Holmes, W. A.}, {Hornstrup, A.}, {Hovest, W.}, {Huang, Z.}, {Huffenberger, K. M.}, {Hurier, G.}, {Jaffe, A. H.}, {Jaffe, T. R.}, {Jones, W. C.}, {Juvela, M.}, {Keih\"anen, E.}, {Keskitalo, R.}, {Kisner, T. S.}, {Kneissl, R.}, {Knoche, J.}, {Knox, L.},
  {Kunz, M.}, {Kurki-Suonio, H.}, {Lagache, G.}, {L\"ahteenm\"aki, A.}, {Lamarre, J.-M.}, {Lasenby, A.}, {Lattanzi, M.}, {Lawrence, C. R.}, {Leahy, J. P.}, {Leonardi, R.}, {Lesgourgues, J.}, {Levrier, F.}, {Lewis, A.}, {Liguori, M.}, {Lilje, P. B.}, {Linden-V\o{}rnle, M.}, {L\'opez-Caniego, M.}, {Lubin, P. M.}, {Mac\'{\i}as-P\'erez, J. F.}, {Maggio, G.}, {Maino, D.}, {Mandolesi, N.}, {Mangilli, A.}, {Marchini, A.}, {Maris, M.}, {Martin, P. G.}, {Martinelli, M.}, {Mart\'{\i}nez-Gonz\'alez, E.}, {Masi, S.}, {Matarrese, S.}, {McGehee, P.}, {Meinhold, P. R.}, {Melchiorri, A.}, {Melin, J.-B.}, {Mendes, L.}, {Mennella, A.}, {Migliaccio, M.}, {Millea, M.}, {Mitra, S.}, {Miville-Desch\^enes, M.-A.}, {Moneti, A.}, {Montier, L.}, {Morgante, G.}, {Mortlock, D.}, {Moss, A.}, {Munshi, D.}, {Murphy, J. A.}, {Naselsky, P.}, {Nati, F.}, {Natoli, P.}, {Netterfield, C. B.}, {N\o{}rgaard-Nielsen, H. U.}, {Noviello, F.}, {Novikov, D.}, {Novikov, I.}, {Oxborrow, C. A.}, {Paci, F.}, {Pagano, L.}, {Pajot, F.}, {Paladini, R.},
  {Paoletti, D.}, {Partridge, B.}, {Pasian, F.}, {Patanchon, G.}, {Pearson, T. J.}, {Perdereau, O.}, {Perotto, L.}, {Perrotta, F.}, {Pettorino, V.}, {Piacentini, F.}, {Piat, M.}, {Pierpaoli, E.}, {Pietrobon, D.}, {Plaszczynski, S.}, {Pointecouteau, E.}, {Polenta, G.}, {Popa, L.}, {Pratt, G. W.}, {Pr\'ezeau, G.}, {Prunet, S.}, {Puget, J.-L.}, {Rachen, J. P.}, {Reach, W. T.}, {Rebolo, R.}, {Reinecke, M.}, {Remazeilles, M.}, {Renault, C.}, {Renzi, A.}, {Ristorcelli, I.}, {Rocha, G.}, {Rosset, C.}, {Rossetti, M.}, {Roudier, G.}, {Rouill\'e d\'{}Orfeuil, B.}, {Rowan-Robinson, M.}, {Rubi\~no-Mart\'{\i}n, J. A.}, {Rusholme, B.}, {Said, N.}, {Salvatelli, V.}, {Salvati, L.}, {Sandri, M.}, {Santos, D.}, {Savelainen, M.}, {Savini, G.}, {Scott, D.}, {Seiffert, M. D.}, {Serra, P.}, {Shellard, E. P. S.}, {Spencer, L. D.}, {Spinelli, M.}, {Stolyarov, V.}, {Stompor, R.}, {Sudiwala, R.}, {Sunyaev, R.}, {Sutton, D.}, {Suur-Uski, A.-S.}, {Sygnet, J.-F.}, {Tauber, J. A.}, {Terenzi, L.}, {Toffolatti, L.}, {Tomasi, M.}, {Tristram,
  M.}, {Trombetti, T.}, {Tucci, M.}, {Tuovinen, J.}, {T\"urler, M.}, {Umana, G.}, {Valenziano, L.}, {Valiviita, J.}, {Van Tent, F.}, {Vielva, P.}, {Villa, F.}, {Wade, L. A.}, {Wandelt, B. D.}, {Wehus, I. K.}, {White, M.}, {White, S. D. M.}, {Wilkinson, A.}, {Yvon, D.}, {Zacchei, A.}, \& {Zonca, A.}}]{Planck2016}
{Planck Collaboration}, {Ade, P. A. R.}, {Aghanim, N.}, {et~al.} 2016, A\&A, 594, A13, \dodoi{10.1051/0004-6361/201525830}

\bibitem[{{Pushkarev} {et~al.}(2012){Pushkarev}, {Hovatta}, {Kovalev}, {Lister}, {Lobanov}, {Savolainen}, \& {Zensus}}]{Pushkarev+2012}
{Pushkarev}, A.~B., {Hovatta}, T., {Kovalev}, Y.~Y., {et~al.} 2012, \aap, 545, A113, \dodoi{10.1051/0004-6361/201219173}

\bibitem[{{Readhead} {et~al.}(1979){Readhead}, {Pearson}, {Cohen}, {Ewing}, \& {Moffet}}]{Readhead+1979}
{Readhead}, A.~C.~S., {Pearson}, T.~J., {Cohen}, M.~H., {Ewing}, M.~S., \& {Moffet}, A.~T. 1979, \apj, 231, 299, \dodoi{10.1086/157193}

\bibitem[{{Sambruna} {et~al.}(2004){Sambruna}, {Gambill}, {Maraschi}, {Tavecchio}, {Cerutti}, {Cheung}, {Urry}, \& {Chartas}}]{Sambruna2004}
{Sambruna}, R.~M., {Gambill}, J.~K., {Maraschi}, L., {et~al.} 2004, \apj, 608, 698, \dodoi{10.1086/383124}

\bibitem[{{Sambruna} {et~al.}(2002){Sambruna}, {Maraschi}, {Tavecchio}, {Urry}, {Cheung}, {Chartas}, {Scarpa}, \& {Gambill}}]{Sambruna2002}
{Sambruna}, R.~M., {Maraschi}, L., {Tavecchio}, F., {et~al.} 2002, \apj, 571, 206, \dodoi{10.1086/339859}

\bibitem[{{Schwartz} {et~al.}(2019){Schwartz}, {Siemiginowska}, {Worrall}, {Birkinshaw}, {Cheung}, {Marshall}, {Migliori}, {Wardle}, \& {Gobeille}}]{Schwartz+2019}
{Schwartz}, D., {Siemiginowska}, A., {Worrall}, D., {et~al.} 2019, Astronomische Nachrichten, 340, 30, \dodoi{10.1002/asna.201913554}

\bibitem[{{Schwartz} {et~al.}(2006){Schwartz}, {Marshall}, {Lovell}, {Murphy}, {Bicknell}, {Birkinshaw}, {Gelbord}, {Georganopoulos}, {Godfrey}, {Jauncey}, {Perlman}, \& {Worrall}}]{Schwartz+2006}
{Schwartz}, D.~A., {Marshall}, H.~L., {Lovell}, J.~E.~J., {et~al.} 2006, \apj, 640, 592, \dodoi{10.1086/500102}

\bibitem[{{Schwartz} {et~al.}(2020){Schwartz}, {Siemiginowska}, {Snios}, {Worrall}, {Birkinshaw}, {Cheung}, {Marshall}, {Migliori}, {Wardle}, \& {Gobeille}}]{Schwartz2020}
{Schwartz}, D.~A., {Siemiginowska}, A., {Snios}, B., {et~al.} 2020, \apj, 904, 57, \dodoi{10.3847/1538-4357/abbd99}

\bibitem[{Shen {et~al.}(2011)Shen, Richards, Strauss, Hall, Schneider, Snedden, Bizyaev, Brewington, Malanushenko, Malanushenko, Oravetz, Pan, \& Simmons}]{Shen+2011}
Shen, Y., Richards, G.~T., Strauss, M.~A., {et~al.} 2011, ApJS, 194, 45, \dodoi{10.1088/0067-0049/194/2/45}

\bibitem[{{Siemiginowska} {et~al.}(2003){Siemiginowska}, {Smith}, {Aldcroft}, {Schwartz}, {Paerels}, \& {Petric}}]{Siemiginowska+2003}
{Siemiginowska}, A., {Smith}, R.~K., {Aldcroft}, T.~L., {et~al.} 2003, \apjl, 598, L15, \dodoi{10.1086/380497}

\bibitem[{{Siemiginowska} {et~al.}(2024){Siemiginowska}, {Burke}, {G{\"u}nther}, {Lee}, {McLaughlin}, {Principe}, {Cheer}, {Fruscione}, {Laurino}, {McDowell}, \& {Terrell}}]{Sherpa2024}
{Siemiginowska}, A., {Burke}, D., {G{\"u}nther}, H.~M., {et~al.} 2024, arXiv e-prints, arXiv:2409.10400, \dodoi{10.48550/arXiv.2409.10400}

\bibitem[{{Simionescu} {et~al.}(2016){Simionescu}, {Stawarz}, {Ichinohe}, {Cheung}, {Jamrozy}, {Siemiginowska}, {Hagino}, {Gandhi}, \& {Werner}}]{Simionescu+2016}
{Simionescu}, A., {Stawarz}, {\L}., {Ichinohe}, Y., {et~al.} 2016, \apjl, 816, L15, \dodoi{10.3847/2041-8205/816/1/L15}

\bibitem[{{Snios} {et~al.}(2021){Snios}, {Schwartz}, {Siemiginowska}, {Sobolewska}, {Birkinshaw}, {Cheung}, {Gobeille}, {Marshall}, {Migliori}, {Wardle}, \& {Worrall}}]{Snios2021}
{Snios}, B., {Schwartz}, D.~A., {Siemiginowska}, A., {et~al.} 2021, \apj, 914, 130, \dodoi{10.3847/1538-4357/abfe64}

\bibitem[{{Snios} {et~al.}(2022){Snios}, {Schwartz}, {Siemiginowska}, {Sobolewska}, {Birkinshaw}, {Cheung}, {Gobeille}, {Marshall}, {Migliori}, {Wardle}, \& {Worrall}}]{Snios+2022}
---. 2022, \apj, 934, 107, \dodoi{10.3847/1538-4357/ac7cf2}

\bibitem[{{Sowards-Emmerd} {et~al.}(2005){Sowards-Emmerd}, {Romani}, {Michelson}, {Healey}, \& {Nolan}}]{Sowards+2005}
{Sowards-Emmerd}, D., {Romani}, R.~W., {Michelson}, P.~F., {Healey}, S.~E., \& {Nolan}, P.~L. 2005, \apj, 626, 95, \dodoi{10.1086/429902}

\bibitem[{{Tavecchio} {et~al.}(2000){Tavecchio}, {Maraschi}, {Sambruna}, \& {Urry}}]{Tavecchio2000}
{Tavecchio}, F., {Maraschi}, L., {Sambruna}, R.~M., \& {Urry}, C.~M. 2000, \apjl, 544, L23, \dodoi{10.1086/317292}

\bibitem[{Team {et~al.}(2022)Team, Bean, Bhatnagar, Castro, Meyer, Emonts, Garcia, Garwood, Golap, Villalba, Harris, Hayashi, Hoskins, Hsieh, Jagannathan, Kawasaki, Keimpema, Kettenis, Lopez, Marvil, Masters, McNichols, Mehringer, Miel, Moellenbrock, Montesino, Nakazato, Ott, Petry, Pokorny, Raba, Rau, Schiebel, Schweighart, Sekhar, Shimada, Small, Steeb, Sugimoto, Suoranta, Tsutsumi, van Bemmel, Verkouter, Wells, Xiong, Szomoru, Griffith, Glendenning, \& Kern}]{CASA}
Team, T.~C., Bean, B., Bhatnagar, S., {et~al.} 2022, Publications of the Astronomical Society of the Pacific, 134, 114501, \dodoi{10.1088/1538-3873/ac9642}

\bibitem[{{Tramacere}(2020)}]{Tramacere2020}
{Tramacere}, A. 2020, {JetSeT: Numerical modeling and SED fitting tool for relativistic jets}, Astrophysics Source Code Library, record ascl:2009.001

\bibitem[{{Tramacere} {et~al.}(2009){Tramacere}, {Giommi}, {Perri}, {Verrecchia}, \& {Tosti}}]{Tramacere+2009}
{Tramacere}, A., {Giommi}, P., {Perri}, M., {Verrecchia}, F., \& {Tosti}, G. 2009, \aap, 501, 879, \dodoi{10.1051/0004-6361/200810865}

\bibitem[{{Tramacere} {et~al.}(2011){Tramacere}, {Massaro}, \& {Taylor}}]{Tramacere+2011}
{Tramacere}, A., {Massaro}, E., \& {Taylor}, A.~M. 2011, \apj, 739, 66, \dodoi{10.1088/0004-637X/739/2/66}

\bibitem[{{Worrall} \& {Birkinshaw}(2006)}]{Worrall+Birkinshaw2006}
{Worrall}, D.~M., \& {Birkinshaw}, M. 2006, in Physics of Active Galactic Nuclei at all Scales, ed. D.~{Alloin}, Vol. 693 (Springer Berlin, Heidelberg), 39, \dodoi{10.1007/3-540-34621-X_2}

\bibitem[{{Worrall} {et~al.}(2020){Worrall}, {Birkinshaw}, {Marshall}, {Schwartz}, {Siemiginowska}, \& {Wardle}}]{Worrall+2020}
{Worrall}, D.~M., {Birkinshaw}, M., {Marshall}, H.~L., {et~al.} 2020, \mnras, 497, 988, \dodoi{10.1093/mnras/staa1975}

\bibitem[{{Yang} {et~al.}(2008){Yang}, {Gurvits}, {Lobanov}, {Frey}, \& {Hong}}]{Yang+2008}
{Yang}, J., {Gurvits}, L.~I., {Lobanov}, A.~P., {Frey}, S., \& {Hong}, X.~Y. 2008, \aap, 489, 517, \dodoi{10.1051/0004-6361:200809846}

\end{thebibliography}
\bibliographystyle{aasjournal}



\end{document}